\documentclass[12pt,reqno]{article}
\usepackage{times}
\usepackage[utf8]{inputenc}
\usepackage{hyperref}
 
\usepackage{amsmath}
\usepackage{amsthm}
\usepackage{amsfonts}
\usepackage{amssymb}
\usepackage{epic}
\usepackage{eepic,epsfig}
\usepackage{times, macros}
\usepackage{graphics}
\usepackage{epsfig,multicol,bbm}
\usepackage{subcaption}
\usepackage{bbm}
\usepackage{eufrak}

\usepackage{xcolor}
\usepackage{graphicx}

\theoremstyle{plain}

\theoremstyle{remark}

\newcommand{\sst}{\scriptscriptstyle}

\newcommand{\ot}{\otimes}
\newcommand{\ra}{\to}
\newcommand{\fr}[2]{{\textstyle \frac{#1}{#2} }}

\renewcommand{\ss}{\mathsf{s}}

\def\beq{\begin{equation}}
\def\ee{\end{equation}}
\def\bea{\begin{eqnarray}}
\def\eea{\end{eqnarray}}

\def\cN{\mathcal{N}}

\def\m{\mathrm{m}}

\newcommand{\al}{\alpha}

\newcommand{\2}{\two}

\newcommand{\bx}{\mathsf{x}}

\newcommand{\cx}{\check{x}}
\newcommand{\ca}{\check{a}}

\newcommand{\bX}{\mathsf{X}}
\newcommand{\cbx}{\check{\mathsf{x}}}

\newcommand{\be}{\beta}
\newcommand{\ga}{\gamma}
\newcommand{\Ga}{\Gamma}
\newcommand{\de}{\delta}

\newcommand{\ep}{\epsilon}
\newcommand{\la}{\hbar}

\newcommand{\si}{\sigma}

\newcommand{\vf}{\varphi}

\newcommand{\pa}{\partial}
\newcommand{\CA}{{\mathcal A}}
\newcommand{\CB}{{\mathcal B}}
\newcommand{\CC}{{\mathcal C}}
\newcommand{\CD}{{\mathcal D}}
\newcommand{\CE}{{\mathcal E}}
\newcommand{\CF}{{\mathcal F}}
\newcommand{\CG}{{\mathcal G}}
\newcommand{\CH}{{\mathcal H}}
\newcommand{\CI}{{\mathcal I}}
\newcommand{\CJ}{{\mathcal J}}
\newcommand{\CL}{{\mathcal L}}
\newcommand{\CM}{{\mathcal M}}
\newcommand{\CN}{{\mathcal N}}
\newcommand{\CO}{{\mathcal O}}  
\newcommand{\CP}{{\mathcal P}}  
  
\newcommand{\CR}{{\mathcal R}}
\newcommand{\CS}{{\mathcal S}}
\newcommand{\CT}{{\mathcal T}}
\newcommand{\CU}{{\mathcal U}}
\newcommand{\CV}{{\mathcal V}}
\newcommand{\CW}{{\mathcal W}}
\newcommand{\CX}{{\mathcal X}}
\newcommand{\CY}{{\mathcal Y}}
\newcommand{\CZ}{{\mathcal Z}}

\newcommand{\sa}{{\mathsf a}}

\newcommand{\sd}{{\mathsf d}}
\newcommand{\se}{{\mathsf e}}

\newcommand{\sv}{{\mathsf v}}

\newcommand{\sx}{{\mathsf x}}

\newcommand{\two}{{\mathfrak 2}}

\newcommand{\BR}{{\mathbb R}}

\newcommand{\BC}{{\mathbb C}}

\newcommand{\BT}{{\mathbb T}}
\newcommand{\BZ}{{\mathbb Z}}

\newcommand{\rf}[1]{(\ref{#1})}

\begin{document}\thispagestyle{empty}
\title{From quantum curves to topological string partition functions II}
\author{Ioana Coman$^{1}$,  Pietro Longhi$^{2}$, J\"org Teschner$^{3,4}$}
\address{
$^{1}$ Institute of Physics, 
University of Amsterdam,\\
1098 XH Amsterdam, the Netherlands\\[2ex]
$^{2}$ Institut f\"ur Theoretische Physik, 
ETH Z\"urich, Switzerland\\
Wolfgang-Pauli-Str. 27, 8093 Z\"urich\\[2ex]
$^{3}$ Department of Mathematics, 
University of Hamburg, \\
Bundesstrasse 55,
20146 Hamburg, Germany,\\[2ex]
$^{4}$ DESY theory, \\
Notkestrasse 85,
20607 Hamburg,
Germany,}
\maketitle

\begin{quote}
\begin{center}{\bf Abstract}\end{center}
{\small
We propose a geometric characterisation of the topological string partition functions associated to the local Calabi-Yau (CY) manifolds used in the geometric engineering of $d=4$, $\mathcal{N}=2$ supersymmetric field theories of class $\mathcal{S}$. A quantisation of these CY manifolds defines differential operators called quantum curves.
The partition functions are extracted from the isomonodromic tau-functions associated to the quantum curves  by  expansions of generalised theta series type. It turns out that the partition functions are in one-to-one correspondence with preferred coordinates on the moduli spaces of quantum curves defined using the Exact WKB method. The coordinates defined in this way jump across certain loci in the moduli space. The changes of normalization of the tau-functions associated to these jumps define a natural line bundle playing a key role in the geometric characterisation of the topological string partition functions proposed here.
}
\end{quote}

\setcounter{tocdepth}{2}
\tableofcontents
\newpage

\section{Introduction}
\setcounter{equation}{0}

The goal of this paper is to propose a non-perturbative geometric definition of the topological string partition functions, 
at least for an interesting family of examples. 
The examples considered in our paper are Calabi-Yau (CY) manifolds $\CY_\Sigma$ defined 
by equations of the form $uv+f_\Sigma^{}(x,y)=0$, with 
$f_\Sigma^{}(x,y)\equiv y^2+q(x)=0$ being the equation defining a curve $\Sigma$ in the 
cotangent bundle of a Riemann surface $C$ equipped with a quadratic differential $q=q(x)(dx)^2$.
This class of local CY appears in the geometric engineering of $d=4$, $\mathcal{N}=2$ supersymmetric field theories
of class $\CS$ from string theory. 
We will in this paper define a global geometric object, a holomorphic line bundle on a certain moduli space 
to be specified in detail below, having a canonical section. 
The functions  locally representing the canonical 
section admit series expansions whose coefficients 
are in some cases shown, and in general 
conjectured to coincide with previously proposed definitions of 
topological string partition function.


\subsection{Previous approaches to topological string partition functions}

We shall begin by briefly reviewing previous approaches to topological string partition functions. 
As the subject has grown very large, we will refrain from offering a guide to the literature, and discuss the most prominent directions of research in this field only very briefly.

\subsubsection{Formal series definitions}

The worldsheet-approach to string theory suggests to define the 
topological string partition functions as a formal series in powers of the topological string coupling $\la$. 
Expanding the coefficients of this series as a formal series in 
exponentials of the complexified K\"ahler parameters yields a formal series admitting a rigorous 
definition as generating function of the Gromov-Witten (GW) invariants, see e.g. \cite{Mir}
for a review.

The holomorphic anomaly equations \cite{BCOV}  yield  relations among the coefficients of this
formal series. By supplying  boundary conditions one may  solve these equations recursively, 
but the explicit form of the solutions is unknown, in general.  
Mirror symmetry suggests an interpretation of the holomorphic anomaly equations in the context of the  quantisation of the 
theory of deformations of complex structures of the 
mirror CY (Kodaira-Spencer theory) \cite{BCOV,Wi93}.

There exist identities of formal power series relating the generating functions of GW invariants
to other formal power series in exponential or trigonometric functions of $\la$. The coefficients of the resulting 
formal series are expected, and in some cases rigorously known to admit an interpretation as other
types of geometric invariants known as Gopakumar-Vafa \cite{GV}, or Donaldson-Thomas (DT) invariants \cite{MNOP,MOOP,PP}. There exist
elaborate mathematical tools for the computation of the DT invariants known as wall-crossing formulae \cite{KSoi}.
Combined with algebro-geometric or other input it is often possible to
compute  topological string partition functions with the help of the above-mentioned relations. 

The resulting series sometimes admit various (partial) summations, depending on the class
of  Calabi-Yau manifolds. For toric CY one can use the
{\it topological vertex} \cite{AKMV} as a tool to derive  power series representations 
having elementary functions formed out of $\la$
and the K\"ahler parameters as coefficients. These series are known to be convergent in 
some cases, but not in general.  The {\it topological recursion} \cite{EO1} yields a definition of the
coefficients in the expansion of these
topological string partition functions as a power series in  $\la$ \cite{BKMP,EO2}.

\subsubsection{Non-perturbative approaches}\label{non-pertTS}

All these approaches yield formal series in $\la$, or in functions of $\la$. Finding actual functions 
of $\la$ 
having the above-mentioned formal series as asymptotic expansions is expected to be very rewarding.
This should offer profound insights into the non-perturbative effects of string theory on the side of physics,
and it could reveal deep relations between different types of geometric invariants on the side of mathematics. 
Such hopes have motivated a long series of efforts aiming at non-perturbative definitions
of topological string partition functions.

A first class of candidates are the matrix models, see \cite{Ma05} for a review. One considers  certain families of 
integrals having $N$ integration variables, regarding $N$ as a variable. These
families of integrals may admit asymptotic expansions in $1/N$ 
having the same form as the topological string partition functions. One may therefore interpret  these families of 
integrals, regarded as functions of $N$, as non-perturbative definitions of the topological string partition 
functions restricted to lattices in the space of K\"ahler parameters with lattice spacing $\la$. 

A more recent research program, often referred to as the TS/ST (topological string/spectral theory) correspondence, 
proposes spectral determinants of certain finite difference operators as non-perturbative topological string partition 
functions for toric CY, see \cite{Ma15} for a review. The finite difference operators are obtained by quantisation of the equations for the algebraic
curves characterising the mirrors of the toric CY.  The spectral determinants forming the basis of this approach
are entire functions of the complex structure moduli of the mirror CY. 

There exists a lot of supporting evidence for both matrix model approach and the TS/ST correspondence. Rigorous proofs
do not seem to be available yet, in general.

Yet another approach starts with the theory of DT-invariants which can be used  as basic data defining certain 
complex-symplectic manifolds from the solutions to problems of Riemann-Hilbert type 
\cite{Br,Br19}. The geometric structures of these 
manifolds  can be encoded in certain generating 
functions. Explicit computations in simple examples indicate that these generating functions can serve as
non-perturbative topological string partition functions
\cite{Br,Br20}. 

\subsubsection{Relations to integrable hierarchies and free fermions}

Local CY manifolds like $\CY_{\Sigma}$ are determined by two-dimensional 
surfaces $\Sigma$. The deformations of the complex structures of such manifolds form the basis of the 
approach taken in \cite{ADKMV}, suggesting relations to integrable hierarchies, and to the theory of 
free fermions on $\Sigma$.

A variant  of these relations of particular relevance both  for the previous paper \cite{CPT}, and for our work
has been proposed in \cite{DHSV}. 
It considers  the dual partition functions 
\begin{align}
Z_{\rm D}(\xi,a;\la)
&=\sum_{p\in H^2(Y,\BZ)}e^{p\xi}Z_{\rm top}(a+p\la,\la),
\label{Ftransf-1}\end{align}
where $Z_{\rm top}(a,\la)$ is the topological 
string partition function associated to $\CY_\Sigma$, the variables $a$ being  
parameters determining the surface $\Sigma$. String-theoretical duality conjectures predict that
$Z_{\rm D}(\xi,a;\la)$ can be identified with the partition function 
of a system of two-dimensional chiral free fermions  on 
a non-commutative deformation of the curve $\Sigma$  \cite{DHSV}. 
It has furthermore been proposed in \cite{DHSV,DHS} that
the deformation of $\Sigma$ can be described 
by an object called quantum curve, in our case represented by a differential operator of the form $\la^2\pa_x^2-q(x)$,
obtained from  the curve $\Sigma$ by replacing $y$ by $-\mathrm{i}\la\pa_x$.

\subsection{Summary of our approach}


Motivation for the proposal made in in the companion paper \cite{CPT} came from the relations to chiral free fermions  proposed
in \cite{DHSV}, as  mentioned above. {
In order  to define topological string partition functions from the quantum curve it turns out to be
useful to consider the dual partition function $Z_{\rm D}(\xi,a;\la)$, for the following reason.
Although the topological string partition function can be defined at large volume by existing techniques, naive analytic continuation to other regions of moduli space would yield incorrect results.
It turns out that the object having fairly simple analytic properties on moduli space is not the topological string partition function itself, 
but rather the dual partition function defined in \rf{Ftransf-1}. 
Different local definitions of the dual partition functions are related by multiplication with holomorphic
transition functions. The study of these issues was initiated in \cite{CPT} and completed in the present paper.

It was proposed in \cite{CPT} to
replace the quadratic differential $q$ defining the quantum curve  by 
a natural $\xi$-dependent deformation $q_{\xi,\la}\equiv q_{\xi,\la}(x)(dx)^2$ such that $q_{\xi,\la}-q=\CO(\la)$.}
Using the formalism of free fermion conformal field theory (CFT) one can 
define partition functions $Z_{\rm D}$ naturally associated to the quantum curves $q_{\xi,\la}$. The conformal 
Ward identities imply that the partition functions $Z_{\rm D}(\xi,a;\la)$ coincide with well-known 
objects called the isomonodromic tau-functions if one identifies the variables $\xi$ and $a$ with 
coordinates on the moduli spaces $\CM_{\rm ch}(C)$ 
of flat $SL(2)$-connections on the underlying Riemann surface $C$.
The isomonodromic tau-functions are important ingredients 
of the theory of mondromy preserving deformations of ordinary differential equations  \cite{JMU}. Such deformations can
often be represented as Hamiltonian flows with respect to a canonical Poisson structure. 
The tau-functions then serve as a generating function for the Hamiltonian functions. 
However, it turns out that this formalism
determines the dependence of the partition functions $Z_{\rm D}$
on the variables $a$ and $\xi$ only incompletely.

The previous paper \cite{CPT} made a first step to fix the remaining freedom. 
A key observation made in \cite{CPT}
was a direct correspondence between certain special choices of 
the coordinates $\xi$ on the one hand, and fully normalised partition functions
$Z_{\rm D}(\xi,a;\la)$ admitting series expansions of the form \rf{Ftransf-1}
on the other hand. This offers a way to fix the freedom remaining after having
identified $Z_{\rm D}(\xi,a;\la)$ as an isomonodromic tau-function by identifiying the 
proper choices of coordinates $a$ and $\xi$.  In \cite{CPT} it was furthermore observed in
the some examples associated to surfaces $C$ of genus zero 
that the  choice of coordinates $(a,\xi)$ must depend on the choice of a chamber
in the space of quadratic differentials, in general.  A natural procedure for fixing this dependence 
can be based on
the formalism called abelianisation in \cite{HN}. The 
functions $Z_{\rm top}(a,\la)$ defined  using \rf{Ftransf-1} from the partition functions
$Z_{\rm D}(\xi,a;\la)$ fully defined in this way 
were shown in some examples to coincide with the topological string partition functions
computed using the topological vertex, chamber by chamber \cite{CPT}.


This paper completes the approach proposed in \cite{CPT}.  
We use the fact  that the so-called exact WKB method allows us to define an distinguished atlas of  systems of 
coordinates $(a,\xi)$ on the moduli spaces of  flat connections on Riemann surfaces. 
Our main result is that to each coordinate system defined in 
this way there exist  fully normalised partition functions
$Z_{\rm D}(\xi,a;\la)$ admitting series representations
of the form \rf{Ftransf-1}. Two 
related types of coordinates are relevant, called coordinates of 
Fenchel-Nielsen (FN) and Fock-Goncharov (FG) type, respectively.

To this aim we observe that the relations between any two  dual partition functions
$Z_{\rm D}(\xi,a;\la)$ and $Z_{\rm D}'(\xi',a';\la)$ 
associated to different choices of coordinates $(a,\xi)$ and $(a',\xi')$, respectively, 
are  very special, represented by the {\it difference generating functions} of the 
changes of coordinates from $(a,\xi)$ and $(a',\xi')$. Difference generating functions
are analogs of the usual generating functions from Poisson geometry 
defined by relations involving finite difference operators instead of derivatives.  
The fact that  $Z_{\rm D}(\xi,a;\la)$ is related to $Z_{\rm D}'(\xi',a';\la)$ by multiplication 
with a difference generating function ensures the both $Z_{\rm D}(\xi,a;\la)$ and 
$Z_{\rm D}'(\xi',a';\la)$ can admit series expansions of the form \rf{Ftransf-1}. 
We are going to outline a proof for the existence of  difference generating functions
for any pair of coordinates $(a,\xi)$ and $(a',\xi')$ on the moduli spaces $\CM_{\rm ch}(C)$  of flat connections
on Riemann surfaces defined by the exact WKB method. 

The
collection of such difference generating functions defines natural line bundles on the 
moduli spaces $\CM_{\rm ch}(C)$,
with $Z_{\rm D}$ being local  holomorphic sections of this line bundle. This geometric 
object is proposed to define  the topological string partition functions 
by means of the expansions \rf{Ftransf-1}. This proposal is supported by 
detailed checks performed in \cite{CPT} and here.

\subsection{Relations to the existing literature}

When we submitted the first version of this paper to the arXiv, the relations between
previous approaches to the definition of non-perturbative partition functions and our approach
have been mostly unclear to us. By combining previously known results with some more recent developments we may now offer 
a brief discussion of some of the relations. As the relations to the isomonodromic tau-functions play an important role for us,
we will begin with a brief discussion of the relations between the literature
on isomonodromic tau-functions and our results. 

\subsubsection{Relations to the literature on isomonodromic tau-functions}

It has been a long-standing problem to fix the dependence of the isomonodromic tau-functions
on the monodromy data in a natural way.  
A set of equations fixing the monodromy dependence of the tau-functions has been
proposed in  \cite{ILP}. The geometric meaning of these equations has been clarified
in the work of Bertola-Korotkin, identifying the tau-functions as generating functions
of the monodromy symplectomorphisms \cite{BK}.

A central  ingredient of our 
approach is another, a priori different, definition of fully normalised isomonodromic tau-functions. 
Our approach in \cite{CPT} and
the present paper is based on the discovery of \cite{GIL} that the tau-functions $Z_{\rm D}(\xi,a;\la)$
can be represented in the form \rf{Ftransf-1}, with $Z_{\rm top}(a,\la)$ identified with the conformal blocks of the 
Virasoro algebra at $c=1$. The conjecture of \cite{GIL} has been proven in \cite{ILT,BS,GL,Ne} by remarkably different methods. 

One of the main points here is to introduce a global object on the space of monodromy data $\CM$ which is locally
represented by the partition functions $Z_{\rm D}(\xi,a;\la)$ admitting expansion of the form \rf{Ftransf-1}
associated to local systems of coordinates $(\xi,a)$ defined on certain chambers 
in $\CM$ forming a cover of this space.  It is crucial  to ensure that changes of coordinates from chamber to chamber 
preserve the existence of expansions of the form \rf{Ftransf-1}. We will explain in our paper how this principle can be used to 
fix the dependence of the tau-functions $Z_{\rm D}(\xi,a;\la)$ on the monodromy data.

Both \cite{ILP,BK} and the approach taken in this paper
define natural line bundles with a holomorphic connection on the space of monodromy data. 
However, the relation between these two line bundles was not clear to us at the time of the 
submission of the first version of our paper. The transition functions characterising the line bundle in the
approach of \cite{ILP,BK} are generating functions of the symplectomorphisms relating different sets of 
Darboux coordinates on $\CM$. Our approach characterises the transition function as {\it difference} generating 
functions of the same symplectomorphisms, functions characterised as solutions to finite difference equations
rather than the differential equations determining ordinary generating functions up to constants.

Remarkably, it turns out that the results of these two, a priori  different looking
approaches are essentially equivalent. One may note  that
the precise relation of the approach of \cite{ILP,BK}  to the 
approach proposed in our paper, based on the series expansions  \rf{Ftransf-1}, follows for the
typical case of $C=C_{0,4}$ from
the results of \cite{Ne,DDG}. These results establish the precise relation between 
an isomonodromic tau-function admitting expansions of the form \rf{Ftransf-1} and the generating function of the 
monodromy symplectomorphism.
One may furthermore check by explicit 
computation 
for elementary coordinate transformations that ordinary and difference generating 
functions of the same symplectomorphism are closely related, as is necessary for the mutual consistency of the 
results mentioned above.  Within the framework described in this paper one can use these results in order to 
establish the equivalence of the approaches  \cite{ILP,BK}  to the one used here. These relations
should be discussed in more detail.\footnote{We started preparing a short note elaborating on this point.} 

\subsubsection{Comparison with Bridgeland's program and the TS/ST-correspondence}

At the time of submission of the first version of this paper it was unclear to us how 
the approach of  T. Bridgeland  is related to our approach. More recent developments
indicate that the approach of Bridgeland, specialised to the spaces of  stability conditions on spaces of 
quadratic differentials defined in \cite{BrS}, 
ultimately leads to the same candidates for non-perturbative 
topological string partition functions as proposed here. The complex-symplectic structures from the theory 
of DT-invariants admit natural generating functions called $\tau$-functions in \cite{Br23}. 
It was furthermore shown 
in \cite{Br23} that specialising these generating functions to the spaces of 
quadratic differentials defined in \cite{BrS}  yields 
the isomonodromic tau-functions.

Comparing our approach with the TS/ST-correspondence is still not  straightforward. 
Both approaches use differential operators called quantum curves as key ingredients. They differ
substantially in the way how the quantum curves are used to define candidates for non-perturbative 
partition functions. While the TS/ST-correspondence is based on the spectral problems of the quantum curves,
our approach studies isomonodromic deformations of the quantum curves. 

There are not very many
cases where a direct comparison of the results is possible. 
A large part of the literature on the TS/ST-correspondence studies the mirrors of toric CY.
Some of these local CY admit a  limit recovering the local CY $\CY_{\Sigma}$
studied here if the underlying Riemann surfaces $C$ has genus $g=0$ or $g=1$. How to recover 
the  local CY $\CY_{\Sigma}$ associated to surfaces $C$ with $g>1$ in this way does not seem to be well-understood
at the moment. The number of cases where a direct comparison with our approach is possible is therefore somewhat limited.

It has been demonstrated in \cite{BGT} that the tau-function of the  Painlev\'e III equation
is related to a limit of the spectral determinants appearing in the TS/ST correspondence by 
specialising a part of the variables the tau-function depends on.
Variants of the  TS/ST-correspondence for 
local CY $\CY_{\Sigma}$ associated to some surfaces of genus $g=0$ and $g=1$ have been studied in \cite{GGM,BGG},
exhibiting relations between 
the spectrum of differential operators of Mathieu and Calogero-Moser type and isomonodromic
tau-functions.

These results indicate  that the TS/ST-correspondence is deeply related to the approach 
proposed in our paper. 
However, whenever one can compare results concretely, there seem to exist different types of
representations of the partition functions as Fredholm determinants, directly related to the isomonodromic 
tau-functions  in some cases, less directly in others.
An important conceptual difference is the feature that the spectral determinants considered in the
TS/ST correspondence are entire functions on the complex structure moduli spaces of the local CY,
whereas our approach considers locally defined partitions functions representing sections of 
a line bundle on a fibration over the complex structure moduli spaces of such CY.

\subsubsection{Non-perturbative effects and resurgence}

The matrix model approach has offered important  insights into non-perturbative effects in topological string theory, 
see \cite{Ma06,MSW} for some early work in this direction, relating them to D-brane effects in some examples.
In \cite{CESV} it has been proposed that ideas from the theory of resurgence could allow one to compute such effects more explicitly 
using trans-series as an ansatz for the solution of the holomorphic anomaly equations.

Ideally one might hope that a canonical definition of non-perturbative topological string partition functions could be found 
by applying a summation method like Borel summation to the formal generating series  of the GW invariants. 
One may expect Stokes phenomena relating different locally defined summations.
An instructive example illustrating this scenario is the resolved conifold. 
The Borel summation of the topological string partition function indeed exhibits Stokes phenomena in this case \cite{PS,HO15,ASTT}.
It has been shown in  \cite{ASTT} that
the Stokes jumps coincide with the difference generating functions of the corresponding changes of
coordinates in the Riemann-Hilbert type problem defined by Bridgeland \cite{Br20} in this case.

The trans-series solutions of the holomorphic anomaly equation 
have recently been used to find exact results for the Stokes jumps in a large class of examples including even 
topological string theories on compact CY \cite{GM22,GKKM}.
Based on these results an explicit general formula for was found in \cite{IM} for the Stokes jumps of the dual partition 
functions. 

Concerning the relation between the results of \cite{IM} and the approach taken in our paper one may note that the coordinates
$(\xi,a)$ defined by exact WKB are known as quantum periods in the literature on topological string theory.
The Stokes jumps of these coordinates are represented by formulae known from the theory of 
cluster algebras \cite{DDP,Al19}.  These  jumps get interpreted as formulae for changes of 
coordinates in our approach. 
The consistency of the expansions  \rf{Ftransf-1} with the Stokes jumps of the quantum periods requires that 
the Stokes jumps of $Z_{\rm D}(\xi,a;\la)$ must be given by the difference generating functions of the corresponding
changes of coordinates. 
The results of  \cite{IM} establish the resulting characterisation of the Stokes jumps more directly. 

The results of \cite{IM} may therefore be taken as support for the conjecture
 that the geometric 
characterisation of dual partition functions based on
the symplectic geometry of the string theory moduli spaces 
proposed here
can be generalised to much larger classes of CY manifolds.

\subsection{Contents}

We start by introducing a somewhat more general class of quantum curves compared to \cite{CPT}
in Section  \ref{Sec:QCurves}, having irregular singularities of the simplest type. 
In the following Section \ref{Sec:Theta} we begin by revisiting known
results on expansions of the form \rf{Ftransf-1} for the examples with regular and irregular singularities
from the perspective of \cite{CPT}.
We show that the variables appearing in some of these expansions admit a geometric 
interpretation as coordinates of Fock-Goncharov (FG) type associated to triangulations of the surface $C$
rather than the coordinates of Fenchel-Nielsen (FN) type associated to pants decompositions of $C$ encountered in 
\cite{CPT}. 

The following Section 4 discusses the relations between the expansions of the form \rf{Ftransf-1}
associated to different types of coordinates in the two basic examples $C=C_{0,2}$ with two 
irregular singularities, and $C=C_{0,4}$ with four regular singularities.
It turns out that the change of coordinates formulae for the case of $C=C_{0,2}$  can serve as 
a building block for more complicated cases like $C=C_{0,4}$, a result that will be crucial later.

Both types of coordinates can be naturally described using abelianisation. 
At the beginning of Section \ref{sec:exact-WKB} we review how the relevant coordinates 
can  be characterised in terms of the Exact WKB method 
using the Borel summation of the WKB expansions of the monodromy data called Voros symbols.
We then demonstrate in a basic example that the coordinates of FN-type can be understood as limits 
of coordinates of FG-type. The following Section \ref{Rev-C04} describes how one
can associate normalised tau-functions to the coordinates defined using the Exact WKB method. 

The experience gathered from the examples studied up to this point then allows us to outline how 
the characterisation of the normalised partition functions should work more generally. 
Sections \ref{sp-qcurves}--\ref{tau-sec} describe the generalisation of our proposal for the cases 
associated to Riemann surfaces $C_{g,n}$ 
of arbitrary genus $g$ and $n$ punctures. 

We conclude by presenting a summary and discussing relations
to several  other directions of research in Section \ref{Sec:summary}.

\section{Quantum curves}\label{Sec:QCurves}
\setcounter{equation}{0}

{In this section we will review how to define the deformed quantum curves 
$\la^2\pa_x^2-q_{\xi,\la}(x)$ playing a basic role in our approach. Apart from clarifying 
some aspects of the framework presented in \cite{CPT} that will become relevant later, 
we will furthermore explain how to generalise this approach to a simple case with 
irregular singularities by taking
a certain collision limit. This case will later exhibit important new features. We will 
then briefly explain how CFT allows one to define the corresponding partition functions both in the 
regular and the irregular cases.}

\subsection{Classical curves}

The classical Seiberg-Witten curve $\Sigma$ for class 
$\mathcal{S}$-theories associated to Riemann surfaces 
$C_{0,n}=\mathbb{P}^1\backslash\{z_1,\ldots ,z_n\}$ of genus zero with 
$n$ regular punctures is
defined as a double cover of the $n$-punctured sphere defined 
 by the equation 
\begin{equation}\label{eq:classical-curve-def}
\Sigma=\{(x,y)\subset T^*C_{0,n} \, | \, y^2+q(x) = 0\} ~,~\text{where}~~ 
q(x)=\sum_{r=1}^n\left(\frac{a_r}{(x-z_r)^2}+\frac{E_r}{x-z_r}\right)~.
\ee
The parameters $E_r$ satisfy three linear equations 
ensuring that the quadratic differential  $q(x)d^2x$ is regular at $x\to\infty$. For the case  $n=4$
one thereby finds that only one 
of these parameters is independent. Taking into account these constraints, the Seiberg-Witten curve $\Sigma$
corresponding to the $SU(2)$ SYM theory with four flavours considered in \cite{CPT} can be represented as 
\begin{align} \label{eq:curveC04}
q(x) =& \frac{a_1^2}{(x-z_1)^2} + \frac{a_2^2}{(x-z_2)^2} + 
\frac{a_3^2}{(x-z_3)^2} - \frac{\kappa}{(x-z_1)(x-z_3)} \\
&\quad+ \frac{(z_2-z_1)(z_2-z_3)}{(x-z_1)(x-z_2)(x-z_3)}H ~,
\notag\end{align}
where $\kappa=a_1^2+a_2^2+a_3^2-a_4^2$, having moved the puncture $z_4$ 
to infinity for later convenience. 

Irregular singularities can then be created by colliding 
regular singular points. To this aim it is convenient to set $z_1=-\frac{1}{2}\epsilon\Lambda^4$, 
$z_2=\frac{1}{2}\epsilon\Lambda^4$. Writing the first two terms in (\ref{eq:curveC04}) as
\begin{align*}
\frac{a_1^2}{(x+\frac{\epsilon}{2}\Lambda^4)^2} + \frac{a_2^2}{(x-\frac{\epsilon}{2}\Lambda^4)^2} 
&=\frac{\frac{\epsilon^2}{4}(a_1^2+a_2^2)\Lambda^8+\epsilon(a_2^2-a_1^2)\Lambda^4x+(a_1^2+a_2^2)x^2}{(x^2-\frac{\epsilon^2}{4}\Lambda^8)^2},
\end{align*}
and considering $\epsilon$-dependent 
parameters $a_1^2=-\frac{1}{2\epsilon\Lambda^2}$, $a_2^2=\frac{1}{2\epsilon\Lambda^2}$ 
and $H=\frac{1}{\epsilon\Lambda^4}U$, 
one finds that the limit $\epsilon\ra 0$ of 
(\ref{eq:curveC04}) exists and can be represented as
\begin{align} \label{eq:curve-deg1}
q'(x) = \frac{\Lambda^2}{x^3} +
\frac{a_3^2}{(x-z_3)^2} - \frac{\kappa}{x(x-z_3)} 
- \frac{z_3}{x^2(x-z_3)}U ~.
\end{align}
In a very similar way one may define a limit $z_3\ra \infty$ further simplifying 
(\ref{eq:curve-deg1}) to 
\begin{equation}\label{classicalSW-PIII-B}
 q''(x) =\frac{\Lambda^2}{x^3} + \frac{U}{x^2} + \frac{\Lambda^2}{x} ~.
\ee
This is known to be the quadratic differential which defines the Seiberg-Witten curve for the pure $SU(2)$ theory.

\subsection{Quantisation}

Having discussed the classical curve, we briefly review its quantisation  following \cite{CPT}. 
A non-commutative deformation of the algebra of functions on the Seiberg-Witten curve is defined
by introducing the commutation relation $[y,x]=-{\mathrm{i}}{\hbar}$ represented by 
$y\to -\mathrm{i}{\hbar}\partial_x$. 
The equation defining $\Sigma$ becomes replaced by a second order differential equation
\begin{equation}\label{eq:SWquantisation}
\mathcal{D}_q\chi(x)=0~,~\text{where}~~ \mathcal{D}_q=\hbar^2\partial_x^2 -q (x)~.
\end{equation}
Such differential equations are closely related to the equations defining flat sections of 
connections of the form 
$dx\left(\hbar \pa_x-\left(\begin{smallmatrix} 0 & q\\ 1 & 0\end{smallmatrix}\right)\right)$
called opers. 
The holonomies of opers define a half-dimensional subspace in the 
character variety 
$\mathcal{M}_{\rm ch}(C)=\mathrm{Hom}(\pi_1(C),\mathrm{SL}(2,\mathbb{C}))/\mathrm{SL}(2,\mathbb{C})$
called the variety of opers.

{ 
As noted above, it is useful to consider deformations 
$q_{\xi,\hbar}(x)=q(x)+\mathcal{O}(\hbar)$ of the function
$q(x)$ in \rf{eq:SWquantisation} 
introducing additional parameters $\xi$.
This can be done by introducing additional poles of a very special form called apparent singularities. }
A second order pole of $q_{\xi,\hbar}^{}(x)$ at $x=u$ is called apparent singularity if 
$q_{\xi,\hbar}(x)=\frac{3\hbar^2}{4(x-u)^2}-\hbar\frac{v}{x-u}+q_{u}+\mathcal{O}(x-u)$, with $q_u$ and 
$v$ related by $v^2=q_u$. {This ensures that  the differential equation 
$(\hbar^2\partial_x^2 -q_{\xi,\hbar}(x))\chi(x)=0$ has 
two linearly independent solutions $\chi_\pm^{}(x)$ of the form $\chi_\pm^{}(x)=(x-u)^{\frac{1}{2}\pm 1}(1+\CO(x-u))$,
implying that the monodromy around $x=u$ is trivial in $\mathrm{PSL}(2,\BC)$. 
{Each apparent singularity introduces one new parameter as 
$u$ and $v$ are related by $v^2=q_u$. The parameters associated to the apparent singularities 
will be related to the parameters $\xi$ by the Riemann-Hilbert correspondence discussed later.
This being understood we will mostly use the notation $q_\la$ for $q_{\xi,\la}$.}

In the  case $C=C_{0,4}$ with $z_1=0$, $z_2=z$, $z_3=1$
one would thereby be led to consider $\hbar$-corrections of the form\footnote{On $C_{0,4}$ just adding one apparent singularity is sufficient, since the complex dimension of the oper locus on $C_{0,n}$ is $(n-3)$, and because each apparent singularity introduces one new complex modulus (more precisely, two moduli subject to one constraint).}
\begin{equation}\label{qla-C04-standard}\begin{aligned}
q_\hbar(x)=&\frac{a_1^2}{x^2} + \frac{a_2^2}{(x-z)^2} + 
\frac{a_3^2}{(x-1)^2} - \frac{a_1^2+a_2^2+a_3^2-a_4^2}{x(x-1)}+ \frac{z(z-1)}{x(x-1)(x-z)}H \\
&-\hbar \frac{u(u-1)}{x(x-1)(x-u)}v+\frac{3}{4}\frac{\hbar^2}{(x-u)^2} ~.
\end{aligned}\end{equation}
{It is elementary to show that any function $q_\la(x)$ defining a quadratic differential $q_\la\equiv q_\hbar(x)(dx)^2$ with 
regular singularities at $x=0,z,1,\infty$ and an apparent singularity at $x=u$ can be represented in the
form \rf{qla-C04-standard}. 
The constraint $v^2=q_u$ determines $H$ as a function of the remaining parameters. 
Considering $a_i$, $i=1,2,3,4$ as fixed parameters one is left with three free parameters $u,v,z$.}

In the case $C=C_{0,2}$ with two poles of third order at $0$ and $\infty$ one
could similarly consider meromorphic quadratic differentials of the form
\begin{equation}
q_\hbar(x) 
=\frac{\Lambda^2}{x^3} + \frac{U}{x^2} + \frac{\Lambda^2}{x} 
-\hbar\, Q_u(x)v 
+\frac{3}{4} \ \frac{\hbar^2}{\left(x - u\right)^2} ~,
\end{equation}
where $Q_u(x)(dx)^2$ is a quadratic differential on $C_{0,2}$ having a simple pole with residue equal to one 
at $x=u$, and
double poles at $0$ and $\infty$. This determines $Q_u(x)$ uniquely up to addition of terms of the 
form $\delta/x^2$. One may fix this freedom by demanding that $\lim_{x\ra u}(Q_u(x)-\frac{1}{x-u})=0$.
The function $Q_u(x)$ uniquely determined by these conditions can be represented as $
Q_u(x)=\frac{  u(2x - u)  }{x^2 (x- u) }=\frac{  1}{x- u}-\frac{x-u}{x^2} $.
The equation $v^2=q_u$ characterising an apparent singularity at $x=u$ is then equivalent to 
\begin{equation}\label{appsingC02}
v^2 
=\frac{\Lambda^2}{u^3} + \frac{U}{u^2} + 
\frac{\Lambda^2}{u} .
\end{equation}
In this way one arrives at a standard form for the function $q_\hbar(x)$ representing the
quantum curve under consideration,  
\begin{equation}\label{eq:P3canonical}
q_\hbar(x) 
= \frac{\Lambda^2}{x^3} + \frac{U(u,v)}{x^2}
+\frac{\Lambda^2}{x} \, 
- \hbar \ \frac{ u \ \left( 2x - u \right) }{x^2 \left(x- u\right) }  \ v \, + 
\frac{3}{4} \ \frac{\hbar^2}{\left(x - u\right)^2} ~,
\end{equation}
with $U(u,v)$ being determined by \rf{appsingC02}. 
{In total one is left with three free parameters $u,v,\Lambda$. 
The parameter $\Lambda$ takes the role of the complex structure parameter  
$z$ of $C_{0,4}$.  }

It is worth noting that the terms which are regular at $x=u$ are $\hbar$-independent in the 
parameterisation (\ref{eq:P3canonical}). From the point of view of quantisation of the 
classical curve $\Sigma$ there is no obvious reason to exclude corrections to (\ref{eq:P3canonical})
which are higher order in $\hbar$. One may, in particular, consider the possibility to
replace $v$ with a function $v(\hbar)$ having a convergent power series expansion 
$v(\hbar)=\sum_{k=0}^\infty v_k\hbar^k$. This would
introduce an infinite set $\{v_k;0\leq k \in\mathbb{Z}\}$ of new parameters.
However, 
the generality gained by this replacement appears to be inessential in the sense that
one can always reach the standard form (\ref{eq:P3canonical}) by a $\hbar$-dependent change of
parameters $v=v(\hbar)$.
We will see shortly, on the other hand, that
alternative choices of parameters for the quantum curves can be useful. The choice of a 
possibly $\hbar$-dependent parameterisation of the quantum curves can be interpreted as
a choice of a particular quantisation scheme. 

\subsection{Representing quantum curves in terms of holomorphic connections}
\label{sec:rep-q-curve-hol-conn}

It is also useful to recall that opers with apparent singularities can be related
to generic holomorphic
$sl_2 (\mathbb{C})$ $\hbar$-connections, in a local trivialisation represented as
\begin{equation}\label{eq:connection1}
\nabla_\hbar = dx\left( \hbar \partial_x - A (x)\right) ~,\quad \text{where}~~ 
A (x) = \left( \begin{matrix} A_0 &A_+ \\ A_- & -A_0 \end{matrix} \right) ~.
\end{equation}
by means of a gauge transformation which is branched at the zeros of $A_-(x)$,
\begin{equation}\label{eq:Opconnection1}
\nabla_\text{Op} = h^{-1} \cdot \nabla \cdot h 
= dx \left(\hbar \partial_x - \left( \begin{matrix} 0 & q_\hbar (x) \\ 1 & 0 \end{matrix} \right)\right) ~ .
\end{equation}
A gauge transformation $h$ which achieves this can be represented as
\begin{equation}
h=\bigg(\begin{matrix} 1/\sqrt{A_-} & 0\\ 0 & \sqrt{A_-}\end{matrix}\bigg)\left(\begin{matrix} 1 & \alpha\\ 0 & 1\end{matrix}\right),
\quad \alpha(x)=\frac{\hbar}{2}\frac{\pa_xA_-}{A_-} + A_0(x).
\end{equation}
The function $q_\hbar(x)$ representing the only nontrivial  matrix element in \rf{eq:Opconnection1}
is found to be
\begin{equation}\label{eq:QquadraticDiff2}
q_\hbar(x) = A_0^2+A_+A_- - \hbar\left(A_0' - \frac{A_0A_-'}{A_-}\right)+
\hbar^2\bigg(\frac{3}{4}\left(\frac{A_-'}{A_-}\right)^2- \frac{A_-''}{2A_-}\bigg) ~.
\end{equation}
For $C=C_{0,n}=\mathbb{P}^1\setminus\{z_1,\dots,z_{n-1},\infty\}$ one considers connections of the form 
\begin{equation}
A(x)=\sum_{r=1}^{n-1}\frac{A_r}{x-z_r}.
\end{equation}
In the case $C=C_{0,2}$ with two irregular singularities 
of the type introduced above  we may consider a connection of the form 
\begin{equation} \label{eq:PIII-conn-folded}
A(x)= 
\left(
\begin{array}{cc}
 \frac{w}{x } & -\Lambda^2 \left(\frac{1}{ux }-1\right) \\
 \frac{1}{x}\left( 1 - \frac{u}{x} \right) & - \frac{w}{x } \\
\end{array}
\right) ~,
\end{equation} 
The resulting function $q_\hbar(x)$ can be brought into the standard form 
(\ref{eq:P3canonical})
by relating the parameters respectively as
\begin{equation} \label{eqdef:parandic}
 U= w^2- \Lambda^2 (u+u^{-1}) ~, \qquad 
v\equiv v(w) =  -\frac{w}{u} - \frac{\hbar}{u}     ~.
\end{equation}

In the case $C=C_{0,4}$ it can also be shown by explicit computations (see Appendix \ref{appendix:quantum-curves}) that
all quantum curves of the form
\rf{qla-C04-standard} can be obtained from holomorphic
$\hbar$-connections by singular gauge transformations of the type 
considered above.

\subsection{Representing quantum curves through the Riemann-Hilbert problem}

Having recalled the relation between quantum curves and holomorphic $\hbar$-connections, this brings us naturally to a third and especially useful way to represent quantum curves, through the Riemann-Hilbert correspondence. 
The Riemann-Hilbert problem is to find a multivalued analytic matrix
function $\Psi(y)$ on $C$ having an analytic continuation $(\ga.\Psi)(y)$ along
$\ga\in\pi_1(C)$ represented in the form $(\ga.\Psi)(y)=\Psi(y)\rho(\ga)$ for a given 
representation $\rho:\pi_1(C)\ra\mathrm{GL}(2,\BC)$.  
The connection to the representation of quantum curves discussed in subsection \ref{sec:rep-q-curve-hol-conn} comes through he observation that  a solution $\Psi(y)$ to the Riemann-Hilbert problem defines a holomorphic connection on $C$ as
\begin{equation}\label{A-from-Psi}
A(y)=(\pa_y\Psi(y))(\Psi(y))^{-1}.
\end{equation}

In the case $C=C_{0,n}$ we may take closed curves $\ga_r$ around $y=z_r$ as
generators for $\pi_1(C)$. In this case we
will
consider the cases where
the matrices $M_r=\rho(\ga_r)$ are diagonalizable, 
$M_r=\mathsf{C}_r^{-1}e^{2\pi \mathrm{i}\mathsf{D}_r}\mathsf{C}_r^{}$, for a fixed choice of
diagonal matrices $\mathsf{D}_r$. 

The solution to this problem is unique up to left multiplication with
single valued matrix functions. In order to fix this ambiguity we need to
specify the singular behaviour of $\Psi(y)$ at $y=z_r$, leading to the following
refined version of the Riemann-Hilbert problem: 
\begin{quote}
{\it Find a matrix
function $\Psi(y)$ such that the following conditions are satisfied.
\begin{itemize}
\item[i)] $\Psi(y)$ is multivalued, analytic and invertible
on $C_{0,n}$, and sarisfies $\Psi(y_0)\,=\,1\,$.
\item[ii)] 
$\Psi(y)$ can be represented in the  neighbourhood of $z_k$  in the form
\begin{equation}\label{asym}
\Psi(y)\,=\,F^{(k)}(y)\cdot(y-z_k)^{\mathsf{D}_k}\cdot \mathsf{C}_k\,,\qquad k=1,\dots,n,
\end{equation}
with $F^{(k)}(y)$  holomorphic and invertible at $y=z_k$,
$\mathsf{C}_k\in \mathrm{GL}(2,\BC)$, and $\mathsf{D}_k$ being diagonal  matrices for $k=1,\dots,n$.
\end{itemize}
}\end{quote}
It is known that generic representations $\rho:\pi_1(C_{0,n})\ra
\mathrm{GL}(2,\BC)$ can be realised as monodromy representation of such a
Fuchsian system, which means that a solution to the Riemann-Hilbert problem
formulated above will generically exist.

If a solution  $\Psi(y)$ to the Riemann-Hilbert problem formulated above
exists, it is for fixed positions $z_1,\dots,z_n$ uniquely determined by the
monodromy data $\mathsf{C}=(\mathsf{C}_1,\dots,\mathsf{C}_n)$ 
and the diagonal matrices $\mathsf{D}=(\mathsf{D}_1,\dots,\mathsf{D}_n)$. 
Changes of the positions $z_1,\dots,z_n$ lead to a change of  $\Psi(y)$ that
can be represented by a system of partial differential equations as follows.
One may  deduce from {\emph{ii)} that the variations of $\Psi(y)$ with respect to $z_r$
with fixed $\mathsf{C}$, $\mathsf{D}$ 
can be represented as
\begin{equation}\label{IsoPsi}
\pa_{z_r} \Psi(y)=A_r(y)\Psi(y),\qquad
A_r(y):=\,-\frac{A_r}{y-z_r} ~.
\end{equation}
The consistency of the differential equations \rf{IsoPsi} for different values of $r$, 
and the consistency with \rf{A-from-Psi} imply
nonlinear partial differential equations for the matrices $A_r$
known as the  Schlesinger equations. Integrating \rf{IsoPsi} allows one to represent a finite variation 
of the positions $z_1,\dots,z_n$ as a gauge transformation acting on $\Psi(y)$ from the left.

Representing quantum curves in terms of $\Psi$ through the Riemann-Hilbert correspondence
offers a convenient 
way to represent the Schlesinger system in Hamiltonian form.
The
Schlesinger equations represent Hamiltonian flows generated by
the Hamiltonians
\begin{equation}
H_r:=\frac{1}{2}\;\underset{x=z_r}{\rm Res}\operatorname{tr}A^2(x)=\sum_{s\neq r}
\frac{{\rm tr}(A_rA_s)}{z_r-z_s}\,,
\end{equation}
using the Poisson structure
\begin{equation}\label{Poissonflat}
\big\{\,A\left(x\right)\,\substack{\otimes\vspace{-0.1cm} \\ ,}\,A\left(x'\right)\,\big\}\,=\,
 \left[\,\frac{\mathcal{P}}{x-x'}\,,\,A\left(x\right)\otimes 1+1\otimes A\left(x'\right)\,\right],
\end{equation}
where $\mathcal{P}$ denotes the permutation matrix.
The generating function of these time-dependent hamiltonians is the isomonodromy tau function, defined by the equation
\begin{equation}\label{eq:tau-function-def}
	H_r = \partial_{z_r} \log {\cal T}(\mu, \mathbf{z}) ~\,,
\end{equation}
whereas $\mu$, the monodromy data, acquires the role of integrals of motion.
An important property of this definition is that the normalisation of ${\cal T}$ is ambiguous, admitting a rescaling by arbitrary functions of $\mu$.

All this has a well-known analog in the case $C=C_{0,2}$ considered above. The Schlesinger 
system will be replaced by the Painlev\'e III equation in this case.\footnote{In fact the two  are related by a confluence limit, as will be illustrated in the following.}

\subsection{Free fermion partition functions as tau-functions}

CFT offers a useful framework for describing how to associate  free fermion partition functions to quantum curves.
We will now quickly 
review and generalise the corresponding discussion in \cite{CPT} before we specialise to another instructive example, 
related to the Painlev\'e III equation.
Let us first recall the general framework underlying the relation between free fermions and tau functions. 
This relation hinges on two key facts: (i) a solution $\Psi$ to the Riemann-Hilbert problem determines a certain state $\mathfrak{f}_\Psi$ in the free fermion Fock space, and (ii) the state $\mathfrak{f}_\Psi$ is characterized by certain relations taking the form of conformal Ward identities, leading to a direct connection with conformal blocks.

Consider a system of $N$ free fermions (useful background suitable for our purposes can be found in \cite{CPT, T17})
\begin{equation}
\begin{split}
	\psi(z) 
	= \sum_{n\in \mathbb{Z}} \psi_n \, z^{-n-1}
	& \qquad
	\bar\psi(z) 
	= \sum_{n\in \mathbb{Z}} \bar\psi_n \, z^{-n}
	\\
	\psi_n = (\psi_{1,n},\dots, \psi_{N,n}) 
	& \qquad
	\bar\psi_n = (\bar \psi_{1,n},\dots, \bar \psi_{N,n})^T 
\end{split}
\end{equation}
and let $\mathcal{F}$ denote the Fock space built from a highest weight vector $\mathfrak{f}_0$.

A solution $\Psi$ to the Riemann-Hilbert problem on $C$ defines a state $\mathfrak{f}\in \mathcal{F}$ as follows.
Let $P$ be a point on $C$ where $\Psi$ is non-singular, and let $1/x$ be a local coordinate 
in a neighbourhood of $P$ vanishing at $P$.
Let $A_{kl}$ be the $N\times N$ matrices defined by the expansion 
\begin{equation}\label{eq:G-psi}
	G_\Psi(x,y) = \frac{\Psi(x)^{-1} \cdot \Psi(y)}{x-y}
	= \frac{1}{x-y} + \sum_{l\geq 0}\sum_{k>0} y^{-l-1} x^{-k} \, A_{kl} \,.
\end{equation}
These uniquely define a state in the Fock space
\begin{equation}\label{eq:f-psi}
	\mathfrak{f}_\Psi =  N_\Psi \, \exp\left(
		-\sum_{k>0}\sum_{l\geq 0} \psi_{-k} \cdot A_{kl}\cdot \bar\psi_l
	\right)\, \mathfrak f_0
\end{equation}
establishing a map 
$\Psi \mapsto \mathfrak{f}_\Psi$
that determines a state uniquely up to normalisation $N_\Psi\in \mathbb{C}$..

By construction $G_\Psi$ is the $N\times N$ matrix-valued correlator
\begin{equation}\label{eq:G-psi-correlator}
	\frac{\langle \mathfrak{f}_0 , \bar \psi(x) \psi(y) \, \mathfrak{f}_\Psi\rangle}{\langle \mathfrak{f}_0 , \mathfrak{f}_\Psi\rangle} = G_\Psi(x,y)\,.
\end{equation}
where $\langle \,\cdot\, , \,\cdot\,\rangle$ denotes the natural pairing $\mathcal{F}^*\otimes \mathcal{F}\to \mathbb{C}$ induced by expectation values in CFT.
Notice that fixing $x$ and varying $y$ around punctures of $C$ induces the right-monodromy action on $G_\Psi$ prescribed by the Riemann-Hilbert problem for $\Psi(y)$. Similarly,  fixing $y$ and varying $x$ induces an (inverse) left-monodromy action for $\Psi(x)^{-1}$. On the other hand, $G_\Psi$ also has a pole with trivial monodromy at $x=y$.

The state $\mathfrak{f}_\Psi$ characterized in this way by $G_\Psi$ satisfies Ward identities of the free fermion VOA
\begin{equation}\label{eq:VOA-CB}
	\left( \frac{1}{2\pi \mathrm{i}} \oint_{\mathcal{C}} \psi(z) \cdot \bar g(z) \right) \mathfrak{f}_\Psi = 0
	\qquad
	\left( \frac{1}{2\pi \mathrm{i}} \oint_{\mathcal{C}}  g(z) \cdot \bar \psi(z)  \right) \mathfrak{f}_\Psi = 0
\end{equation}
where $g(z)$ and $\bar g(z)$ are row and column vectors of the form
\begin{equation}
	g(z) = v(z) \cdot \Psi(z)
	\qquad
	\bar g(z) =  \Psi(z)^{-1}\cdot \bar v(z)
\end{equation}
and $v(z), \bar v(z)$ are row and column vectors of arbitrary meromorphic functions having poles only at $z=\infty$.
It follows that $g(z), \bar g(z)$ solve the monodromy constraints imposed by the Riemann-Hilbert problem at punctures, but may have poles at infinity. 

Conditions (\ref{eq:VOA-CB}) define the fermion VOA conformal blocks, and uniqueness of $\mathfrak{f}_\Psi$ implies that the vector space of the VOA conformal blocks has dimension one. 
{ 
On the other hand, the free fermion extended symmetry includes in particular a Virasoro subalgebra 
generated by the modes $L_n$ of the energy-momentum tensor $T(x)=\sum_{n\in\BZ}x^{-n-2}L_n$
constructed as a bilinear expression of the free fermion fields.}
It follows that free fermion conformal blocks are special conformal blocks for the Virasoro algebra.

Taking $C=C_{0,n}$ a sphere with $n$ punctures, we may consider variations of the complex structure on $C$ parameterized by variation of positions of punctures $z_r$. 
Virasoro conformal blocks are defined to be flat sections of a connection on a 
vector bundle over $\mathcal{M}_C$, the moduli space of complex structures on $C$, defined by a parallel transport equation of the form \cite{FS}
\begin{equation}\label{eq:flatness}
	\left(\partial_{z_r}  - \mathsf{H}_r \right) \, \mathfrak{f}_\Psi(\mathbf{z}) = 0,
\end{equation}
with operators $\mathsf{H}_r$ represented by specific linear combinations of Virasoro generators.
The space of free fermion VOA conformal blocks is preserved by this parallel transport, 
defining  a line sub-bundle with fibre spanned by $\mathfrak{f}_\Psi$.

The chiral partition functions $Z_C(\Psi)=\langle \mathfrak{f}_0^{} , \mathfrak{f}_\Psi^{}\rangle$ 
associated to a flat section $\mathfrak{f}_\Psi$ of the connection defined in \rf{eq:flatness} 
will for $C=C_{0,n}$ satisfy the differential equation 
\begin{subequations}\label{CWI-tau}
\begin{equation}
\left(\sum_{r=1}^n\left(\frac{\Delta_r}{(x-z_r)^2}+\frac{1}{x-z_r}\frac{\pa}{\pa z_r}\right)\!\right)Z_C(\Psi)
=\langle\!\langle T(x)\rangle\!\rangle_\Psi^{} \,Z_C(\Psi), 
\end{equation}
with $\langle\!\langle T(x)\rangle\!\rangle_\Psi^{}$ being the normalised expectation value of the energy-momentum tensor,
\begin{equation}
\langle\!\langle T(x)\rangle\!\rangle_\Psi^{} =
	\frac{\langle \mathfrak{f}_0 , T(x) \mathfrak{f}_\Psi^{}\rangle}
	{\langle \mathfrak{f}_0 , \mathfrak{f}_\Psi^{}\rangle}.
\end{equation}
Finally, it can be shown using Ward identities and the above definition of $\mathfrak{f}_\Psi$, that the normalized expectation value of the energy-momentum tensor is related to the generating function $\mathrm{tr}(A^2)$ 
of the isomonodromic deformation Hamiltonians as follows,
\begin{equation}
	\langle\!\langle T(x)\rangle\!\rangle = 
	\mathrm{tr}\big(A^2(x)\big)=\sum_{r=1}^n\left(\frac{\Delta_r}{(x-z_r)^2}+\frac{H_r}{x-z_r}\right).
\end{equation}
\end{subequations}
It then follows immediately from \rf{CWI-tau} that the free fermion 
partition function satisfies $\frac{\pa}{\pa z_r}Z_C(\Psi)=H_r Z_C(\Psi)$,
relating $Z_C(\Psi)$ to the isomonodromic tau-function $\CT(\mathbf{z})$.

By considering the collision limit of \rf{CWI-tau} one finds in the case of Painlev\'e III
\begin{equation}\label{eq:PIII-tau-function-def}
\Lambda\frac{\partial}{\partial \Lambda}Z_C(\Psi)=
4\,U\, Z_C(\Psi).
\end{equation}
This is equivalent to the defining equation for the tau-function of Painlev\'e III (see e.g. \cite{GL17}).

Equations (\ref{CWI-tau}) do not entirely fix the relation between free fermion partition functions and the tau function.
There are in fact two important problems that are left open by the above discussion. 
{The first is to find the precise relation between the variables  $(\xi,a)$  that appear in the free fermion partition functions $Z_{\rm ff}(\xi,a;\hbar)$, and the parameters of the quantum curves.
One may furthermore note that
the relation between $Z_{\rm ff}$ and the isomonodromic tau-function only fixes the dependence on
the variables $\mathbf{z}=(z_1,\dots,z_n)$. This leaves the problem open how to fix the dependence on  the 
monodromy data $\mu$. We will soon see that the variables  $(\xi,a)$ are naturally identified with suitable functions of the monodromy data.}

\section{Theta series}\label{Sec:Theta}
\setcounter{equation}{0}

It has been observed in \cite{CPT} that there exists a small family of normalisations for the
tau-functions associated to the four-punctured sphere $C_{0,4}$  distinguished by the property
that the normalised tau-functions for the Painlev\'e VI equation admit series expansions having the form 
\rf{Ftransf-1}. It will be convenient to replace $(\xi,a)$ by variables $(\eta,\si,z)$, with $\eta$ and $\si$ 
being functions $\eta(\mu)$ and $\si(\mu)$ of the monodromy data $\mu$, allowing us to represent 
\rf{Ftransf-1} in the form
\begin{equation}\label{eq:gen-theta-exp}
	\CT(\mu,z)=N(\mu) \,
	\sum_{n\in\BZ}e^{2\pi\mathrm{i}\,n\,\eta(\mu)}\CZ(\si(\mu)+n,z) \,.
\end{equation}
The functions 
$\si(\mu)$ and $\eta(\mu)$ of the monodromy data that may appear in expansions 
of the  theta series form \rf{eq:gen-theta-exp} turned out to be very special. All such functions 
can be conveniently defined within a framework called abelianisation in \cite{HN}, leading to 
a one-to-one correspondence between pairs of functions $(\si,\eta)$ and 
certain graphs $\Ga$ on $C=C_{0,4}$ called Fenchel-Nielsen (FN) networks in \cite{HN}.
The functions $(\si,\eta)$ are coordinate functions for the 
character variety $\CM_{\rm ch}(C)$ parameterising the monodomy data having 
a simple relation to complexifications of the Fenchel-Nielsen coordinates 
on the Teichm\"uller spaces, therefore referred to as Fenchel-Nielsen type coordinates.  

In order to gain insight on how to generalise this description to other Riemann surfaces
it turns out to be instructive to compare the resulting picture to 
the case of the two-punctured sphere $C=C_{0,2}$ with 
two irregular singularities of the type considered above. The equation describing isomonodromic
deformations of holomorphic $\mathrm{SL}(2)$-connections is then equivalent to the 
Painlev\'e III equation.
Using the results
of \cite{GL17,ILTy2} we will again find a correspondence between normalised 
tau-functions having expansions of theta series form and graphs $\Ga$ 
defining coordinates on the space of monodromy data through abelianisation.
The results of \cite{ILTy2} will allow us to reveal an interesting new feature.
There exist theta series expansions of the 
tau-function for Painlev\'e~III
in which the role of the Fenchel-Nielsen type coordinates
is taken by Fock-Goncharov coordinates.

\subsection{Factorisation of free fermion partition functions}\label{Sec:Factor}

Representing the Riemann surface $C$ of interest by gluing two Riemann surfaces $C_i$, $i=1,2$
of simpler topology allows one to construct conformal blocks on $C$ from conformal blocks 
associated to the surfaces $C_i$. This construction is a powerful tool, leading
in particular to Fredholm determinant representations for the free fermion 
conformal blocks associated to solutions of the Riemann-Hilbert problem \cite{CPT}.  

Let $C_i$, $i=1,2$, be two Riemann surfaces with punctures $P_i$
embedded in discs $D_i$ having 
coordinates $x_i$ vanishing at $P_i$ such that 
$D_i=\{Q_i\in C;0<|x_i(Q_i)|<1\}$, for $i=1,2$, respectively.
Out of these data one may define families of Riemann surfaces $C_z$, $z\in\mathbb{C}$ being a 
complex parameter,  by
identifying pairs of points 
$(Q_1,Q_2)$ in the annuli $A_i=\{Q_i\in D_i;|z|<|x_i(Q_i)|<1\}$
satisfying $x_1(Q_1)x_2(Q_2)=z$
to get an annulus $A\subset C_z$ with coordinate $x\equiv x_2$.

It is easy to extend the gluing operation to punctured surfaces $C_i$ with flat bundles. 
Let $\rho_i:\pi_1(C_i)\ra\mathrm{SL}(2,\mathbb{C})$
be representations of $\pi_1(C_i)$ characterising flat bundles on 
$C_i$ for $i=1,2$, respectively. We will only consider representations $\rho_i$ having the 
property that the holonomies around 
$P_i$ represented by the diagonal matrices $\mathrm{diag}(e^{2\pi\mathrm{i}\si_i},e^{-2\pi\mathrm{i}\si_i})$
with $\si_2=-\si_1\equiv\si-\frac{1}{2}$. For generic $\rho_i$ we may characterise the flat bundles 
by the corresponding solutions to the Riemann-Hilbert problems, concretely represented
by matrix-valued function $\Psi_i(Q_i)$ on $D_i$ having for each $\gamma\in\pi_1(C_i)$ an analytic continuation 
$(\gamma.\Psi)(Q_i)$ satisfying $(\gamma.\Psi)(Q_i)=\Psi(Q_i)\rho_i(\gamma)$. The coordinates 
$x_i$ 
allow us to represent the functions $\Psi_i$ by functions $\Phi_i(x_i)$ on the punctured 
disc $D=\{x\in\mathbb{C};|x|<1\}$. It will for our goals be sufficient to consider only
the case that $P_i$ is a regular singularity of the form
$\Phi_i(x_i)=F_i(x_i)x_i^{\mathsf{d}_{i}}$, with $\mathsf{d}_i=\mathrm{diag}(\si_i,-\si_i)$, 
and $F_i(x_i)$ holomorphic at $x_i=0$.
The gluing condition $x_1x_2=z$ furthermore
allows us to represent the functions $\Phi_i(x_i)$ by functions $\Xi_i(x)$ on the annulus $A$, for $i=1,2$, and where $x$ is a local coordinate on $A$.
One may then define a flat bundle on $C$ by
imposing the relation
\begin{equation}\label{Xi-TA}
\Xi_2(x)=\Xi_1(x) T_A,\qquad T_A\in\mathrm{SL}(2,\mathbb{C}).
\end{equation}
Assuming that $\Xi_i(x)$ both have diagonal monodromy around $A$ for $i=1,2$ one generically has to require that
$T_A$ is diagonal, and can therefore be represented as $T_A=\mathrm{diag}(e^{-\pi\mathrm{i}\eta},
e^{\pi\mathrm{i}\eta})$.
 
The parameters $(\si,\eta)$ introduced in this way define coordinates for the character 
variety $\mathcal{M}_{\rm ch}(C)$. It is clear that the origin of the coordinate $\eta$ depends on the 
choice of the functions $\Xi_i$, $i=1,2$. Multiplying either $\Xi_1$ or $\Xi_2$ by a diagonal matrix from the right will shift 
$\eta$. Fixing a choice for $\Xi_1$ and $\Xi_2$ determines the coordinate $\eta$ uniquely.

The gluing construction of conformal blocks can then be used to define 
a conformal block on $C$ from two conformal blocks
$\mathfrak{f}_{\Psi_i}^{}$ associated to $C_i$ for $i=1,2$, respectively. This leads 
to the following representation of the partition function 
\begin{equation}\label{partfactor-A}
Z_C(\Psi)\,=\,\langle\,\mathfrak{f}_{\Psi_2}^{\ast}\,,\,\mathsf{T}_\eta z^{L_0}\mathfrak{f}_{\Psi_1}^{}\rangle_{\CF_{\si}}^{}\,,
\end{equation}
with $\mathsf{T}_\eta$ being an operator satisfying $\psi(x)\mathsf{T}_\eta=\mathsf{T}_\eta\psi(x)T_A$ and
$\bar{\psi}(x)\mathsf{T}_\eta=\mathsf{T}_\eta T_A^{-1}\bar{\psi}(x)$.
{The insertions  $z^{L_0}$ and $\mathsf{T}_\eta$ implement the gluing 
conditions $x_1x_2=z$ and \rf{Xi-TA}, respectively.}

Other useful objects are the normalised partition functions
\begin{equation}\label{partfactor-B}
\mathcal{T}_0\big(\,\si,\eta\,;\Psi_1,\Psi_2;\,q\big)\,=\,\frac{\langle\,\mathfrak{f}_{\Psi_2}^{\ast}\,,\,\mathsf{T}_\eta z^{L_0}\mathfrak{f}_{\Psi_1}^{}\rangle_{\CF_{\si}}^{}}{\langle\,\mathfrak{f}_{\Psi_2}^{\ast}\,,\,\mathfrak{f}_{\si}^{}\rangle_{\CF_{\si}}^{}
\langle\,\mathfrak{f}_{\si}^\ast\,,\, \mathfrak{f}_{\Psi_1}^{}\rangle_{\CF_{\si}}^{}}\,.
\end{equation}
There is a double grading\footnote{The definition of $\mathfrak{f}_\sigma, \CF_\sigma$ and a discussion of their properties can be found in \cite{CPT}.} for the Fock space  $\CF_{\si}$, defined by the eigenvalues of $\mathsf{T}_\eta$ and 
$L_0$, 
implying that  $\CT_0$ can be represented by  an expansion of  the following form
\begin{equation}\label{eq:ff-tau-series}
\mathcal{T}_0\big(\,\si,\eta\,;\Psi_1,\Psi_2;\,z\big)=
\sum_{n\in\BZ} \,e^{2\pi\mathrm{i}n\,\eta}\,\CD_n(\,\si\,;\Psi_1,\Psi_2)
\CF^{}(\,\si+n\,;\Psi_1,\Psi_2;\,z\,),
\end{equation}
with $\CF^{}(\si,\Psi_1,\Psi_2;z)$ represented by a power series in $z$. 

As explained in \cite{CPT}, one may represent the right hand side of \rf{partfactor-B}
as a Fredholm determinant of the following form
\begin{equation}\label{Fredholm}
\mathcal{T}_0\big(\,\si,\eta\,;\Psi_1,\Psi_2;\,q\big)
=\mathrm{det}_{\mathcal{H}_+}^{}(1+\mathsf{B}_{\Psi_2}\mathsf{A}_{\Psi_1}),
\end{equation}
where $\mathsf{A}_{\Psi_1}$ and $\mathsf{B}_{\Psi_2}$ are integral 
operators constructed from solutions $\Psi_1$ and $\Psi_2$ to the Riemann Hilbert problems
defined by flat bundles on the surfaces $C_1$ and $C_2$ used in the gluing construction, 
respectively. It is in this way becoming clear that the compositions 
appearing in the gluing construction of free fermion conformal blocks are analytically 
well-defined. 

Whenever there are explicit solutions to the 
Riemann Hilbert problems
on $C_1$ and $C_2$, one may use \rf{partfactor-A} or \rf{Fredholm} to get combinatorial expansions 
of the isomonodromic tau-functions. The resulting expansions have been 
worked out for the cases $C=C_{0,4}$ and $C_{0,2}$ of our interest 
in the papers \cite{GL} and \cite{GL17}, respectively.

\subsection{Fenchel-Nielsen type coordinates}\label{FN-gen}

The gluing construction described in the previous Section \ref{Sec:Factor} can be used to 
introduce a pair of coordinates $(\si,\eta)$ for any Riemann surface $C$ that can be represented 
by gluing two surfaces $C_1$ and $C_2$.
Such coordinates will be called coordinates of Fenchel-Nielsen (FN) type 
in the following. It has been noted in  Section \ref{Sec:Factor} that 
the precise definition of the coordinate $\eta$ depends on some choices. 
We are now going to discuss two useful ways to fix these choices.

\subsubsection{Normalisation factors}

One may recall, on the one hand, that $\eta$ can be defined by fixing a pair of
functions $\Phi_i(x)$ representing the solutions of the Riemann-Hilbert problem on $C_i$
in the discs $D_i$. Given  that $\Phi_i(x)$ have a regular singularity at $x=0$ for $i=1,2$,
we may fix the freedom to multiply $\Phi_i(x)$ by constant diagonal matrices from the right by imposing
the condition
\begin{equation}\label{Phiform}
\Phi_i(x_i)=
\bigg(\,\begin{matrix} \tilde{\xi}_+^{(i)}(x_i) & \tilde{\xi}_-^{(i)}(x_i) \\ \xi_+^{(i)}(x_i) & \xi_-^{(i)}(x_i)\end{matrix}\,\bigg)\,,
\qquad \xi_\pm^{(i)}(x_i)=\nu_\pm^{(i)}\, x_i^{\mathsf{d}_i}\, (1+\CO(x_i))\,.
\end{equation} 
Choosing for each $i=1,2$ a pair $(\nu_+^{(i)},\nu_-^{(i)})$ of complex 
numbers fixes the freedom to multiply~$\Phi_i(x_i)$ by constant diagonal matrices from the right.
The residual gauge freedom to multiply~$\Phi_i(x_i)$ from 
the left by matrix valued holomorphic functions of $x_i$ can be fixed, for example, by demanding that
$\tilde{\xi}_s^{(i)}(x_i)=\pa_{x_i}\xi_s^{(i)}(x_i)$, which is equivalent to demanding that the 
connection one-forms $A_i(x_i)=(\pa_x\Phi_i(x_i))(\Phi_i(x_i))^{-1}$ have oper form. It is then natural to further
reduce the normalisation freedom by imposing the condition $\mathrm{det}(\Phi_i(x_i))=1$ for $i=1,2$. 
This implies $(2\si-1)\nu_+^{(i)}\nu_-^{(i)}=1$. The left-over normalisation freedom is then parameterised 
by the normalisation factors\footnote{The relative inversion of the signs is related to the choice of conventions in subsection \ref{Sec:Factor}.}
\begin{equation}\label{eq:n-nu-def}
\mathsf{n}^{(1)}:=\frac{\nu_-^{(1)}}{\nu_+^{(1)}}\,,
\qquad
\mathsf{n}^{(2)}:=\frac{\nu_+^{(2)}}{\nu_-^{(2)}}\,.
\end{equation}
We may thereby use the complex numbers $\mathsf{n}^{(i)}$ as parameters allowing us to distinguish 
different choices of the FN coordinate $\eta$ corresponding to the same pants decomposition.
A distinguished choice is to consider $\mathsf{n}^{(1)} =\mathsf{n}^{(2)} = 1$, which we'll denote by $\eta_0$.
Then from (\ref{Xi-TA}) one may easily deduce that other choices of normalisation lead to a coordinate $\eta$ related to $\eta_{0}$ as follows
\begin{equation}\label{eta-eta0}
{e^{2\pi\mathrm{i}\eta_0^{}}}=e^{2\pi\mathrm{i}\eta}\,{\mathsf{n}^{(2)}\mathsf{n}^{(1)}}\,.
\end{equation}
The direct relation between choices of normalisation for the solutions \rf{Phiform} and choices of coordinates 
$\eta$ expressed in \rf{eta-eta0} will often be used in the following.

\subsubsection{Trace functions}\label{tracegluing}

The traces of holonomies $\mathrm{tr}(\rho(\ga))$, $\ga\in\pi_1(C)$ 
generate the ring of algebraic functions on the character variety. The relations in $\pi_1(C)$
together with identities satisfied by traces of $SL(2,\BC)$-matrices imply algebraic relations
between the trace functions. 
In the case $C=C_{0,4}$, for example, one can consider the
seven trace functions
\begin{subequations}
\begin{align}\label{Mk}
&L_k=\operatorname{Tr} M_k=2\cos2\pi \theta_k,\qquad k=1,\ldots,4,\\ \label{Mk-stu}
&L_s=\operatorname{Tr} M_1 M_2,\qquad L_t=\operatorname{Tr} M_1 M_3,\qquad L_u=\operatorname{Tr} M_2 M_3,
\end{align}
\end{subequations}
generating the algebra of invariant polynomial functions on $\CM_{\rm ch}(C_{0,4})$, with $M_k$ being 
the monodromy around $z_k$, for $k=1,2,3,4$.
These trace functions satisfy the quartic equation
\begin{align}
 \label{JFR}
& L_1L_2L_3L_4+L_sL_tL_u+L_s^2+L_t^2+L_u^2+L_1^2+L_2^2+L_3^2+L_4^2=\\
 &\nonumber \quad=\left(L_1L_2+L_3L_4\right)L_s+\left(L_1L_3+L_2L_4\right)L_t
+\left(L_2L_3+L_1L_4\right)L_u+4.
\end{align}
FN type coordinates can alternatively be defined by picking a certain 
parameterisation of the trace functions solving the algebraic relations they satisfy. We will now 
explain how the definition of FN type coordinates by means of the gluing construction 
induces a parameterisation of the trace functions. 

Considering surfaces $C$ represented by gluing surfaces $C_1$ and $C_2$ as described in 
Section \ref{Sec:Factor} one may find a set of generators for $\pi_1(C)$ by picking 
sets of generators for $\pi_1(C_i)$, $i=1,2$, and adding sufficiently many curves $\ga$ traversing the 
annulus $A$ created in the gluing construction. It suffices to consider curves $\ga$ which can be 
represented as a composition $\varpi_2^{}\circ\varpi_A^{}\circ\varpi_1^{}\circ\varpi_A^{-1}$, 
where $\varpi_i$ are paths
supported in $C_i$, for $i=1,2$, respectively, and  $\varpi_A$ is a path traversing the annulus $A$
from $C_2$ to  $C_1$. 
The holonomy $M_{\ga}$ can then be factorised in the form 
$T^{-1}_AM_1^{}T_A^{}M_2^{}$, with $M_i$ being the holonomies along $\varpi_i$, for $i=1,2$, respectively,
and $T_A$ being the transition function across $A$. 
For generic $\rho$ one may diagonalise simultaneously both the
holonomy 
$M_A=-\left(\begin{smallmatrix} e^{2\mathrm{i}\pi\si} & 0 \\ 0 & e^{-2\mathrm{i}\pi\si} \end{smallmatrix}\right)$
around the non-contractible cycle of the annulus, and the matrix 
$T_A=\left(\begin{smallmatrix} e^{\mathrm{i}\pi\eta} & 0 \\ 0 & e^{-\mathrm{i}\pi\eta} \end{smallmatrix}\right)$
describing
the parallel transport from one 
boundary of the annulus to the other.
The variables $(\si,\eta)$ defined in this way can be used to complete
coordinate systems for the moduli spaces of flat connections 
on $C_i$, $i=1,2$, to a coordinate system for the moduli space
of flat connections on $C$. 

One may note that the 
trace functions
$\mathrm{tr}(M_{\ga})
=\mathrm{tr}(T^{-1}_AM_1^{}T_A^{}M_2^{})$
will then have a simple $\eta$-dependence of the  form
\begin{equation}\label{eq:FN-def-general}
\begin{aligned}
&\mathrm{tr}(M_{\ga})=t^+(\si)e^{2\pi \mathrm{i} \eta}+t^0(\si)+t^-(\si)e^{-2\pi \mathrm{i} \eta},\\
&
t^+(\si)=b_2 c_1, \quad 
t^0(\si)= a_1 a_2 + d_1 d_2,\quad
t^-(\si)= b_1 c_2,\end{aligned}
 \qquad M_i=\bigg(\begin{matrix} a_i & b_i \\ c_i & d_i\end{matrix}\bigg).
\end{equation}
Fixing a particular definition of the coordinate $\eta$ is thereby seen to be equivalent to specifying 
the functions $t^{\pm}(\si)$ appearing in the expressions \rf{eq:FN-def-general} for the trace functions.
We will next illustrate this relation by some examples which will be used later.

\subsubsection{Case $C=C_{0,4}$}\label{etaC04}

We have just illustrated two ways to parametrise ambiguities in the definition of $\eta$: either in terms of normalisation of flat sections in the factorization limit of the Riemann-Hilbert correspondence, or by a specific parametrisation of trace functions.
The two possibilities are of course related to each other, we will now explain this with the specific example of $C_{0,4}$. 

In the case $C=C_{0,4}\simeq\mathbb{P}^1\setminus\{0,z,1,\infty\}$ 
one may express the solutions to the Riemann-Hilbert problem on the 
surfaces $C_i\simeq C_{0,3}\simeq\mathbb{P}^1\setminus\{0,1,\infty\}$, $i=1,2$,  
in terms of the Gauss hypergeometric function. 
The solution $\Phi_2$ to the Riemann-Hilbert problem on $C_2$ can be represented in the  
form \rf{Phiform} with  
\begin{equation}\label{Psioutdef}
\begin{aligned}
\xi_{s}^{(2)}(x)=\nu_s^{(2)}\, x^{s(\si-\frac{1}{2})}(1-x)^{s\theta_3}F(A_s,B_s,C_s;x),
\end{aligned}
\end{equation} 
for $s=\pm 1$, where $F(A,B,C;x)$ is the Gauss hypergeometric function and
\begin{equation}\label{ABC}
\begin{aligned}
A_+=A, \quad A_-=1-A,\\
B_+=B, \quad B_-=1-B,\\
\end{aligned}\qquad
\begin{aligned} &C_+=C,\\ & C_-=2-C,
\end{aligned}
\qquad
\begin{aligned}
A&=\theta_3+\theta_4+\si,\\
B&=\theta_3-\theta_4+\si,
\end{aligned}\qquad C=2\si.
\end{equation}
$\Phi_1$, on the other hand, may be chosen by
defining 
 $\xi_\pm^{(1)}(x)$  from $\xi_\pm^{(2)}(x)$ by the replacements 
$x\ra x^{-1}$, $\theta_4\ra \theta_1$, $\theta_3\ra\theta_2$ and $s\ra -s$.

As an example for the computation of trace functions
let us then consider the trace $\mathrm{tr}(M_{z1})$ of the holonomy around $z$ and $1$.
According to the discussion in Section \ref{tracegluing}
we can represent $M_{z1}$ in the form $T^{-1}_AM_1^{}T_A^{}M_2^{}$, 
with $M_2$ being the monodromy around $x=1$ on $C_2$
and $M_1$  being the monodromy around the point representing $x=z$ on $C_1$.
The well-known formulae for the monodromies of the hypergeometric function then yield formulae
for  $M_2$, 
\begin{equation}\label{M3out}
M_2=\left(\begin{matrix} a_2 & b_2 \\ c_2 & d_2\end{matrix}\right),\qquad
\begin{aligned}
&b_2=- \, 
\mathsf{n}^{(2)}
\frac{2\pi\mathrm{i}\,\Ga(C_+)\Ga(C_+-1)}
{\Ga(A_+)\Ga(B_+)\Ga(C_+-A_+)\Ga(C_+-B_+)},\\
&c_2=+
\frac{1}{\mathsf{n}^{(2)}}
\frac{2\pi\mathrm{i}\,\Ga(C_-)\Ga(C_--1)}
{\Ga(A_-)\Ga(B_-)\Ga(C_--A_-)\Ga(C_--B_-)},
\end{aligned}
\end{equation}
A similar formula for monodromy $M_1$ is found by the replacements defining 
 $\xi_\pm^{(1)}$ from
 $\xi_\pm^{(2)}$. 

Two choices for $\mathsf{n}^{(i)}$, $i=1,2$, appear to be fairly natural. One may, on the one hand, simply 
choose $\mathsf{n}^{(i)}=1$ 
for $i=1,2$. In that case we easily
see that $t^{\pm}(\si)=t^{\pm}_0(\si)$ with
\begin{align}\label{Cepdef1}
t^{\pm}_0(\si)& = \frac{(2\pi)^2\,\big[\Ga(1\pm(2\si-1))\Ga(\pm(2\si-1))\big]^2}{ 
\prod_{s,s'=\pm 1}\Ga\big(\frac{1}{2}\pm\big(\si-\frac{1}{2}\big)+s\theta_1+s' \theta_2\big)\Ga\big(\frac{1}{2}\pm\big(\si-\frac{1}{2}\big)+s\theta_3+s' \theta_4\big)}.
\end{align}
This corresponds to the definition of $\eta_0$ given in subsection \ref{tracegluing}.
The normalisation factors $\mathsf{n}^{(i)}$, $i=1,2$,  can alternatively be chosen such that
$b_2c_1=1$, defining a coordinate $\eta$ such that 
\begin{align}\label{Cepdef2}
t^+(\si) = 1,\quad\;\;
t^{-}(\si)&=\prod_{s,s'=\pm 1} 
\frac{2\sin\pi(\si+s\theta_1+s' \theta_2)\,2\sin\pi(\si+s\theta_3+s' \theta_4)}{(2\sin 2\pi\si)^4}.
\end{align}
One many now check that the coordinates $\eta$ and $\eta_0^{}$ are related precisely as in (\ref{eta-eta0}), 
if 
$\mathsf{n}^{(2)}$  is 
defined by choosing $b_2=1$ in \rf{M3out}, 
and $\mathsf{n}^{(1)}$ defined by the replacements above.

\subsection{Normalised tau-functions for $C=C_{0,4}$}\label{normTauC04}

Series expansions for the isomonodromic tau-functions of theta
series type
have been derived in \cite{GIL,ILT,BS,GL}. It is quite remarkable that these expansions can be represented in a
form resembling the series defining the theta-functions,
\begin{align}\label{thetaser-gen}
&\CT^{(\eta)}(\,\si,\eta\,,\,\underline{\theta}\,;\,z\,)=\sum_{n\in\BZ} \,e^{2\pi\mathrm{i}n\,\eta}\,
\CN^{(\eta)}(\si+n\,;\,\underline{\theta}\,)\CF^{}(\,\si+n\,,\,\underline{\theta}\,;\,z\,),
\end{align}
with functions $\CF$ represented by power series of the form 
\[
\CF^{}(\,\si\,,\,\underline{\theta}\,;\,z\,)=z^{\si^2-\theta_1^2-\theta_2^2}\bigg(1+\sum_{k=1}^\infty
\CF_k(\,\si\,,\,\underline{\theta}\,) \, z^k\bigg).
\]
As observed in \cite{CPT}, there is a fairly small family of distinguished choices for the coordinate $\eta$ for which 
one can recast the series expansions for the tau-functions in the form \rf{thetaser-gen}. We anticipate in the notations that the choice of  functions $\CN^{(\eta)}$ is related to
the choice of $\eta$.
The function $\CN^{(\eta)}$ corresponding to the coordinate $\eta$ previously defined in 
Section \ref{etaC04}  is
\begin{subequations}\label{CN-formula}
\begin{align}\label{CN-NN}
&\CN^{(\eta)}(\,\si\,,\underline{\theta}\,)=
{N(\si,\theta_2,\theta_1)N(\si,\theta_3,\theta_4)}\\
&
N(\vartheta_1,\vartheta_2,\vartheta_3) = 
 \frac{
		\prod_{\epsilon,\epsilon' = \pm}
		G(1+\vartheta_{1}+\epsilon \vartheta_{2}+\epsilon' \vartheta_{3})
		}{(2\pi)^{\vartheta_1}G(1)
		\prod_{r=1}^{3}G(1+2\vartheta_r)
		}\,.
\end{align}
\end{subequations}
In Appendix \ref{appendix:consistency-check} it is shown how to derive this form of the expansion from the results of \cite{ILT}.

In order to clarify how $\CN^{(\eta)}$ varies with the choice of $\eta$ 
it will be useful to note
 that \rf{thetaser-gen} is equivalent to 
an expansion for the normalised tau-function
of the following form
\begin{equation}\label{tau0-exp}
\mathcal{T}^{(\eta_0)}\big(\,\si,\eta\,;\,\underline{\theta}\,;\,z\big)=
\sum_{n\in\BZ} \,e^{2\pi\mathrm{i}n\,\eta_0^{}}\,\CC_n(\,\si\,,\underline{\theta}\,)
\CF^{}(\,\si+n\,,\,\underline{\theta}\,;\,z\,),
\end{equation}
with $\eta_0^{}$ being the FN-type coordinate 
defined in Section \ref{tracegluing}, and $\CC_n(\si,\underline{\theta})$, $n\in\BZ$,   being a family of 
functions which can be represented 
in terms of a single function $\CN^{(\eta)}(\si,\underline{\theta})$ as
\begin{equation}\label{CNn-CNeta}
\CC_n(\,\si\,,\underline{\theta}\,)=
\left(\frac{\CN^{(\eta)}(\si-1\,;\,\underline{\theta}\,)}{\CN^{(\eta)}(\si\,;\,\underline{\theta}\,)}\right)^n
\frac{\CN^{(\eta)}(\si+n\,;\,\underline{\theta}\,)}{\CN^{(\eta)}(\si\,;\,\underline{\theta}\,)}.
\end{equation}
Indeed, by observing that $\CN^{(\eta)}(\si,\underline{\theta})$ is related to the normalisation factors 
$\mathsf{n}^{(i)}$ defining the coordinate $\eta$ through the identity
\begin{align}
\frac{\CN^{(\eta)}(\si-1\,;\,\underline{\theta}\,)}{\CN^{(\eta)}(\si\,;\,\underline{\theta}\,)}=
\frac{1}{\mathsf{n}^{(1)}\mathsf{n}^{(2)}},
\end{align}
one may easily check the equivalence of 
\rf{thetaser-gen} and \rf{tau0-exp} using 
the relation between $\eta$ and $\eta_0$ from (\ref{eta-eta0}), 
together with
\begin{equation}
	\CT^{(\eta)}=\CN^{(\eta)} \, \CT^{(\eta_0)}\,.
\end{equation}

At this point, it should be stressed that while (\ref{thetaser-gen}) has the form of a generalized theta-series expansion, (\ref{tau0-exp}) does not.
Nevertheless the latter is somewhat canonical, in the sense that choosing coordinate $\eta_0$ corresponds to $\mathsf{n}^{(i)} = 1$. 
Moreover the relation (\ref{CNn-CNeta}) provides an interesting way to find additional choices of $\eta$ inducing a theta-series expansion like (\ref{thetaser-gen}).
One may note, in fact, that the representation 
of the coefficient functions $\CC_n(\si;\underline{\theta})$ in the form 
\rf{CNn-CNeta} is ambiguous. It is possible to 
replace the function
$\CN^{(\eta)}(\si;\underline{\theta})$ on the right side of \rf{CNn-CNeta} by other functions 
$\CN^{(\eta')}(\si;\underline{\theta})$
without changing 
the functions $\CC_n(\si;\underline{\theta})$.
In order to see this, let us consider the possibility to replace $\CN^{(\eta)}$ by a function $\CN^{(\eta')}$ of the
form
$\CN^{(\eta')}(\si;\underline{\theta})=\CE(\si;\underline{\theta})
\CN^{(\eta)}(\si;\underline{\theta})$. 
It is easy to see that this replacement will leave the functions $\CC_n(\si;\underline{\theta})$
unchanged iff the function
\begin{equation}\label{eq:e-E}
\mathsf{e}\,(\,\si\,,\underline{\theta}\,):=\frac{\CE(\,\si+1\,,\underline{\theta}\,)}{\CE(\,\si\,,\underline{\theta}\,)},
\end{equation}
is periodic in $\si$, $\mathsf{e}(\si+1,\underline{\theta})=\mathsf{e}(\si,\underline{\theta})$. Such a replacement will 
leave $\CT^{(\eta_0)}$ unchanged, but it will allow us to define another normalisation for the isomondromic
tau-function 
\begin{equation}\label{eq:E-transition-tau}
	\CT^{(\eta')}=\CE\, \CT^{(\eta)}
\end{equation}
admitting an expansion of theta series type,
\begin{align}
&\CT^{(\eta')}(\,\si\,,\,\underline{\theta}\,;\,z\,)=\sum_{n\in\BZ} \,e^{2\pi\mathrm{i}n\,\eta'}\,
\CN^{(\eta')}(\si+n\,;\,\underline{\theta}\,)\CF^{}(\,\si+n\,,\,\underline{\theta}\,;\,z\,)
\end{align}
with $\eta$ related to $\eta'$ such that
\begin{equation}\label{eta'-eta}
e^{2\pi\mathrm{i}\,\eta}=e^{2\pi\mathrm{i}\,\eta'}\mathsf{e}(\,\si\,,\underline{\theta}\,).
\end{equation}
We thereby see that the function $\CN^{(\eta')}$ is associated to a new FN type coordinate $\eta'$ defined using 
\rf{eta'-eta}. 

This argument leads to a vast class of possibilities for generalised theta-series expansions of the tau function, since the only requirement we imposed is periodicity of $\mathsf{e}$, which can be expressed as the following step-two relation for $\cal E$
\begin{equation}
	\CE(\sigma) \CE(\sigma+2) = \CE(\sigma+1)^2 \,.
\end{equation}
Among this large class of possibilities, however, there turns out to be a rather small distinguished subset.

As an example let us consider the function 
\begin{align}\label{CN'def}
&\CN^{(\eta')}(\,\si\,,\underline{\theta}\,)=
{N(\si,\theta_2,\theta_1)N'(\si,\theta_3,\theta_4)}\\
&
N'(\vartheta_1,\vartheta_2,\vartheta_3) =
 \frac{
		\prod_{\epsilon,\epsilon' = \pm}
		G(1+\vartheta_{2}+\epsilon \vartheta_{3}+\epsilon' \vartheta_{1})
		}{(2\pi)^{\vartheta_1}G(1)
		\prod_{r=1}^{3}G(1+2\vartheta_r)
		}\,,
\end{align}
This would correspond to the choice
\begin{equation}\label{CEchoice}
\CE(\,\si\,,\underline{\theta}\,)=S(\theta_3-\si+\theta_4)S(\theta_3-\si-\theta_4),
\qquad S(x):=\frac{G(1+x)}{G(1-x)}.
\end{equation}
Noting that $S(x)$ satisfies the functional relation 
\begin{equation}
S(x\pm 1)=\mp(2\sin\pi x)^{\mp 1}S(x),
\end{equation}
one easily sees that the function $\mathsf{e}(\si,\underline{\theta})$ associated to the choice \rf{CEchoice}
is periodic in $\si$. 
The coordinate $\eta'$ associated to 
$\CN^{(\eta')}$ by means of \rf{eta'-eta} can  be identified with an element of the 
family of coordinates defined in Section \ref{FN-gen} 
by replacing the normalisation factor 
$\mathsf{n}^{(2)}$ defined in Section \ref{etaC04} by $\mathsf{n}'{}^{(2)}=\mathsf{e}\,\mathsf{n}^{(2)}$.

In a similar way one can show that replacing one of the functions $N$ in 
\rf{CN-NN} by any one of the functions $N_i$, $i=1,2,3$, defined as 
\begin{align}\label{Ni-def}
N_i(\vartheta_1,\vartheta_2,\vartheta_3) =
 \frac{
		\prod_{\epsilon,\epsilon' = \pm}
		G(1+\vartheta_{i}+\epsilon \vartheta_{i+1}+\epsilon' \vartheta_{i+2})
		}{(2\pi)^{\vartheta_1}G(1)
		\prod_{r=1}^{3}G(1+2\vartheta_r)
		}\,,\qquad \vartheta_{i+3}\equiv \vartheta_i,
\end{align}
will leave 
the functions $\CC_n(\si;\underline{\theta})$ unchanged. This means that the functions 
$\CC_n(\si;\underline{\theta})$ have symmetries under permutations of the arguments that
are not manifest in the factorisation \rf{CNn-CNeta}.

To conclude this overview of the case of $C_{0,4}$, let us recollect a few of the most relevant facts learned so far.
On the one hand, not all choices of $\eta$ induce expansions of $\CT$ having the form of generalised theta-series, $\eta_0$ being a counter-example.
On the other hand, different choices of $\eta$ inducing such an expansion involve changes in the normalisation $\CN^{(\eta)}$ of the tau function, compatible with ambiguities in its  definition (\ref{eq:tau-function-def}). In turn, the \emph{relative} normalisation coefficients~$\CE$ for $\CT$ are related in a precise way to \emph{relative} normalisation coefficients $\mathsf{e}$ for $\eta$,  as described by (\ref{eq:e-E}).
It is also worth noting that the gluing parameter $z$ for $C_1, C_2$ turns out to play the role of the isomonodromy `time' coordinate, as well as the series-expansion parameter in the theta-series coefficients $\CF$. 
Finally, we should stress that the discussion so far concerns only a \emph{fixed} choice of pants decomposition for $C_{0,4}$, this is directly related to the fact that the theta-series coefficients~$\CF$ appearing in (\ref{thetaser-gen}) are the \emph{same} for all these choices of $\eta$.

\subsection{Normalised tau-functions for $C=C_{0,2}$}\label{sec:weak-coupling-expansion}

A first expansion of theta series form for the 
tau-function associated to Painlev\'e III has been found in \cite{GL17}.
This expansion takes the following form
\begin{equation}\label{GL-exp}
\CT^{(w)}(\si',\eta';\Lambda)=
	\sum_{n\in\BZ}e^{4\pi\mathrm{i}\,n\,\eta'}\, 
	\CN^{(w)}(\sigma'+n) \, \CF(\sigma'+n,\Lambda)
\end{equation}
where 
\begin{equation}
\CN^{(w)}(\si) =\prod_{s=\pm}\frac{1}{G(1+2s\si)},\qquad
\CF(\si,\Lambda)=\Lambda^{4\si^2}\bigg(1+\sum_{k=1}^\infty\CF_k(\si)\Lambda^{4k}\bigg),
\end{equation}
with 
$G(x)$ being the Barnes G-function satisfying $G(x+1)=\Ga(x)G(x)$.
A formula for the coefficients $\CF_k(\si)$ can be found in \cite{GL17}. 
Comparing with the notations used in \cite{GL17} one should note that the parameters
we denote by $\si'$ and $\eta'$ correspond to the parameters denoted $\si$ and $\eta$ in \cite{GL17}, respectively. 
We will see that $\si'$ and $\eta'$  are closely related, but not quite identical to coordinates
$(\si,\eta)$ of FN-type as defined in this paper.
It may also be noted that the role of $z$ from the case of $C_{0,4}$ has been taken by $\Lambda$, which indeed plays the role of isomonodromy time and expansion parameter for $\CF$.


The functions $(\si(\mu),\eta(\mu))$ appearing in \rf{GL-exp} have been defined  in \cite{GL17} 
by parameterising the Stokes data $\mu$ associated to the irregular 
singularities at $z=0$ and $z=\infty$ as follows.
A pair of solutions $Y^{(0)}(z)$, $Y^{(\infty)}(z)$
of the equation $(\pa_z-A(z))Y(z)=0$ can be defined having asymptotic behaviour 
at $z\ra 0$  and $z\ra \infty$  of the form 
\begin{equation}\label{eq:asymptotic-solutions}
\begin{split}
	Y^{(0)}(z)
	& \simeq 
	G^{(0)}(\sqrt{z})  
	\big[ 1+\CO(\sqrt{z})\big]e^{+2\Lambda^2\frac{1}{\sqrt{z}}\si_z},
	\\
	Y^{(\infty)}(z)
	& \simeq 
	G^{(\infty)}(\sqrt{z})\big[ 1+\CO(\sqrt{z^{-1}})\big]e^{-2{\sqrt{z}}\si_z},
\end{split}
\quad
\begin{split}
	& -\pi < \arg(z \,\Lambda^{-4} )< 3\pi,\\
	& -\pi < \arg(z)< 3\pi \,,
\end{split}
\end{equation}
where 
$G^{(0)}(\sqrt{z})$ and $G^{(\infty)}(\sqrt{z})$ are 
gauge transformations. 
The Stokes data have been parameterised in  \cite{GL17} in terms of a pair of 
variables here denoted
by $({\si}',{\eta'})$
in such a way that the  monodromy around $x=0$ gets represented in the following form:
\begin{equation}
\begin{aligned}
Y^{(0)}(e^{2\pi\mathrm{i}}z)&=Y^{(0)}(z) M_0,\\
Y^{(\infty)}(e^{2\pi\mathrm{i}}z)&=Y^{(\infty)}(z)  M_\infty
\end{aligned}
\qquad
 M_0=\si_x M_\infty\si_x=\left(\begin{matrix} 0  & \mathrm{i} \\ 
\mathrm{i} & -2\cos 2\pi{\si'} \end{matrix}\right), 
\end{equation}
with $\si_x=\big(\begin{smallmatrix} 0 & 1 \\ 1 & 0 \end{smallmatrix}\big)$, together with the 
fact that  $Y^{(0)}(z)$ and $Y^{(\infty)}(z)$ are  related 
as\begin{equation}\label{eq:conn-mat}
Y^{(\infty)}(z)=Y^{(0)}(z)E,\quad E=\frac{1}{\sin 2\pi{\si}'}
\left(\begin{matrix} \sin 2\pi{\eta'}  & -\mathrm{i}\sin 2\pi({\eta'}+{\si'}) \\ 
\mathrm{i}\sin 2\pi({\eta'}-{\si'}) & \sin 2\pi{\eta'} \end{matrix}\right).
\end{equation}
We will determine the relation between the parameters $\si'$ and $\eta'$ and  
the FN-type variables $(\si,\eta)$
defined in Sections \ref{Sec:Factor} and \ref{FN-gen} in the following subsection.

\subsection{Fenchel-Nielsen type coordinates for $C_{0,2}$}\label{sec:FN-for-PIII}

We will now see that the coordinates $({\si'},{\eta'})$ 
appearing in the expansion \rf{GL-exp}
are closely related to coordinates of FN type for the Riemann sphere with 
two irregular singularities of the mildest type. 
To this aim one may begin by noting  that 
the monodromy in an annulus $A$ separating the punctures at $z=0$ 
and $z=\infty$ is diagonalised by the following change of basis
\begin{equation}\label{Diagonaliz}
	\begin{aligned}
		Y^{(0)}(z)& =Y_1(z) W_1,\\
		Y^{(\infty)}(z)& =Y_2(z) W_2,
	\end{aligned}
 	\qquad 
	W_1=W_2\si_x,
\qquad 
	W_2=
		\left(\begin{matrix} 1 & -\mathrm{i}\,e^{-2\pi\mathrm{i}{\si'}} \\ 
1 & - \mathrm{i}\,e^{+2\pi\mathrm{i}{\si'}} \end{matrix}\right)\,,
\end{equation}
so that we have for both $i=1,2$,
\begin{equation}
	Y_i(e^{2\pi\mathrm{i}}z) =Y_i(z) M_A\,,
	\qquad 
	M_A = 
	- \left(\begin{matrix} 
		e^{2\pi\mathrm{i}\si'} & 0 \\ 
		0 & e^{-2\pi\mathrm{i}{\si'}} 
	\end{matrix}\right)\,.
\end{equation}
$Y_1$ and $Y_2$ are related
by a diagonal matrix,
\begin{equation}\label{Y2Y1TA}
	Y_2(z)=Y_1(z) \, T_A\,,
	\qquad
	T_A = WE \si_x W^{-1}
	=-\mathrm{i}\,
	\left(\begin{matrix} 
	e^{-2\pi\mathrm{i}{\eta'}}  & 0 \\ 
	0 & e^{2\pi\mathrm{i}{\eta'}}  
	\end{matrix}\right).
\end{equation}
This needs to be compared to the definition of the FN-type coordinates 
used in this paper.

To this aim one may note, on the one hand, that
a sphere with two irregular punctures can be represented
by gluing two spheres $C_i$  having a regular and an irregular puncture each.
Solutions to the Riemann-Hilbert problem on 
the two-punctured spheres  $C_i$ can be obtained 
by isomonodromic deformations of the solutions $Y_i(z)$ on $C$ in the factorisation 
limit $\Lambda\ra 0$.

In the gluing construction described in Sections \ref{Sec:Factor} and 
\ref{FN-gen} one needs a pair of functions $(\Phi_1,\Phi_2)$ representing solutions to the 
Riemann-Hilbert problems on $C_i$. We need to find the pair of functions
$(\Phi_1,\Phi_2)$ related to  the bases $(Y_1,Y_2)$ defined above by the limit $\Lambda\ra 0$.
 Let us first consider the solution $Y_2(x)$ of 
$(\pa_x-A(x))Y_2(x)=0$, with $A(x)$ given in \rf{eq:PIII-conn-folded}. It is related
to a solution $\Phi_2(x)$ of  $(\pa_x-A_0(x))\Phi_2(x)=0$, with $A_0(x)=A(x)\big|_{\Lambda=0}$, 
by a gauge transformation, as
follows from the equation replacing  \rf{IsoPsi} in the case $C=C_{0,2}$.

 It next follows from \rf{Diagonaliz} that the function $\Phi_2(x)$ is related to a function
$\Phi^{(\infty)}(x)=\Phi_2(x)W_2$ having the same 
asymptotic behaviour  \rf{eq:asymptotic-solutions} for $x\ra\infty$ as $Y^{(\infty)}(x)$. 
The function $\Phi^{(\infty)}(x)$ is thereby uniquely 
defined up to gauge transformations. 
We claim that
\begin{equation}
\Phi^{(\infty)}(x)=g^{(\infty)}(x)\left(\begin{matrix} \ast & \tilde{\ast} \\ K_{2\si-1}(2\sqrt{x}) & \mathrm{i}K_{2\si-1}(2e^{\pi \mathrm{i}}\sqrt{x})
\end{matrix}\right),
\end{equation}
with $K_\al(x)$ being the modified Bessel function of the second kind, and 
matrix elements in the upper row left unspecified, being dependend on the choice 
of gauge transformation $g^{(\infty)}(x)$. In order to check
that $\Phi^{(\infty)}(x)$ represents the limit $\Lambda\ra 0$ of $Y^{(\infty)}(x)$
one may use the formula
\begin{equation}
K_{\al}(z)\underset{z\ra\infty}{\sim} \sqrt{\frac{\pi}{2z}}e^{-z}(1+\CO(z^{-1})).
\end{equation} 
The functions $K_{\al}(x)$ are related to the modified Bessel functions  $I_\al(x)$ of the first 
kind as
\begin{equation}
K_{\al}(z)=\frac{\pi}{\sin\pi\al}(I_{-\al}(z)-I_{\al}(z)). 
\end{equation}
Noting that the functions $I_{\al}(z)$ have diagonal monodromy
 $I_{\al}(e^{2\pi\mathrm{i}}z)=e^{2\pi\mathrm{i}\al}I_{\al}(z)$  around $z=0$, 
it becomes easy to show that $\Phi^{(\infty)}(x)=\Phi_2(x)W_2$, with
\begin{equation}\label{Phi-Bessel}
\Phi_2(x)=g^{(\infty)}(x)\left(\begin{matrix} \ast' & \ast'' \\ I_{2\si-1}(2\sqrt{x}) & 
I_{1-2\si}(2\sqrt{x})
\end{matrix}\right)\!,\quad
W_2=\frac{\pi}{\sin2\pi\si}
		\left(\begin{matrix} 1 & -\mathrm{i}\,e^{+2\pi\mathrm{i}{\si}} \\ 
1 & - \mathrm{i}\,e^{-2\pi\mathrm{i}{\si}} \end{matrix}\right)\!.
\end{equation}
Comparing with \rf{Diagonaliz} we conclude that ${\si}'=-\si$. 
There is a similar relation between $\Phi_1(x)$ and the limit 
$\Lambda \ra 0$ of $Y_1(x)$. Comparing \rf{Y2Y1TA} to the equation \rf{Xi-TA}
defining $\eta$ yields the relation $\eta=-2\eta'-\frac{1}{2}$.

%
%
%

\subsection{Strong coupling expansion}\label{sec:LDE-review}

We are now going to observe that the tau-function for Painlev\'e III  admits an expansion of 
theta series type which does not involve Fenchel-Nielsen type coordinates as before. 
The role of the Fenchel-Nielsen type will be taken by the coordinates for the character variety 
$\CM_{\rm ch}(C)$ introduced by 
Fock and Goncharov.

In addition to the weak coupling expansion reviewed in Section \ref{sec:weak-coupling-expansion}, there is another expansion of the Painlev\'e III tau function, conjectured in \cite{ILTy2}:
\begin{equation}\label{eq:long-distance}
	\mathcal{T}^{(s)}(\nu,\rho;\Lambda) 
	= \sum_{n\in\mathbb{Z}} e^{4\pi\mathrm{i} \rho n}\, 
	\CN^{(s)}(\nu+\mathrm{i}n,\Lambda) \, \CG(\nu+\mathrm{i}n,\Lambda)
\end{equation}
where 
$\CG(\nu,\Lambda)=1+\sum_{k=1}^\infty\CG_k(\nu)\Lambda^{-k}$ is a normalised series in powers of $\Lambda^{-1}$ and the prefactor $\CN^{(s)}(\nu,\Lambda) =e^{\frac{\mathrm{i}\pi \nu^2}{4}} 2^{\nu^2} (2\pi)^{-\frac{\mathrm{i}\nu}{2}} G(1+\mathrm{i}\nu ) (8\Lambda)^{\frac{\nu^2}{2}+\frac{1}{4}} e^{4\Lambda^2 + 8\nu\Lambda}$.
Compared to the expansion \rf{GL-exp} one may note, in particular, that negative powers of 
$\Lambda$ appear in this expansion.

The parameters $\nu$ and $\rho$ have been defined in \cite{ILTy2} 
as the following functions of the parameters 
$\sigma,\eta$ introduced previously, 
\begin{equation} \label{eq:FN-FG-relation}
	e^{\pi\nu} 
	= \frac{\sin 2\pi \eta'}{\sin 2\pi\sigma'}
	\,,
	\qquad
	e^{4\pi \mathrm{i}\rho} 
	= \frac{\sin 2\pi\eta'}{\sin 2\pi (\sigma'+\eta')}
	\,.
\end{equation}
We will now show that the parameters $\nu$, $\rho$ are particular Fock-Goncharov coordinates.

\subsection{Fock-Goncharov coordinates}\label{sec:FG}

Fock-Goncharov coordinates are assigned to triangulations of the Riemann surface \cite{FG06}. In the case of Painlev\'e III we have a triangulation of the annulus with 
two irregular punctures characterized by a single Stokes sector each (see Section \ref{sec:exact-WKB} for the necessary background regarding Stokes graphs). 
Such a triangulation $T$ therefore has exactly one vertex $v_0$ at $z=0$ and one vertex $v_\infty$ at $z=\infty$. 
Two edges of $T$ correspond to boundary edges: one connects $v_0$ to itself encircling the singularity at $z=0$, the other one has a similar role at $v_\infty$.
There are also two internal edges marked by letters $X$ and $Y$ in Figure \ref{fig:annulus-labels}. 
The Fock-Goncharov coordinate $X_e$ 
associated to an edge $e$ separating two triangles of an ideal triangulation
is  defined as the cross-ratio
\begin{equation}\label{eq:FG-cross-ratio}
X_e=	-\frac{(s_1\wedge s_2)(s_3\wedge s_4)}{(s_2\wedge s_3)(s_4\wedge s_1)} \,,
\end{equation}
where $s_i$ are flat sections satisfying $(\partial_z-A(z))s_i(z)$ vanishing at the puncture 
representing the $i$-th corner of the 
quadrilateral formed by the two triangles, with corners labeled counter-clockwise 
starting from one of the ends of $e$. 
The cross-ratio \rf{eq:FG-cross-ratio} does not depend on the point where
all sections $s_1,\dots,s_4$ appearing in \rf{eq:FG-cross-ratio}
are evaluated. 

\begin{figure}[h!]
\begin{center}
\includegraphics[width=0.25\textwidth]{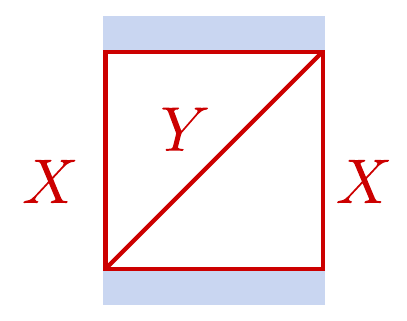}
\caption{Triangulation of the annulus of the Painlev\'e III system.}
\label{fig:annulus-labels}
\end{center}
\end{figure}


In order to compute the Fock-Goncharov coordinates in our case, 
we first need to identify the small flat section at each vertex. 
This can be inferred from the explicit form of their asymptotic expansions (\ref{eq:asymptotic-solutions}).
For convenience, let $e_1 = (1,0)^t$ and $e_2 = (0,1)^t$, then the flat section is $ Y^{(0)}\cdot e_2$.\footnote{\label{foot:small-sec}
An equivalent way is to use the general theory of Stokes phenomena, which identifies the small section with that multiplying the Stokes factor featuring in the jump.
By declaring a choice of small flat section near $z=0$, we implicitly fix a trivialization for the universal covering of the cylinder. Then, the small flat section corresponding to the clockwise image of $v_0$ will be $ Y^{(0)}(z\, e^{-2\pi \mathrm{i}})\cdot e_2 =  Y^{(0)}(z)\cdot  M_0^{-1} \cdot e_2 $.}
A similar analysis yields the small flat section at $v_\infty$, it is  $ Y^{(\infty)}\cdot e_1$.

Computing cross-ratios (\ref{eq:FG-cross-ratio})  is now straightforward. 
Let us start from edge $Y$, the sections associated to the four corners of the corresponding quadrilateral are\footnote{Here $s_1$ is associated to the bottom-left vertex, $s_2$ to the bottom-right vertex, $s_3$ to the upper-right vertex, and $s_4$ to the top-left vertex in Figure \ref{fig:annulus-labels}. All four are evaluated at $z_*$ near the bottom-left vertex.}
\begin{equation}
\begin{split}
	s_1 & =  Y^{(0)} \cdot e_2 \\
	s_3 & 
	=  Y^{(0)}\cdot E\cdot   M^{-1}_\infty \cdot e_1
\end{split}
\qquad \quad
\begin{split}
	s_2 & =  Y^{(0)}\cdot   M_0^{-1} \cdot e_2 \\
	s_4 & 
	=  Y^{(0)}\cdot E \cdot e_1 
\end{split} ~.
\end{equation}
Here we used monodromy data to express $s_{i}$ evaluated at $z_*$ chosen in the bottom-left corner of Figure \ref{fig:annulus-labels}, in terms of the local asymptotic solution $ Y^{(0)}$, see footnote \ref{foot:small-sec} for details.
Thanks to this, all dependence on $ Y_+^{(0)}(z_*)\wedge  Y_-^{(0)}(z_*)$ drops out of the cross ratio, leaving
\begin{equation}\label{eq:Y-FG}
	Y = -\left(\frac{\sin 2 \pi \sigma'}{\sin 2\pi \eta'}\right)^2 = -e^{-2\pi\nu}\,.
\end{equation}

Similarly, to compute $X$ we consider the quadrilateral crossed by the left-most vertical edge in Figure \ref{fig:annulus-labels}. The four sections are therefore\footnote{Here  $s_1$ corresponds to  the bottom-left vertex, $s_2$ to the top-right vertex, $s_3$ with to the top-left vertex, and $s_4$ is the image of $s_1$ after transporting clockwise around $z=0$.}
\begin{equation}
\begin{split}
	s_1 & =  Y^{(0)} \cdot e_2 \\
	s_3 & 
	=  Y^{(0)}\cdot E \cdot e_1 
\end{split}
\qquad\quad
\begin{split}
	s_2 & 
	=  Y^{(0)}\cdot E\cdot    M_\infty^{-1} \cdot e_1 \\
	s_4 &=  Y^{(0)}\cdot   M_0 \cdot e_2
\end{split}
\end{equation}
plugging these into (\ref{eq:FG-cross-ratio}) gives the coordinate associated to edge $X$
\begin{equation}\label{eq:X-FG}
	X = -\left(\frac{\sin 2\pi (\sigma'+\eta')}{\sin 2\pi\sigma'}\right)^2
	= - e^{-8\pi \mathrm{i}\rho+2\pi\nu}\,.
\end{equation}
We have shown that coordinates $(\nu,\rho)$ appearing in the strong coupling generalized theta-series expansion of the Painlev\'e III tau function are particular Fock-Goncharov coordinates.

\section{Relations between theta series}\label{sec:relations-theta-series}
\setcounter{equation}{0}

By multiplying the isomonodromic tau-functions with certain monodromy-dependent normalisation factors
one can define free fermion partition functions having expansions of generalised theta series type. 
Such a partition function is essentially uniquely characterised by the set of Darboux coordinates
for the space $\CM_{\rm ch}(C)$ of monodromy data appearing in the series expansions.
There are many such expansions, in general. Any two of them must be related by a change of Darboux
coordinates. It turns out that the ratios of the corresponding free fermion partition functions
have an elegant characterisation in terms of the change of Darboux coordinates defining the 
two partition functions, respectively. They will turn out to be analogs of the generating functions
for the relevant changes of Darboux coordinates with taking derivatives being replaced by 
forming finite differences.  
{The reader may notice that we already came across this structure in the previous section: studying different Darboux coordinates associated with a fixed pants decomposition we observed that the discrete derivative (\ref{eq:e-E}) of the ratio of tau functions (\ref{eq:E-transition-tau}) is directly related to the change of coordinates (\ref{eta'-eta}).
In this section we will significantly extend this picture, by showing that a similar structure describes changes of the tau function induced by changes of Darboux coordinates in greater generality.
We will focus on the two cases $C=C_{0,2}$ and $C=C_{0,4}$ representing our main examples.
}

\subsection{Relation between strong and weak coupling expansions}

The two expansions of theta series \rf{GL-exp} and  \rf{eq:long-distance} 
quoted above are related by multiplication with 
a function of the monodromy data,
\begin{equation}\label{weak-to-strong}
\CT^{(w)}(\si,\eta;\Lambda)=\CX(\si,\nu)\CT^{(s)}(\nu,\rho;\Lambda),
\end{equation}
asuming that $(\si,\eta)$ and $(\nu,\rho)$ are related by \rf{eq:FN-FG-relation}.
The main result of \cite{ILTy2} is an explicit formula for the function $\CX(\si,\nu)$.

To find this formula, it has been noted in \cite{ILTy2} that
the theta series expansions for \rf{GL-exp} and  \rf{eq:long-distance} imply 
the difference equations
\begin{equation}
\CT^{(w)}(\si+1,\eta;\Lambda)=e^{-4\pi\mathrm{i}\,\eta}\,\CT^{(w)}(\si,\eta;\Lambda),\quad
\mathcal{T}^{(s)}(\nu+\mathrm{i},\rho;\Lambda) = e^{-4\pi\mathrm{i}\,\rho}\,\mathcal{T}^{(s)}(\nu,\rho;\Lambda). 
\end{equation}
Consistency of \rf{weak-to-strong} with these difference equations implies that 
the normalisation factor $\CX(\si,\nu)$ must satisfy the pair of relations
\begin{equation}\label{eq:PIII-difference-rel}
	\CX(\sigma+1,\nu) = e^{-4\pi\mathrm{i}\eta} \CX(\sigma,\nu)\,,
	\qquad
	\CX(\sigma,\nu+\mathrm{i}) = e^{4\pi\mathrm{i}\rho} \CX(\sigma,\nu)\,.
\end{equation}
Both $(\si,\eta)$ and $(\nu,\rho)$ 
are Darboux coordinates for the space of monodromy data 
parameterising solutions of the  Painlev\'e III equation.
This follows from our results above relating $(\si,\eta)$ and $(\nu,\rho)$ 
to coordinates of FN- and FG-type, respectively, and has been verified directly in \cite{ILTy2}.
Equations 
\rf{eq:PIII-difference-rel} resemble the equations characterising 
the generating function of the change of Darboux variables from 
$(\si,\eta)$ to  $(\nu,\rho)$, with derivatives being replaced by
finite difference operators. 
{In comparing with the discussion of subsection \ref{normTauC04}, one may note that 
(\ref{eq:PIII-difference-rel}) and (\ref{eq:e-E}) play analogous roles.}}
Such functions will be called 
 \emph{difference generating functions} in the following.

\subsection{Changes of pants decomposition}

{Besides changes of FN-type coordinates associated to different normalisations for $\eta$ in a fixed pants decomposition (as considered in subsection {\ref{normTauC04}}) one will have to discuss the changes of pants decomposition. }
The following summarises
some key points needed for such a discussion.

Within the CFT framework  it is straightforward to derive expansions similar to 
\rf{thetaser-gen} corresponding to other pants decompositions of $C$. In the 
case $C=C_{0,4}$ studied in this section, it  suffices to consider 
the pants decomposition defined by cutting along a curve $\ga_u$ 
separating  $z$, $1$ from $0$, $\infty$, in analogy with standard terminology from 
high energy physics referred to as the $u$-channel. This pants decomposition 
allows one to define FN-type coordinates $(\si_u,\eta_u)$ 
in the same way as described above. It is then natural to regard the tau-function as
a function of the coordinates $(\si_u,\eta_u)$. The analog of \rf{thetaser-gen}
corresponding to the $u$-channel will then have the form 
\begin{equation}\label{thetaser-u}
\mathcal{T}^{(u)}\big(\si_u,\eta_u\,;\,\underline{\theta}\,;\,z\big)=
\sum_{n\in\BZ} \,e^{2\pi\mathrm{i}n\eta_u}\,
\CG^{(u)}(\,\si_u+n\,,\,\underline{\theta}\,;\,z\,),
\end{equation}
where
\begin{align}
&\CG^{(u)}(\,\si_u\,,\,\underline{\theta}\,;\,z\,)=\CN^{(u)}(\,\si_u\,,\,\underline{\theta}'\,)
\CF^{}_{}(\,\si_u\,,\,\underline{\theta}'\,;\,1-z\,),
\end{align}
with $\underline{\theta}'=(\theta_3,\theta_2,\theta_1,\theta_4)$ if $\underline{\theta}=
(\theta_1,\theta_2,\theta_3,\theta_4)$.\footnote{{
It is understood that, just like for the case of the $s$-channel, there should be more than one choice of $\eta_u$. Each of these induces a different expansion of the tau function, but all will be related as discussed in subsection \ref{normTauC04}, \emph{mutatis mutandis}.}}

A simple but important consequence of the generalised theta-series form of the expansions in \rf{thetaser-gen},
\rf{thetaser-u} are
the difference equations
\begin{equation}\label{diffrel}\begin{aligned}
\CT^{(s)}\big(\si_s+1,\eta_s\,;\,\underline{\theta}\,;\,z\big)&=
e^{-2\pi\mathrm{i}\eta_s}\,\CT^{(s)}\big(\si_s,\eta_s\,;\,\underline{\theta}\,;\,z\big),\\
\CT^{(u)}\big(\si_u+1,\eta_u\,;\,\underline{\theta}\,;\,z\big)&=
e^{-2\pi\mathrm{i}\eta_u}\,\CT^{(u)}\big(\si_u,\eta_u\,;\,\underline{\theta}\,;\,z\big),
\end{aligned}
\end{equation}
having set $\CT^{(\eta)}\equiv \CT^{(s)}$, $\si\equiv\si_s$ and $\eta\equiv \eta_s$ to make the notation more uniform.
It follows from the Painlev\'e VI equation that $\CT^{(s)}\big(\si_s,\eta_s\,;\,\underline{\theta}\,;\,z\big)$
must be proportional to $\CT^{(u)}\big(\si_u,\eta_u\,;\,\underline{\theta}\,;\,z\big)$ if the coordinates
$(\si_s,\eta_s,\underline{\theta})$ and $(\si_u,\eta_u,\underline{\theta})$ are evaluated on the same
point of the character variety,
\begin{equation}\label{ILT-rel}
\CT^{(s)}\big(\si_s,\eta_s\,;\,\underline{\theta}\,;\,z\big)\,=\,
\CH(\,\si_s,\si_u\,,L_t\,;\,\underline{\theta}\,)\,
\CT^{(u)}\big(\si_u,\eta_u\,;\,\underline{\theta}\,;\,z\big).
\end{equation}
Choosing $\si_s$, $\si_u$ and $L_t$ as arguments of the function $\CH$ is of course somewhat redundant. 
Given $\si_s$ and  $\si_u$ one may use \rf{JFR} together with $L_s=2\cos 2\pi \si_s$ and
$L_u=2\cos 2\pi \si_u$ to determine $L_t$ up to a sign.
The factor $\CH(\,\si_s,\si_u\,,L_t\,;\,\underline{\theta}_s\,)$ in \rf{ILT-rel} has been calculated explicitly in the remarkable
work \cite{ILTy,ILP}. The basis for the approach taken in \cite{ILTy} are the functional relations 
\begin{subequations}\label{ILT-funrel}\begin{align}\label{ILT-funrel-s}
\CH(\si_s+1,\si_u\,,L_t\,;\,\underline{\theta}\,)&=
e^{-2\pi\mathrm{i}\eta_s(\si_s,\si_u,L_t)}\,\CH(\si_s,\si_u\,,L_t\,;\,\underline{\theta}\,),\\
\label{ILT-funrel-u}
\CH(\si_s,\si_u+1\,,L_t\,;\,\underline{\theta}\,)&=
e^{+2\pi\mathrm{i}\eta_u(\si_s,\si_u,L_t)}\,\CH(\si_s,\si_u\,,L_t\,;\,\underline{\theta}\,).
\end{align}
\end{subequations}
The relations \rf{ILT-funrel} identify the function $\CH$ as a difference generating function.

One should notice that the changes of normalisation of the tau-functions induced by changes
for Darboux coordinates for $\CM_{\rm ch}(C)$ can be composed. In this way one may represent the 
relations \rf{ILT-rel} involving a rather complicated difference generating function as the 
composition of similar relations involving the difference generating function for the 
change from FG to FN type coordinates considered above.

\subsection{Transitions between two sets of Fock-Goncharov coordinates}\label{FG-to-FG}

Having discussed difference generating functions for transitions between FG and FN coordinates, as well as between FN and FN coordinates associated to different pants decompositions, we now complete the picture with a discussion of transitions between FG and FG coordinates associated to different choices of triangulations.

Among all these cases, the FG-FG transition enjoys a special status.
In fact it should be noted that the transitions considered here and those considered above are not all independent: in particular, the FN-FG transition can be understood as an infinite sequence of FG-FG transitions (as part of the \emph{juggle} phenomenon, see Section \ref{WKBnetworks} for details). Moreover a change of pants decomposition, i.e. an FN-FN transition, can be usually obtained by an FN-FG transition followed by an FG-FG transition and an FG-FN transition of a different type.
In view of this, the FG-FG transition may be regarded as the building block underlying all other cases.

\subsubsection{Dehn twist}
To illustrate the features of FG-FG difference generating functions, we consider the composition of a flip and a relabeling of coordinates defined by
\begin{equation}\label{Dehn}
X'=Y^{-1},\qquad Y'=X(1+Y^{-1})^{-2}.
\end{equation}
This change of variables can be seen to correspond to a Dehn twist for the triangulation of the annulus shown in Figure \ref{fig:annulus-labels}.
While specific, this case nevertheless exhibits all the features of flips of more general triangulations, it is then transparent how the upcoming analysis extends to the more general setting. 

To proceed, we switch to logarithmic variables $x,y,x',y'$,
\begin{equation}\label{Logs}
X=e^{2\pi\mathrm{i}\,x},\qquad Y=-e^{2\pi\mathrm{i}\,y},\qquad
X'=-e^{2\pi\mathrm{i}\,x'},\qquad Y'=e^{2\pi\mathrm{i}\,y'}.
\end{equation}
The equations \rf{Dehn} can be solved for $Y$ and $Y'$,
\begin{equation}
Y(x,x')=-e^{-2\pi\mathrm{i}\,{x'}},\qquad Y'(x,y)=e^{2\pi\mathrm{i}\,x}(1-e^{2\pi\mathrm{i}\,x'})^{-2}.
\end{equation}
The difference generating function $\CJ(x,x')$ 
associated to the change of variables \rf{Dehn} has the following defining properties,
\begin{equation}
\frac{\CJ(x+1,x')}{\CJ(x,x')}=-(Y(x,x'))^{-1},\qquad
\frac{\CJ(x,x'+1)}{\CJ(x,x')}=Y'(x,x').
\end{equation}
A function satisfying these properties can be constructed in the following form
\begin{equation}
\CJ(x,x')={e^{2\pi\mathrm{i}xx'}}{(E(x'))^{2}},
\end{equation}
provided that the function $E(z)$ satisfies the functional equation
\begin{equation}\label{Eshiftrel}
E(z+1)=\frac{1}{1-e^{2\pi\mathrm{i}\,z}}E(z).
\end{equation}
A function $E(z)$ satisfying \rf{Eshiftrel} can be constructed as follows
\begin{equation}
E(z)=(2\pi)^{-z}e^{-\frac{\pi\mathrm{i}}{2}z^2}\frac{G(1+z)}{G(1-z)},
\end{equation}
where $G(z)$ is the Barnes $G$-function satisfying $G(z+1)=\Ga(z)G(z)$.
{
An alternative representation of the same function is 
\begin{equation}
	E(z)=\exp\left( \frac{1}{2\pi\mathrm{i}} \mathrm{Li}_2(1-e^{2\pi \mathrm{i}z}) \right)\,.
\end{equation}
}

\subsection{Relations between Fock-Goncharov and Fenchel-Nielsen coordinates}\label{Sec:FG-FN}
In the case $C=C_{0,2}$ 
we have shown above that the coordinates $(\eta,\sigma)$ and $(\nu,\rho)$ introduced in \cite{ILTy}
are coordinates of Fenchel-Nielsen and  Fock-Goncharov type, respectively. The relation (\ref{eq:FN-FG-relation}) 
thereby acquires an interpretation as a change of coordinates on the character variety.
It will be useful to rewrite (\ref{eq:FN-FG-relation}) in the following form
\begin{equation}\label{eq:XYtoUV-first}
	X = \left(\frac{ V+  V^{-1}}{ U-  U^{-1}}\right)^2,
	\qquad
	Y = \left(\frac{ U- U^{-1}}{{UV}+{U}^{-1} {V}^{-1}}\right)^2,
	\qquad
\begin{aligned}
	& U 
	= e^{2\pi \mathrm{i}\sigma}\,,\\
	& \mathrm{i}  V 
	=e^{\pi \mathrm{i}(- \eta-2\sigma  - 1/2)}
	\,,
	\end{aligned}	
\end{equation}
We see that the change from Fenchel-Nielsen to Fock-Goncharov coordinates becomes rational in these variables.
Quite remarkably, we are now going to show that similar relations between these two types of coordinates
exist also for more general surfaces $C$.


In the  case $C=C_{0,4}$ one may, for example,  consider the triangulation 
dual to the fat graph depicted in Figure \ref{FG4a}.

\begin{figure}[h!]
\centering
\includegraphics[width=0.6\textwidth]{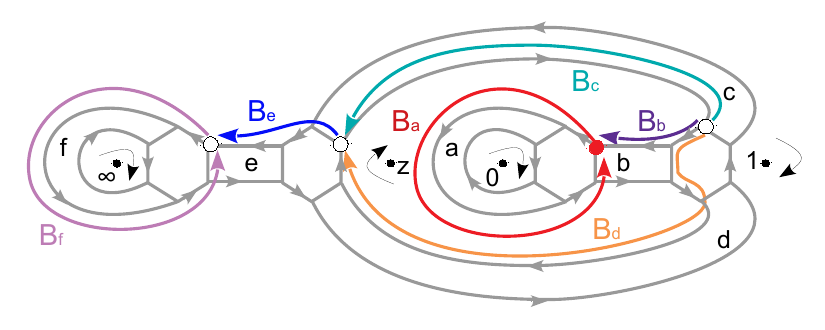}
\caption{\it Fat graph on $C_{0,4}$ with six basic paths.}
\label{FG4a}
\end{figure}

One may note that the part of the  triangulation dual to this fat graph which is contained 
in an annulus separating the pairs of punctures $(0,z)$ from $(1,\infty)$ is of the same form 
as considered in the case $C=C_{0,2}$ above. The edges labelled $c$ and $d$ in Figure 
\ref{FG4a} will correspond to the edges labelled with letters $X$ and $Y$ in Figure 
\ref{fig:annulus-labels}.
We are going to show that substituting $c$ and $d$ 
by the expressions on the right sides of equations (\ref{eq:FN-FG-relation}) yields a parameterisation 
of the monodromy data through FN type coordinates $\si$ and $\eta$.

The Fock-Goncharov coordinates allow us to represent the monodromies on $C_{0,4}$ 
as compositions 
of elementary building blocks as follows:
\begin{equation}
\begin{split} 
	M_1& =B_a^{-1}~,\\
	M_3 &= B_b B_d^{-1} B_e^{-1} B_f B_e B_c B_b^{-1} ~,
\end{split}
\qquad
\begin{split} 
	M_2 &= B_b B_c^{-1} B_d B_b^{-1} B_a ~,\\
	M_4 &= B_b B_d^{-1} B_e^{-1} B_f^{-1} B_e B_d B_b^{-1}~,
\end{split}
\end{equation}
using the following building blocks
\begin{equation} 
\begin{aligned}
&B_a=\ss\sv\se (a)\sv~,~ B_b = \se(b)\sv ~,~ B_c = \se(c) ~,~ \\
&B_d = \sv\se(d)\sv\ss\sv ~,~ B_e = \se(e)\sv\ss ~,~ B_f = \ss\sv\se(f)\sv ~,
\end{aligned}
\end{equation}  
with $\ss$, $\sv$, $\se$ being the following matrices
\begin{equation}
\mathsf{s}=  \left( \begin{matrix} 0 & 1 \\ -1 & 0\end{matrix} \right)~,  \quad 
\mathsf{v}=   \left( \begin{matrix} 1 & 0 \\ 1 & 1\end{matrix} \right)~, \quad 
\mathsf{e}(x)= \left( \begin{matrix} 1/\sqrt{x} & 0 \\ 0 & \sqrt{x} \end{matrix} \right)~.
\end{equation}
The monodromies are
defined choosing the point marked in red in Figure \ref{FG4a} as a base point. The resulting
expressions for the trace function $L_u$ is
\begin{align} \label{eq:tracesFG}
L_u=-\frac{1}{b d e \sqrt{af}} (&a b^2 d^2 e^2 f+a b^2 d^2 e f+a b^2 d^2 e+a b^2 d^2+a b d^2 e^2 f+a b d^2 e f+
a b d^2 e  \notag\\[-1ex]
 & + a b d^2+a b d+b d^2 e^2 f+b d^2 e f+b d^2 e+d^2 e^2 f+d^2 e f \notag\\
 & +d^2 e+b d^2+b d+d^2+d e f+d e+2 d+1) ~.
\end{align}
The eigenvalues $\m_i$
of the matrices $M_i$,  are for $i=1,2,3,4$ given respectively as
\begin{equation}
	\m_1=\sqrt{a}\,,\quad
	\m_2=-b\sqrt{ad/c}\,,\quad
	\m_3=-e\sqrt{df/c}\,,\quad
	\m_4=\sqrt{f}\,. 
\end{equation}
It is thereby possible to express the variables $a,b,e$ and $f$
as
\begin{equation}
	a=\m_1^2\,,\quad
	b=-\frac{\m_2}{\m_1}\sqrt{\frac{c}{d}}\,,\quad
	e=-\frac{\m_3}{\m_4}\sqrt{\frac{c}{d}}\,,\quad
	f=\m_4^2\,. 
\end{equation}
Using these relations together with $X=d$,  $Y=c^{-1}$
one obtains a parameterisation of all monodromy matrices in terms of 
$X$, $Y$, and $\m_1,\dots,\m_4$. By further using 
\rf{eq:XYtoUV-first} one obtains a parameterisation 
of the monodromies on $C_{0,4}$ in terms of the variables $\si$ and $\eta$.

In order to compare the coordinates $\si$ and $\eta$ defined in this 
way with the coordinates defined  by means of abelianisation in 
\cite{CPT} one may note that the resulting expressions for the 
trace functions $L_{12}= {\rm Tr} (M_1M_2)$ and $L_{23}= {\rm Tr}( M_2M_3)$ are found to be of the form 
\begin{equation}
\label{eq:tracesFG-modif}\begin{aligned}
L_{12} &= 2\cos 2\pi \sigma' ~,\\
L_{23} &= e^{4\pi\mathrm{i}\eta'}C_+(\si')+C_0(\si')+e^{-4\pi\mathrm{i}\eta'}C_-(\si'),
\end{aligned}
\end{equation}
with $C_{\pm}(\si)$ given as
\begin{equation}
C_s(\si) = 4\,\frac{e^{-4s \pi\mathrm{i}\sigma}} {(\sin 2\pi \sigma)^2}  \prod_{s'=\pm}
{\sin\pi(\sigma +s'\theta_1-s\theta_2) 
\sin\pi(\sigma+s'\theta_4-s\theta_3)} . 
\end{equation}
In this way it becomes easy to check that the coordinates $\si'$, $\eta'$ introduced above\footnote{{The relation between $(\sigma,\eta)$ and $(\sigma',\eta')$ was given in Section \ref{sec:FN-for-PIII}.}} 
as a 
parameterisation of the Fock-Goncharov coordinates $c$ and $d$ 
coincide with one of the systems of coordinates on $\CM_{\rm ch}(C_{0,4})$ 
defined using abelianisation in \cite{CPT}\footnote{
To be fully precise, there exists a relative normalisation factor between the coordinates $\eta_\text{\ CPT}$ and $\eta'$ here,  
$e^{2\pi\mathrm{i}\eta_\text{\ CPT}}=e^{2\pi\mathrm{i}\eta'}  \frac{\CE(2(\sigma +1))}{\CE(\sigma )}$, like in equation \eqref{eq:e-E}, 
with $\CE(x)=(2\pi)^{-x} e^{\frac{\pi\mathrm{i}}{2}x^2}\frac{G(1+x)}{G(1-x)}$. This corresponds to a change of normalisation of the 
corresponding normalised tau-functions $\CT^{(\eta')}$ and $\CT^{(\eta_\text{\ CPT})}$ like in equation 
\eqref{eq:E-transition-tau}.
}.

It is clear that 
a similar relation between FG and FN coordinates will be found for 
all fat graphs on $C_{0,4}$. 
In order to see this, it suffices to note that an arbitrary fat graph of $C_{0,4}$ is related to
the one in Figure \ref{FG4a} by a sequence of flips. 
As the flips are represented in terms of the Fock-Goncharov coordinates by rational changes
of coordinates, it suffices to compose changes of coordinates induced by the flips relating
a generic fat graph to the one depicted in Figure \ref{FG4a} with the change of coordinates 
described above.

We had noted above that there exists
a difference generating function for the change of coordinates \rf{eq:XYtoUV-first}. 
As essentially the same change of variables can be  used in the case of $C=C_{0,4}$,
we may conclude that
there exist normalised tau-functions having expansions of theta series type involving 
FG coordinates also  in the case $C_{0,4}$.

\section{Coordinates from exact WKB}\label{sec:exact-WKB}
\setcounter{equation}{0}

{In the previous section we have illustrated how the theta-series expansion of the tau function changes with the choice of Darboux coordinates, both of FG and of FN type.
Recall that the coefficients of FN-type theta series expansions, denoted above by $\CF$, can be interpreted as topological string partition functions.
This prompts the question of whether certain kinds of expansions can be more natural than others in 
a given region of the moduli space of quantum curves $\mathcal{Z}$. 
Reformulating this question in terms of Darboux coordinates, 
}
this section discusses how exact WKB can be used to associate distinguished 
coordinates of both FG- and FN-type to generic pairs $(\Sigma,\la)$, where
$\Sigma$ is the classical curve, and $\la\in\BC^\times$. 

\subsection{WKB networks} \label{WKBnetworks}

The double cover $\Sigma=\{(y,x);y^2+q(x)=0\}$ defined by a quadratic differential $q(x)$
comes equipped with a canonical differential $\sqrt{q}$. We will assume $q$ to have at 
least one pole of order at least two. A trajectory of $(q,\la)$ is a leaf of the 
foliation defined
by the equation 
\begin{equation}\label{w-def}
\mathrm{Im}(w(x))=\mathrm{const.},\qquad w(x)=e^{-\mathrm{i}\arg(\la)}\int^x\!dx'\,\sqrt{q(x')}.
\end{equation}
A Stokes curve is a trajectory having at least one end at a branch point. 
The Stokes graph $\CS_{q,\la}$ is the graph formed by the Stokes curves of $(q,\la)$.
$\CS_{q,\la}$ has vertices at
the branch points of $\Sigma$, and edges represented 
by trajectories emanating from the branch points. 

Saddle trajectories are Stokes curves which either connect two branch points, or closed trajectories which 
encircle double poles of $q(x)$.
For generic $(q,\la)$ one finds Stokes graphs which do not have saddle trajectories. This means that
all Stokes curves emanating from a branch point must end at poles of $q$. The complement of 
$\CS_{q,\la}$ is a collection of contractible regions foliated by trajectories connecting poles of $q$. 
Choosing a generic representative trajectory from each region defines an ideal triangulation of $C$ 
called the WKB triangulation. 
Generically, regions come in two types, called strips and half-planes. 
Strips are quadrangles with two double poles of $q$ on the boundary, and all trajectories stretch between these poles.
Half-planes arise when $q$ has a pole of degree higher than two, then trajectories sufficiently close to it have both endpoints on the same pole, but end on it along different (anti-)Stokes rays.

It is useful to study the families of Stokes graphs $\CS_{q,\la}$ obtained by
varying the phase $\arg(\la)$. These families depend on the choice of $q$ in an interesting way
which has attracted a lot of attention in view of applications to the spectrum 
of BPS-states in $\CN=2$, $d=4$ supersymmetric field theories of class $\CS$ \cite{GMN}. 
For generic values of $\arg(\la)$ one finds Stokes graphs defining a WKB triangulation.
The topological type of the WKB triangulation will not change for $\arg(\la)$ in a wedge of the 
complex $\la$-plane. Any two such wedges are separated by a ray through the origin 
of the $\la$-plane associated to a change of the topological type of Stokes graph. 

\subsubsection{Ring domains}\label{sec:ring-domains}

Non-generic cases of particular importance for us are pairs $(q,\la)$ where 
the Stokes graph has non-degenerate ring domains, annular regions embedded in
$C$ which have empty intersection with $\CS_{q,\la}$. 
Within a ring domain one 
can use the function $w$ defined in \rf{w-def} as a local coordinate. 
The ring domain is foliated by closed trajectories defined
by the equation $\mathrm{Im}(w(x))=\mathrm{const}$. A ring domain is called 
degenerate if it consists of closed 
curves encircling a double pole of $q$. 
The boundary of a non-degenerate ring domain consists of unions of saddle trajectories 
connecting a finite, non-empty set of branch points. 
A collection of Stokes graphs illustrating the transition between Stokes graphs with 
and without ring domains is depicted in Figure \ref{fig:ring-domains} for 
quadratic differentials on $C=C_{0,2}$ of the form 
\begin{equation}\label{eq:SU2-quad-diff}
q(x)=\frac{\Lambda^2}{z^3}+\frac{U}{z^2}+\frac{\Lambda^2}{z}.
\end{equation}

\begin{figure}[h!]
\begin{center}
\fbox{\includegraphics[width=0.23\textwidth]{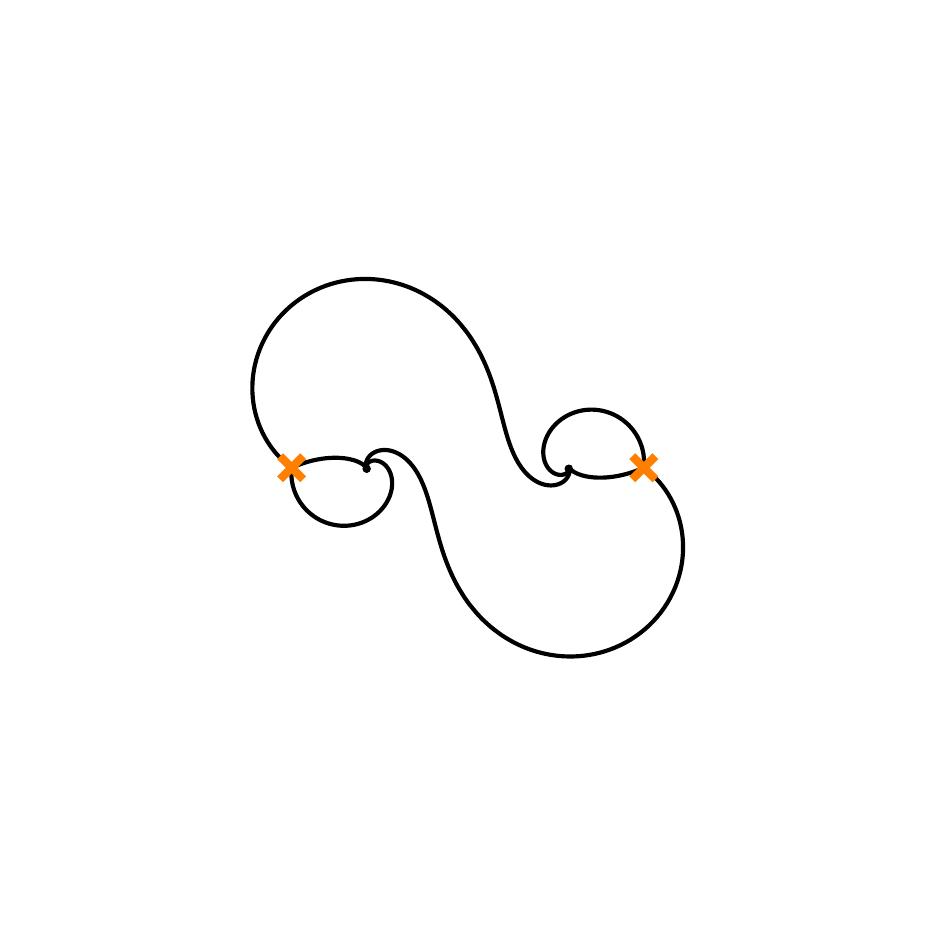}}
\fbox{\includegraphics[width=0.23\textwidth]{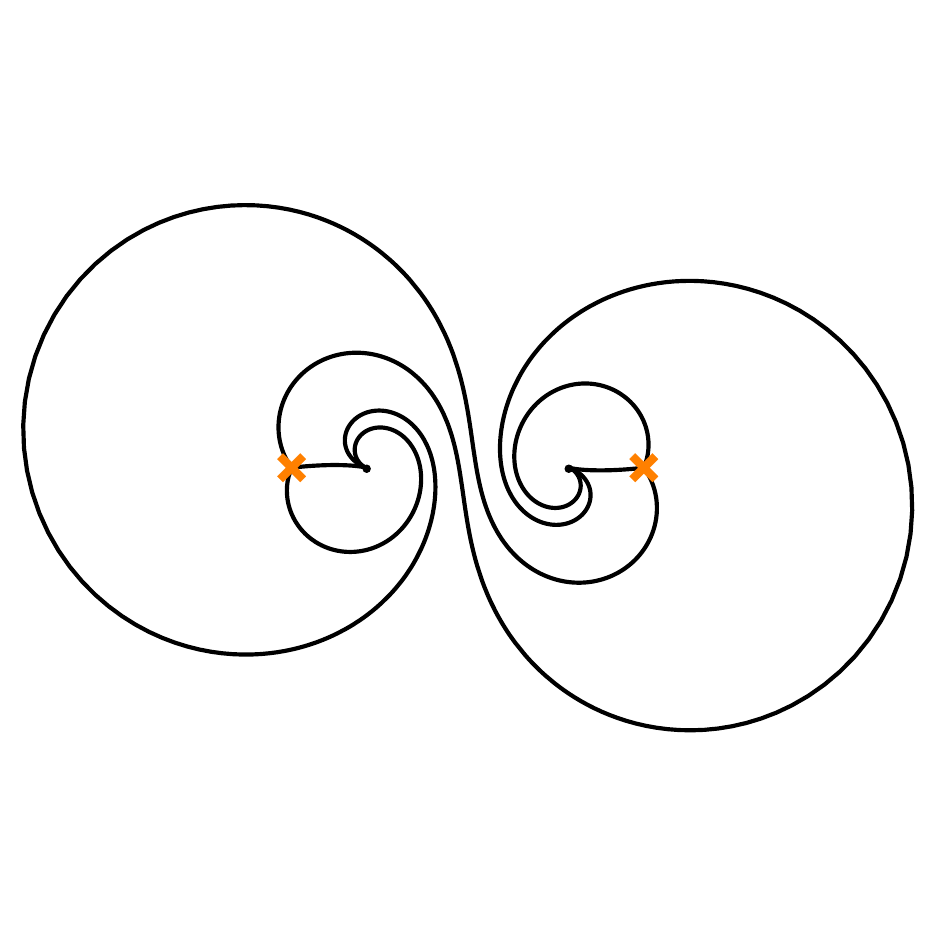}}
\fbox{\includegraphics[width=0.23\textwidth]{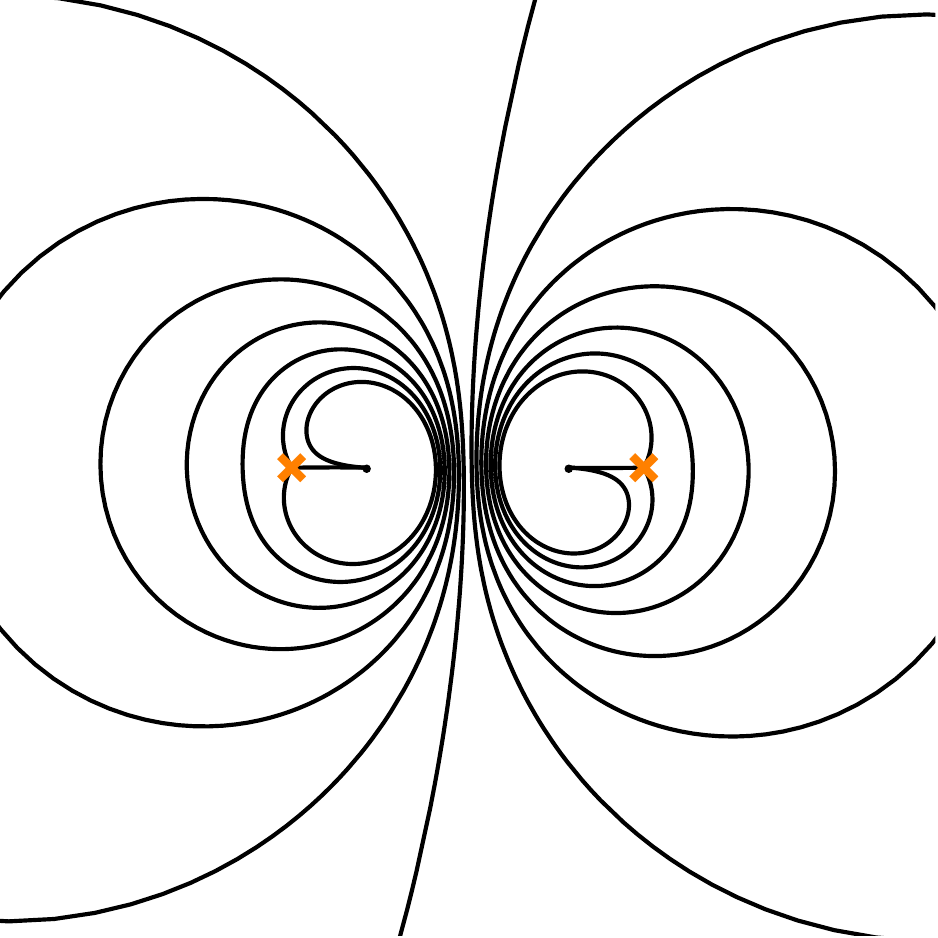}}
\fbox{\includegraphics[width=0.23\textwidth]{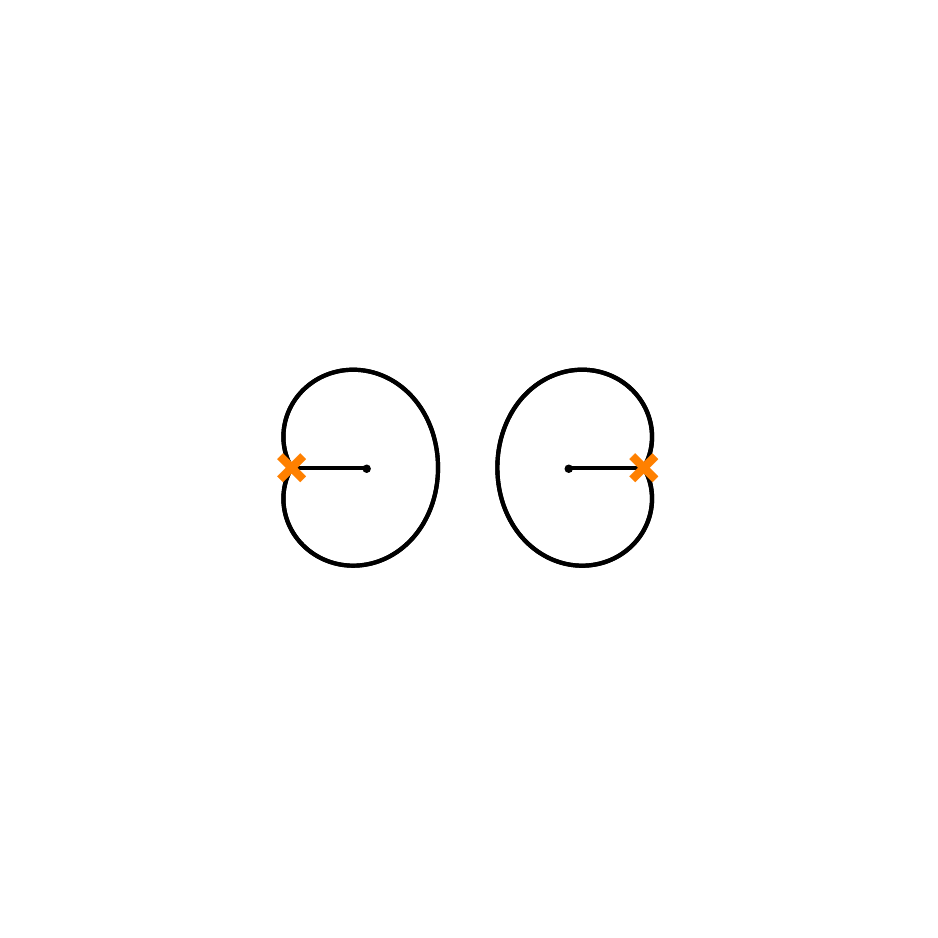}}
\caption{Stokes graphs for the differential (\ref{eq:SU2-quad-diff}) with $\Lambda=1$ and $u=2$, varying $\arg(\hbar)$ until a non-degenerate ring domain appears. 
}
\label{fig:ring-domains}
\end{center}
\end{figure}

The existence of a ring domain in the complement of $\CS_{q,\la}$ 
is unstable under small perturbations of 
$q$ or the value $\la_\ast$ of $\la$ at which the 
ring domain appears.
In order to understand the nature of this instability one may consider a 
branch point $b$ on the boundary of a ring domain, and the two saddle trajectories
emanating from it that bound the ring domain. 
A small perturbation $(q,\la_\ast e^{\mathrm{i}\ep})$ of $\la_\ast$ will yield a Stokes graph 
$\CS_{q,\la_\ast,\ep}$ having two Stokes lines emanating from $b$
rotated in the same direction relative to the  two saddle 
trajectories of $\CS_{q,\la_\ast}$ 
emanating from $b$ by an angle proportional to $\ep$.
One of these Stokes lines must traverse the ring domain, the other one will not, see Figure \ref{fig:ring-domains}.
An arc connecting 
two branch points on different boundaries of one of the ring domains will intersect any Stokes line
traversing the ring domain a number to times which diverges when $\ep\ra 0$.
This phenomenon
is closely related to the phenomenon called ``juggle'' in \cite{GMN}, featuring two infinite towers of saddles as one takes $\epsilon \to 0^\pm$.
For this reason we will adopt the name ``accumulation ray'' for the ray $\hbar_\ast \mathbb{R}_{>0} \subset \mathbb{C}$ where the ring domain appears.

 The neighbourhood of a pair $(q,\la)$ having 
a Stokes graph with ring domains displays certain universal features 
which play an important role in our story. 
 It will turn out that, roughly speaking,  there is a finite neighbourhood of $\ep=0$
in which the Stokes graphs $\CS_{q,\la_\ast,\ep}$ vary
substantially only 
within ring domain occurring at $\ep=0$, and only very little outside. 
We will in the following describe this behaviour more precisely in the case $C=C_{0,4}$.
As a preparation it will be useful  to consider the case $C=C_{0,3}$ first.

\subsubsection{Stokes graphs on the pair of pants}\label{Stokespants}

The unique family of quadratic differentials with double poles at punctures on $C_{0,3}$ is:
\begin{equation}
	q_0(x) = \frac{- a_3^2 x^2 +  (a_3^2 + a_1^2 - a_2^2 ) x - a_1^2}{x^2(x-1)^2}.
\end{equation}
The 
classification of the  topological types of Stokes graphs is fairly simple
in this case \cite{AT}: The Stokes graphs can be classified by the number of Stokes lines
ending in the punctures. There exist four possibilities, depicted\footnote{These graphs have been plotted using the Mathematica package [Npl].} 
in Figure \ref{Stokesgraphs}. In three of the 
cases there are four Stokes lines ending in one puncture, and only one Stokes line ending in each 
of the remaining punctures. In the fourth case there are two Stokes lines ending in each puncture. 

 In the cases where four Stokes lines end in a puncture, this puncture is distinguished. 
It is natural to denote the Stokes graphs having distinguished puncture $z_i$ by $\CS_i$, for $i=1,2,3$, 
respectively. The remaining one is more symmetric than the other three, suggesting the 
notation $\CS_s$.

%
\begin{figure}[h]
\centering
\includegraphics[width=0.8\textwidth]{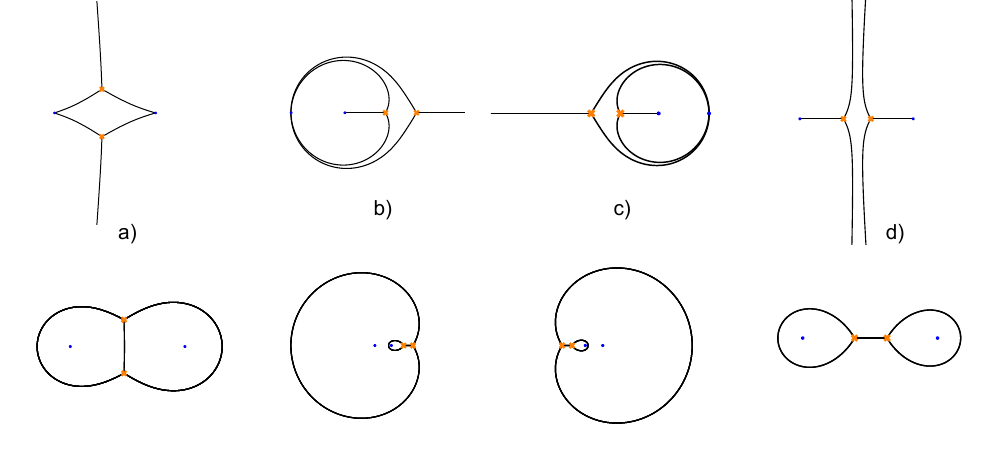}
\caption{\it Top row: Stokes graph examples from \cite{AT}. Bottom row: Corresponding Fenchel-Nielsen networks, in this case identical with Anti-Stokes graphs.}
\label{Stokesgraphs}
\end{figure}
The topological type of Stokes graphs is determined in a simple way by the parameters 
$\vartheta_i=\mathrm{i} a_i/\hbar$, $i=1,2,3$, with $a_i$ corresponding to residues of $\sqrt q$ at the three punctures. 
The type $\CS_i$ is found when 
\begin{subequations}\label{chamberdef}
\begin{equation}
\mathrm{Re}(\vartheta_i+s\,\vartheta_{i+1}+s'\,\vartheta_{i+2})>0,\qquad \forall \;\,s,s'=\pm 1,
\end{equation}
while $\CS_s$ will be realised if   
\begin{equation}
\begin{aligned}
\mathrm{Re}(\vartheta_1+\vartheta_{2}+\vartheta_{3}-2\vartheta_i)&>0,
\quad \,i\in\{1,2,3\},\\
\text{and}\quad\mathrm{Re}(\vartheta_1+\vartheta_{2}+\vartheta_{3})&>0.\\
\end{aligned}
\end{equation}
\end{subequations}
Saddle trajectories are found at the boundaries of the regions defined by the inequalities defining $\CS_s$
and $\CS_i$, $i=1,2,3$.

\emph{Comment on conventions:} 
The quadratic differential near a puncture is $q_0 = -\frac{a^2}{x^2} + \dots$. 
The Stokes lines of $\hbar^2\partial^2_x - q_0(x)$ for $\hbar \in \mathbb{R}$ correspond to horizontal trajectories emanating
from $x = 0$ for $a\in \mathrm{i}\mathbb{R}$, and to closed trajectories encircling $x = 0$ for $a\in\mathbb{R}$. 
We thus adopt the 
definition of the parameter $\hbar = {\rm i}\lambda$, with $\lambda$ being the string coupling in \cite{CPT}. 

It will furthermore be useful to keep in mind that
parameters like 
$\vartheta_i$ in \cite{CPT} coincide with parameters used here. They were
naturally real in \cite{CPT}, being related to K\"ahler parameters $t_i$ as  
$t_i=R\lambda\vartheta_i$, 
assuming that $R$ and the topological string coupling $\lambda$ 
are real.
The parameters $\vartheta_i$ in \cite{CPT} were related to the residue parameters 
$a_{i}$ in the classical curve as 
$a_{i}=\vartheta_i\lambda$. Real values of the K\"ahler parameters $t_i$ therefore correspond
to real values of the variables $a_i$ used in this paper.
}

\subsubsection{Cutting of  Stokes graphs}

Let us now investigate the case of a Stokes graph $\CS_{q,\la}$ on $C=C_{0,4}$ 
having a ring domain. Cutting 
along two distinct closed trajectories foliating the ring domain 
decomposes $C$ into an annular region and two spheres having two punctures and a hole each. 
One can fill punctured discs into the holes producing two three-punctured spheres $C_1$ and $C_2$
with Stokes graphs defined by $\CS_{q,\la}$. The quadratic differentials $q^{(1)}$ and $q^{(2)}$
defined by this procedure
on $C_1$ and $C_2$, respectively, 
have a double pole at the puncture contained in the discs glued into holes
with a residue given by the period of $\sqrt{q}$ along the closed trajectories used to decompose $C$.
The construction ensures that two out of the four branch points of $q$ will be found on each of the 
two surfaces $C_1$ and $C_2$.
The Stokes graphs $\CS_{q,\la_\ast,\ep}^{(1)/(2)}$ 
defined on $C_1$ and $C_2$ 
using $q^{(1)}$ and $q^{(2)}$
are  obtained
from $\CS_{q,\la}$ by cutting $C$ as described above, and deleting all trajectories 
which do not emerge from the branch points located in $C_1$ and $C_2$, 
respectively.

We may then discuss  using this decomposition 
of $C$ how $\CS_{q,\la_\ast,\ep}$ varies with $\ep$.  
A change of the topological type of the Stokes graph occurs exactly when a saddle trajectory occurs. 
This happens when two Stokes lines emanating from branch points meet each other.
The decomposition of $C$ into $C_1$,  $C_2$ and an annulus allows one to 
distinguish saddle trajectories located entirely in $C_1$ or  $C_2$ from 
saddle trajectories traversing the annulus.  
For $\ep\neq 0$ one finds a pair of Stokes graphs $\CS_{q,\la_\ast,\ep}^{(1)}$ and
$\CS_{q,\la_\ast,\ep}^{(2)}$
on $C_1$ and $C_2$ which each must have
one of the four topological types described in Section \ref{Stokespants}, respectively.\footnote{%
The range of $\epsilon$ must be taken sufficiently small, for the pants decomposition to hold.
This is because in some cases another non-degenerate ring domain may appear at finite $\epsilon\neq 0$, inducing a different pants decomposition.
}
Importantly there is a finite range of values for $\ep$ in which the topological type of the 
Stokes graphs $\CS_{q,\la_\ast,\ep}^{(1)/(2)}$ 
does not change under variations of $\ep$. 
This easily follows from the well-known fact that there is a finite number of saddles for $\CS_{q,\hbar_\ast,\epsilon}^{(1)/(2)}$
as one varies $\epsilon$ for a fixed $q$.\footnote{There are precisely eight saddles, corresponding to BPS states of the $T_2$ trinion theory, the free theory of a half-hypermultiplet in the tri-fundamental representation of $SU(2)^{\times 3}$.}
Any other boundary between two 
regions $\CS_i$ for $i=1,2,3,s$ can only be found at finite non-vanishing values of $\ep$.
In any neighbourhood of $0$ 
there will, on the other hand, always exist infinitely many values of $\ep$ at which there
exist saddle trajectories traversing the annulus.

\subsection{WKB expansion}\label{Sec:Exact WKB}

The solutions of the basic ODE $(\la^2\pa_x^2-q_{\la}^{}(x))\chi(x)=0$ 
can be represented as an expansion 
in 
$\la$, usually referred to as the WKB-expansion. We will here give a very brief summary of the 
relevant results, referring to \cite{IN} for a more detailed discussion and references to the original 
literature.

The WKB-expansion can conveniently be described 
by first representing the solution ${S}$ of the Riccati equation $q_\la=\la^2({S}^2+{S}')$ as a formal series
\begin{equation}\label{WKBseries}
{S}(x)\equiv{S}(x;\la)=\sum_{k=-1}^{\infty} \la^k{S}_k(x).
\end{equation}
The family of functions $\{{S}_k(x);k\geq -1\}$ must satisfy the recursion relations 
\begin{equation}\label{WKBRiccati}
({S}_{-1}(x))^2=q^{(0)}(x)~, \qquad
2{S}_{-1}{S}_{n+1}+{S}_n' +\sum_{\substack{k+l=n\\0\leq k,l\leq n}} {S}_k{S}_l=
q^{(n+2)}\quad \text{for}\;\,n\geq -1,
\end{equation}
where $q^{(n)}(x)$ is defined by $q_\la(x)=\sum_{n=0}^{\infty}\la^nq^{(n)}(x)$.
The series \rf{WKBseries} is usually not convergent, but turns out to be Borel summable under certain 
conditions,\footnote{The relevant results are in the literature
attributed to a paper in preparation by Koike and Sch\"afke. {N. Nikolaev has announced an alternative proof in upcoming work \cite{Ni}.}} see \cite{IN} for 
a careful discussion of the relevant results. Let ${S}^{(\pm)}$ be the solutions to the 
Riccati equation with leading terms $\pm\sqrt{q_0}$, and let ${S}_{\rm\sst odd}=\frac{1}{2}({S}^{(+)}-{S}^{(-)})$.

One natural pair  $\chi^{(b)}_{\pm}$
of solutions to $(\la^2\pa_x^2-q_{\la}^{}(x))\chi(x)=0$ is normalised at a branch point $b$ of $\Sigma$.
The solutions can be represented in the form
\begin{equation}\label{WKBsol}
\chi_\pm^{(b)}(x)=\frac{1}{\sqrt{{S}_{\rm\sst odd}(x)}}\exp\bigg[\pm\int_{\CC_{b,x}}dx'\;{S}_{\rm\sst odd}(x')\bigg].
\end{equation}
The contour $\CC_{b,x}$  can be chosen to be one-half of a path $\be_{b,x}$ on $\Sigma$ 
starting at the pre-image of $x$
on the second sheet, encircling the branch point $b$, and ending at the pre-image of 
$x$ on the first sheet. Such solutions clearly depend on the choice of the branch point $b$ and 
satisfy a simple normalisation condition at this point.

Another  pair $\chi_{\pm}^{(p)}$ of solutions  is normalised at a double pole $p$ of the quadratic differential $q$.
Assuming that $x$ is a local coordinate such that $x(p)=0$ we may define $\chi_{\pm}^{(p)}$ as
\begin{equation}
\chi_{\pm}^{(p)}(x)=\frac{x^{\pm(\si-\frac{1}{2})}}{\sqrt{{S}_{\rm\sst odd}}}\exp\left(\pm\int_0^x dx' 
\left({S}_{\rm\sst odd}-\frac{2\si-1}{2x'}\right)\!\right),
\quad\si=\frac{1}{2}+\mathop{\rm Res}_{x=0}({S}_{\rm\sst odd}).
\end{equation}
Note that on $C_{0,3}$ this implies the following relation $\sigma = \frac{1}{2} + \frac{a_i}{\hbar}$, if the puncture $p$ is identified with the puncture with residue $a_i$.

Another important object defined using the WKB expansion is the Voros-symbol $e^{V_{\ga}}$ associated to the 
cycle $\ga$ on $\Sigma\setminus \hat{P}$, with $\hat{P}$ being the pre-image of the set of poles and zeros of
$q$ under the covering projection $\Sigma\ra C$.
The formal series $V_\ga$ is defined as \cite{DDP,IN}
\begin{equation}
V_{\ga}=\int_{\ga}dx'\;{S}_{\rm\sst odd}(x').
\end{equation}
The definition of the Voros symbols can be generalised to paths starting and ending at poles of~$q$. 
Assuming that $p$ is a double pole of $q$ as considered above, and $b$ is a branch point lying in the 
same chart  with coordinate $x$ one can define \cite{IN,ATT1}
\begin{equation}\label{Voros-split}
V^{(pb)}=V^{(pb)}_{>0}+V^{(pb)}_{\leq 0},\qquad
\begin{aligned}
&V^{(pb)}_{\leq 0}=\frac{1}{2}\lim_{x\ra 0}\left(\int_{\ga_x}dx'\;{S}^{^{\sst\rm odd}}_{\leq 0}(x')+(2\si-1)\log(x)\right),\\
&V^{(pb)}_{> 0}=\frac{1}{2}\int_{\ga_0}dx'\;{S}^{^{\sst\rm odd}}_{>0}(x'),
\end{aligned}
\end{equation}
using the notations ${S}^{^{\sst\rm odd}}_{>0}=\sum_{k>0}\la^k{S}^{^{\sst\rm odd}}_{k}$,
${S}^{^{\sst\rm odd}}_{\leq 0}=\sum_{k\leq 0}\la^k{S}^{^{\sst\rm odd}}_{k}$, and $\ga_x$ being a contour starting at
$x$, encircling $b$ before returning to $x$.

Under certain conditions it turns out that the series for $\chi_{\pm}^{(b)}$,  and ${V}_\ga$ are Borel summable. In particular this holds if the foliation has at most one saddle connection, ensuring that each domain is either a strip, a half-plane, or a degenerate ring domain. 
We refer to \cite{IN} for a review of the known results on Borel summability. 

\subsection{Fock-Goncharov coordinates from exact WKB}

If the formal series defining the Voros symbols $V_{\ga}$ are Borel summable one may
use the functions defined thereby as local coordinate functions for the space $\CZ$ of 
quantum curves. Whenever the Stokes graph associated to $(q,\la)$ is saddle-free
one may  naturally assign  cycles $\ga$ defining $V_{\ga}$ to edges of the
WKB triangulation associated to $(q,\la)$ as follows. An edge of the WKB triangulation
may either be a boundary edge corresponding to a generic leaf trajectory near a higher-order 
pole of $q$, or it may separate two triangles containing different branch points of $q$, or it 
may be an edge in the interior of a self-folded triangle obtained by deforming an
ordinary triangle until two corners and two sides lie on top of each other, being surrounded 
by the third side. 
We will not assign cycles to boundary edges.
In the case of internal edges, one may define $\ga$
as the homology class of 
the anti-invariant lift of a segment dual to the edge, 
connecting the pair of branch points on either side.
The orientation may be fixed by demanding ${\rm Re}\hbar^{-1} \oint_\gamma\sqrt{q_0} >0$.  
The definition of the path for the case of a self-folded triangle can be found in \cite[Figure 11]{IN}.

The Borel sum of the Voros symbol then defines coordinate functions on $\CZ$. 
It has been shown in \cite{Al19} that these coordinate functions coincide with the 
coordinates obtained from the 
FG coordinates associated to the WKB triangulation defined by $(q,\la)$
by composition with the holonomy map from $\CZ$ to the character variety.
It seems natural to conjecture that the relation between Voros symbols and 
Fock-Goncharov extends from $\CM_{\rm op}^\la(C)$ to 
$\CM_{\rm conn}^\la(C)$.

The relation between 
Fock-Goncharov coordinates and Voros symbols proven in \cite{Al19}
offers a direct explanation of the previous results from
\cite{DDP,IN} showing that the  transformations of the Voros symbols 
associated to a flip of WKB triangulation take the form of 
cluster mutations familiar from the theory of FG coordinates.
This implies that the transformations of the 
FG coordinates induced by a change of Stokes graph dual to a flip of the triangulation represent
examples of the Stokes phenomenon studied in \cite{Br}. 
The FG coordinates associated to a particular
type of Stokes graph admit analytic continuations beyond the region of $\CZ$ characterised 
by the given Stokes graph. The cluster transformations give the expressions for the 
analytic continuation of the set of coordinates associated to one type of WKB triangulation 
in terms of the coordinates associated to a WKB triangulation related to the first one by
a flip.

\subsection{Fenchel Nielsen type coordinates from exact WKB}\label{FNfromWKB}

Exact WKB can be used in a rather different way to define coordinates of Fenchel-Nielsen
type. 
A pair $(q,\la)$ having an associated Stokes graph with a maximal number of non-degenerate ring domains defines a pants decomposition of $C$.\footnote{%
For example, this is the case if $q_0$ is a Strebel differential (this is the case considered in \cite{HN} to construct Fenchel-Niesen coordinates). 
However one should keep in mind that the Strebel condition is sufficient but not necessary: having a maximal number of ring domains is a weaker condition. 
} 
When $C$ is a sphere with four regular punctures, this amounts to having one ring domain.
We may first recall from Section \ref{FN-gen}
that the FN type coordinates associated to a given pants decomposition 
are uniquely defined by fixing the relative normalisation of the solutions $\Phi_i(x_i)$
on the two pairs of pants $C_i$, $i=1,2$, arising in such a decomposition.

It remains to be noticed that the exact WKB method offers natural
ways to fix the normalisation of each of the solutions $\Phi_i$, $i=1,2$, 
{by constructing them as Borel sums $\Psi_i$ of formal WKB series.}
We choose a pair $(q,\hbar_*)$ whose Stokes graph features a non-degenerate ring domain $A$, and assume for simplicity that the corresponding non-contractible cycle on $C$ is separating. 
We then choose $\hbar\neq \hbar_*$ and generic, meaning that $(q,\hbar)$ is saddle-free on  $C$, within a sufficiently small wedge centered around $\hbar_*$. 
Picking branch points $b_i$, $i=1,2$ on distinct components of the boundary of $A$, we may define
\begin{equation}\label{Yfrompsi}
	\Psi_i(x)
	=\left(\begin{matrix} \pa_x\psi^{(b_i)}_{+}(x) & \pa_x{\psi}^{(b_i)}_-(x) \\ \psi^{(b_i)}_+(x) & 
{\psi}^{(b_i)}_-(x) \end{matrix}\right),
\end{equation}
where $\psi^{(b_i)}_\pm(x)$, $i=1,2$, are defined on the annulus $A$, respectively in neighbourhoods of $b_i$, as follows:
\begin{itemize}
\item
$\psi^{(b_i)}_+(x)$, $i=1,2$ are the solutions to $(\la^2\pa_x^2-q_{\la}^{}(x))\psi(x)=0$
defined by Borel summation of the WKB series \rf{WKBsol}
with the following behavior along a Stokes line $x(\tau)$ traversing the annulus,%
\footnote{More precisely, Borel resummation is being invoked separately within $C_i$, $i=1,2$. Then by `Stokes lines traversing the annulus' we mean the Stokes lines emanating from $b_i$ which flow into the puncture on $C_i$ at which the annulus is glued.}
\begin{equation}\label{psi+def}
\psi^{(b_i)}_+(x)=\left\{
\begin{aligned}
\chi_+^{(b_i)}(x) \quad\text{if}\quad
\frac{d}{d\tau}\left( e^{-\mathrm{i}\,\arg{\la}}\int_{b_i}^x\sqrt{q_0}(x) \right) <0,\\
\chi_-^{(b_i)}(x) \quad\text{if}\quad
\frac{d}{d\tau}\left(  e^{-\mathrm{i}\,\arg{\la}}\int_{b_i}^x\sqrt{q_0}(x) \right) >0,
\end{aligned}\right.
\end{equation} 
\item $\psi_{-}^{(b_i)}(x)$, $i=1,2$, are the unique solutions to $(\la^2\pa_x^2-q_{\la}^{}(x))\psi(x)=0$
with diagonal monodromy around $A$
satisfying the  condition
\begin{equation}\label{Wronski}
\mathrm{det}(\Psi_i(x))=\psi_{-}^{(b_i)}(x)\pa_x\psi_{+}^{(b_i)}(x)-\psi_{+}^{(b_i)}(x)\pa_x\psi_{-}^{(b_i)}(x)=1.
\end{equation}
The functions $\psi_{-}^{(b_i)}(x)$ can be expressed in terms of $\psi_{+}^{(b_i)}(x)$ by the formula
\begin{equation}\label{psi-frompsi+}
\psi_{-}^{(b_i)}(x)=\frac{1}{1-e^{2\pi\mathrm{i}\,\si}}\,\psi_{+}^{(b_i)}(x)\int_{\ga_x}dx' \,{(\psi_{+}^{(b_i)}(x'))^{-2}},\qquad
i=1,2,
\end{equation} 
where  $\si$ is defined by representing
the monodromy of $\psi_+^{(b_i)}(x)$ around $A$ as 
$(\ga_x\psi_+^{(b_i)})(x)=e^{-2\pi\mathrm{i}\,\si}\psi_+^{(b_i)}(x)$, and $\ga_x$ is a closed contour 
starting at $x$ encircling $A$ exactly once.  
\end{itemize}

This definition is ``as canonical as possible'' in the sense that it only depends on the choices of 
branch points $b_i$, $i=1,2$, whenever boundaries of $A$ have more than one branch point on either component.\footnote{It may also happen that a single branch point occurs more than once along a boundary component of $A$. An example of this can be found in Figure \ref{Stokesgraphs} b), with $A$ glued into the left-most puncture. As the bottom frame shows, the boundary of $A$ features two branch points, each of them occurs twice as one travels along the boundary. This kind of degeneracy introrduces a finite degree of additional freedom of choice.} 
The definition of $\psi_+^{(b_i)}(x)$ invokes Borel resummation for the wavefunctions, which is defined piecewise in domains bounded by Stokes lines. 
However the choice of a recessive WKB solution, which is preserved by Stokes jump across the corresponding Stokes line, is well defined in a neighbourhood of $b_i$ that extends across the Stokes line.
This is a direct consequence of the connection formula derived in \cite[Section 6]{Vo}.
The definition of $\psi_+^{(b_i)}(x)$ is then extended to a universal covering of the annulus $A$ by analytic continuation away form this neighbourhood.
On the other hand, the definition of $\psi_-^{(b_i)}$ does not invoke Borel resummation at all, instead it relies entirely on the definition of $\psi_+^{(b_i)}$. Thus both these definitions are well-posed, for generic $\hbar$.



\subsection{Voros symbols as normalisation factors}\label{Voros-norm}

In the above we have been considering surfaces $C$ which 
can be represented by gluing two surfaces $C_i$ having punctures $P_i$ with coordinate
neighbourhoods $D_i$ surrounding $P_i$. This has been done by identifying
points in annuli $A_i=D_i\setminus D'_i$, $D'_i$ being smaller discs containing $P_i$.
The construction described in the previous subsection defines functions $\Psi_i(x)$
in the annulus $A\subset C$ constructed by identifying $A_1\subset C_1$
and $A_2\subset C_2$. 

Conversely, let $C$ be a Riemann surface, and let $q_0$ be a holomorphic quadratic differential on $C$
such that there exists $\la\in\BC^\ast$ such that $\la^{-2}q_0$ has a ring domain $A$.
The function $w(x)=\int^x \sqrt{q}$ defines a coordinate in $A$ making $q_0$ constant,
$(dw)^2=q(x)(dx)^2$. If $a=\frac{1}{2\pi\mathrm{i}}\int_\ga\sqrt{q_0}$ is the period around $A$, one can define 
a function
$y(x)=\exp\big(\pm
\frac{1}{a}\int^x_b \sqrt{q}\big)$ such that $\sqrt{q_0}=\pm a\frac{dy}{y}$.
If $b$ is a point on the boundary of $A$ one may choose
the sign in the definition of $y(x)$ such that $|y(x)|>1$ for $x\in A$, giving a
function $y(x)$ from $A$ to the annulus $A_h=\{y\in\BC;1<|y|<h\}$.

By cutting $C$ along the two boundaries of $A$ one can define two surfaces $\check{C}_{i}$, $i=1,2$, 
each having a hole and  containing annuli $A_i$ mapping to $A$ under the canonical embedding
$\check{C}_i\hookrightarrow C$.
The coordinate $y$ canonically defines  coordinates $y_i$ on the annuli $A_i$ for $i=1,2$. 
Gluing punctured discs to the inner or outer boundaries of $A_i$ defines, for $i=1,2$, respectively,
punctured surfaces $C_i$ with quadratic differentials $q_i$ 
having punctures $p_i$ contained in discs $D_i$ equipped with coordinates $y_i$ 
vanishing at $p_i$ in which 
the quadratic differentials $q_i$ are represented simply  as $q_i=\frac{a^2}{y^2}(dy_i)^2$.

In this set-up one has a natural one-to-one correspondence between flat sections $\Psi$ on $C$
and pairs of flat sections $\Psi_i$ on $C_i$. The Borel summation of the WKB expansion
can be used to define two types of 
flat sections in $C_i$, $i=1,2$. The flat section $\Psi_i$,
is of the form \rf{Yfrompsi} with 
$\psi_\pm^{(b_i)}$ normalised at the branch points $b_i$ at the
boundary of $D_i$. 
Another flat section denoted $\Psi_i^{(0)}$ is defined by replacing in \rf{Yfrompsi} 
$\psi_\pm^{(b_i)}$ by $\psi_{\pm}^{(p_i)}$, 
normalised at the punctures $p_i$, for $i=1,2$, respectively. 
Note that the functions
$\psi_{\pm}^{(p_i)}(y)$ behave near $y=0$ as
\begin{equation}\label{punctasym1}
	\psi^{(p_i)}_\ep(y)=\frac{y^{\ep(\si-\frac{1}{2})}}{\sqrt{{2\si-1}}} (1+\CO(y)),
\end{equation}
{corresponding to having $\nu_+^{(i)} = \nu_-^{(i)}$ in \eqref{Phiform}},
whereas the behavior of the functions $\psi^{(b_i)}_\ep(x)$ is of the form
\begin{equation}\label{punctasym2}
	\psi^{(b_i)}_\ep(y)=\nu^{(b_i)}_\ep \frac{ y^{\ep(\si-\frac{1}{2})}}{\sqrt{{2\si-1}}}(1+\CO(y)).
\end{equation}
The condition \rf{Wronski} implies the relation $\nu^{(b_i)}_+\nu^{(b_i)}_-=1$. 
It follows that the solutions 
$\psi^{(b_i)}_\ep(x)$ are  uniquely characterised by the normalisation 
factors $\nu^{(b_i)}_+$. It is not hard to show that 
$\nu^{(b_i)}_+$ are given by the Voros symbols\footnote{{Here and in (\ref{eq:normalization-factors-voros}), the appearance of opposite signs is related to the convention adopted in (\ref{eq:n-nu-def}).}}
\begin{equation}
\nu^{(b_1)}_+=e^{V^{(p_1b_1)}}\,,
\qquad
\nu^{(b_2)}_+=e^{-V^{(p_2b_2)}}\,.
\end{equation}

As explained in Section \ref{FN-gen}, each of the 
flat sections $\Psi_i$ and $\Psi^{(0)}_i$ leads to an unambiguous definition of  twist coordinates $\eta$
and $\eta_0$ of FN type, respectively.
Useful quantities characterising the relation between $\eta$
and $\eta_0$ are the normalisation factors 
\begin{equation}\label{eq:normalization-factors-voros}
	\mathsf{n}^{(1)}
	:=\frac{\nu^{(1)}_-}{\nu^{(1)}_+}=\big(\nu^{(1)}_+\big)^{-2}=e^{-2V^{(p_ib_i)}} \,,
	\qquad
	\mathsf{n}^{(2)}
	:=\frac{\nu^{(2)}_+}{\nu^{(2)}_-}=\big(\nu^{(2)}_+\big)^2=e^{-2V^{(p_ib_i)}} \,.
\end{equation}
The normalisation factors $\mathsf{n}^{(i)}$ allow us to represent the 
relation between $\eta$
and $\eta_0$ in the form
\begin{equation}\label{eq:eta-normalization-factors}
e^{2\pi\mathrm{i}\,\eta}\,\mathsf{n}^{(1)}\mathsf{n}^{(2)}=e^{2\pi\mathrm{i}\eta_0}.
\end{equation}
Indeed, $\Psi_i$ and $\Psi^{(0)}_i$ are related by right multiplication with 
the diagonal matrix $\mathrm{diag}(\nu^{(b_i)}_+,\nu^{(b_i)}_-)$. It follows that 
the matrices representing the monodromies of $\Psi_i$ and $\Psi^{(0)}_i$
have off-diagonal elements differing by factors of $(\mathsf{n}^{(i)})^{\pm 1}$.

\subsection{FN-type coordinates as limits of FG-coordinates}\label{sec:juggle}

We have now seen how Exact WKB can be employed in different situations to produce coordinates of two different types.
On the one hand, FG coordinates can be defined for generic choices of $(q,\hbar)$. On the other hand, the definition of FN-type coordinates relies on certain properties of the Stokes graph that only arise in particular circumstances.
The relevant class of Stokes graphs for which one may define FN-type coordinates corresponds to those choices of $q$ for which, upon specialization of $\hbar$ to a distinguished value $\hbar_*$, the Stokes graph develops a non-degenerate ring domain (cf. Section \ref{WKBnetworks}).
We henceforth refer to the regions in the moduli space of quadratic differentials characterized by this property as `weak coupling' regimes.\footnote{In the context of class ${\cal S}$ theories, these correspond to chambers of the Coulomb branch where the BPS spectrum includes a vectormultiplet. Due to wall-crossing, ring domains may be present only in certain chambers.}

Weak coupling regions have the special property that they admit simultaneously the definition of \emph{both} FG and FN-type coordinates. 
In these situations it turns out that FN-type coordinates are often preferable over FG coordinates, for the purpose of expanding the tau function as a theta series.
It should be noted that not all quadratic differentials belong to weak coupling regions, in fact in some cases it is clear that there are no such differentials at all.\footnote{For example, if $C$ is a sphere with only one irregular singularity there cannot be a ring domain because $\pi_1(C)$ is trivial.} 
For this reason, FG coordinates are often necessary to cover other regions of the moduli space.
It may also occur that multiple ring domains appear for a given $q$ for different values of $\hbar$, corresponding to cycles inducing different factorizations of $C$. 
Clearly, such situations allow to transition between FN-type coordinates associated to different factorizations.

Overall, our goal is to determine a set of distinguished coordinates associated to each region of the moduli space, with the property that they induce theta series expansions of the tau function.
Taken together, FG and FN-type coordinates cover the whole moduli space, and each is canonically assigned to a different region. 
Weak coupling regions are those where the two domains of definition overlap. We will now describe how the two sets of coordinates patch together.

\subsubsection{}

In the neighbourhood of an accumulation ray, as  described in Section \ref{WKBnetworks},
one finds an infinite sequence of wedges in the $\la$-plane having associated networks
allowing us to define simultaneously coordinates of FG and FN type.
It turns out that two types of 
coordinates defined in this way have a simple relation. Remarkably one finds the same relation
both in the case $C=C_{0,2}$ (Painlev\'e III) and $C=C_{0,4}$ (Painlev\'e VI), which can be represented
as follows:
\begin{equation}\label{XY-to-UV}
Y(U,V)=\left(\frac{U-U^{-1}}{V+V^{-1}}\right)^2,
\qquad
X(U,V)=\left(\frac{VU+(UV)^{-1}}{U-U^{-1}}\right)^2, 
\end{equation}
using the notations $U=e^{2\pi\mathrm{i}\,\si'}$ and 
$V=\mathrm{i}\,e^{2\pi\mathrm{i}\,\eta'}$. In both cases one can identify 
$X$ and $Y$ with (part of) the FG type coordinates associated to the WKB networks 
appearing in this context.

To sharpen the statement of relation (\ref{XY-to-UV}), one must specify a choice of chart both for FG coordinates and for FN-type coordinates.
FG coordinates are associated to edges of an ideal triangulation, which in our case is the WKB triangulation defined by $(q,\hbar)$ as explained above.
To fix a triangulation and determine which edges correspond to $X,Y$, we start by fixing $q$ so that there is an accumulation ray containing $\hbar_*$.
Varying the phase of $\hbar$ away from $\hbar_*$ induces infinite sequences of flips in either direction.
Under good conditions, the infinite sequence of rays lies within a wedge of finite angular width `centered' at the accumulation ray.\footnote{There may be cases in which one has ``accumulation of accumulation rays'' (or higher iterations thereof, measured by the \emph{Cantor-Bendixson rank} of the differential) where this does not always hold. Examples of this include the torus with one puncture, and compact surfaces.}
Choosing $\hbar$ anywhere within this wedge, such that the Stokes graph is saddle-fee, 
gives a triangulation with two edges traversing the annulus, as in Figure \ref{fig:annulus-flip}. 
Then $X$ and $Y$ are the FG coordinates associated to these edges, for more details we refer to \cite{GMN}. 
FN-type coordinates, on the other hand, are defined by a choice of non-contractible cycle on $C$ induced by the ring domain appearing at $\hbar_*$.\footnote{%
Standard FN coordinates have an ambiguity  by arbitrary integer shifts $V\to V U^k$ with $k\in\mathbb{Z}$, which is fixed in the definition of FN-type coordinates via the Exact WKB approach. One may ask whether this way of fixing $k$ is compatible with the definition of $(U,V)$ via (\ref{XY-to-UV}) and the construction of $(X, Y)$ corresponding to the WKB triangulation. In the following we argue that this is the case.}

As emphasized, there are infinitely many wedges in the $\hbar$-plane near the accumulation ray of~$\hbar_*$. We have not specified any particular choice of wedge, except for the requirement that one must stay ``sufficiently close'' to the accumulation ray.
Switching from a wedge to another involves crossing a finite number of active rays\footnote{Active rays are rays from the origin to infinity in the plane $\mathbb{C}^\times$ (parametrised by $\hbar$), which separate wedges where the spectral network has a fixed topological type (see also Section \ref{sec:monodromy}).}, which induces a change in the topology of the Stokes graph, which in turn changes \emph{both} FG coordinates $(X,Y)\to (X',Y')$ and FN-type coordinates $(U,V)\to (U',V')$.
The former jump can be understood via the duality between a Stokes graph and a WKB triangulation (which undergoes a flip).
Likewise, the change in FN-type coordinates can be understood through their definition based on curve factorization and Exact WKB on each pair of pants.
We will show below that the two jumps are \emph{compatible} with relation (\ref{XY-to-UV}), in the sense that applying this formula to the FN-type coordinates $(U',V')$ gives precisely the FG coordinates $(X',Y')$. 
For this reason, the validity of (\ref{XY-to-UV}) holds separately in each sector within a sufficiently narrow wedge across an accumulation ray.
In the remainder of this subsection we give a simple derivation of this property.

\subsubsection{}

We are now going to describe the family of WKB triangulations occurring in sufficiently small 
neighbourhoods of an accumulation ray. (A definition of ``sufficiently small'' will be given below).

It is easy to see that crossing an active ray
will lead to a change of topological types of the Stokes graphs which can be represented by 
the diffeomorphisms from $A$ to itself representing the effect of a Dehn twist. The
induced changes of coordinates describe a discrete evolution that we will now describe in 
more detail. 

We fix a triangulation $T$ of standard form in an annulus $A$ associated to a non-contractible curve.\footnote{The local form of $T$ that we consider here is not fully generic, since in general $A$ may be tessellated by more than two triangles. On the one hand, the present discussion can be suitably generalized to such cases. On the other hand, if one is merely interested in a topological ideal triangulation (as opposed to a WKB triangulation),
it should be possible to reduce any case to the one we consider here, by a finite sequence of flips. 
The change of FG coordinates induced by these additional flips is known to be a rational transformation, 
which should be pre-composed with the change of coordinates we derive. } 
We may represent $A$ as the quotient $\{x\in\BC;|\mathrm{Im}(x)|\leq 1\}/\BZ$,
with $1\in\BZ$ acting as $x\ra x+1$. A triangulation of $A$ is defined by the left side $e=i[0,1]$
and the diagonal $d=\{x\in\BC,0\leq \mathrm{Im}(x)=\mathrm{Re}(x)\leq 1\}$.
\begin{figure}[h!]
\begin{center}
\includegraphics[width=0.65\textwidth]{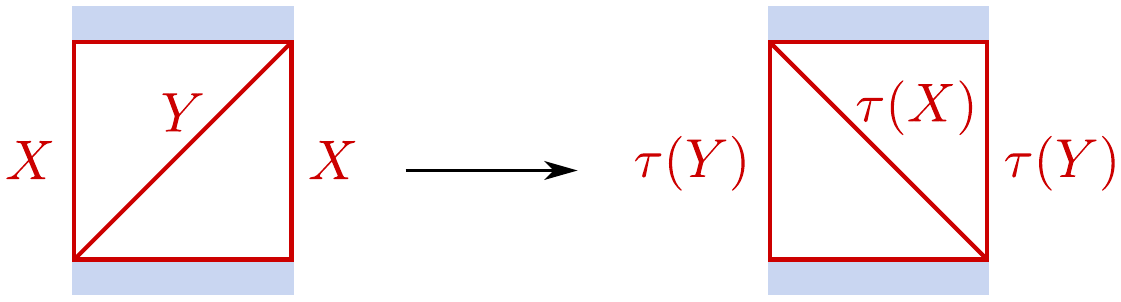}
\caption{The flip of a triangulation on the annulus.}
\label{fig:annulus-flip}
\end{center}
\end{figure}
We can assign FG-coordinates $X$ and $Y$ to the edges $e$ and $d$, 
respectively. A flip changes the triangulation above to another triangulation defined by $e'=e$ 
and $d'=\{x\in\BC,0\leq \mathrm{Im}(x)=1-\mathrm{Re}(x)\leq 1\}$. The flip
induces a change of coordinates
\begin{equation}
X'=X (1+Y^{-1})^{-2}, \qquad Y'=\frac{1}{Y}.
\end{equation}

A diffeomorphism representing a Dehn twist is equivalent to
a flip changing the diagonal to the opposite one, followed 
by a relabelling of the edges $X'\ra Y'$, $Y'\ra X'$. The automorphism of the Poisson-algebra $A$ 
induced by the Dehn twist will be denoted $\tau$. It can be represented  as
\begin{equation}\label{eq:FG-Dehn-twist}
\tau(X)=\frac{1}{Y},
\qquad
\tau(Y)=X(1+Y^{-1})^{-2}.
\end{equation}

One may note, on the other hand, that the Dehn twist induces the simple change of 
FN-type coordinates 
\begin{equation}\label{eq:FN-Dehn-twist}
	U'\equiv\tilde{\tau}(U)=U\,, 
	\qquad
	V'\equiv \tilde{\tau}(V)=VU^{-1}\,.  
\end{equation}
To see this, recall the definition of $V$ in terms of $\eta$ given below (\ref{XY-to-UV}), together with the relation between $\eta$ and Voros symbols in (\ref{eq:normalization-factors-voros})-(\ref{eq:eta-normalization-factors}).
As the Voros symbols are defined by a choice of integration contour starting from a branch point $b\in\partial A$ and progressing into the annulus $A$, the effect of a Dehn twist is to 
modify this contour by adding a detour along the non-contractible cycle associated to $A$.
As a result, the Voros symbol shifts precisely by $\sigma$, since $e^{\pm 2\pi {\rm i} \sigma}$ are the eigenvalues of the transport matrix $M_A$ around $A$ (cf. Section \ref{FN-gen}).
Recalling the definition of $U$ in terms of $\sigma$ given below (\ref{XY-to-UV}), explains (\ref{eq:FN-Dehn-twist}).

Having established how each set of coordinates transforms under Dehn twists corresponding to phase-rotations of $\hbar$ near an accumulation ray, it remains to show that the change of coordinates \rf{XY-to-UV} maps the discrete evolution $\tilde{\tau}$ to $\tau$. This is a simple exercise which we leave to the interested reader. 
We will henceforth rename $\tilde{\tau}$ into $\tau$. 

We can now give a definition for what  it means to take a sufficiently small wedge around an accumulation ray.
We define this to be a wedge in $\hbar$-plane corresponding to a family of WKB triangulations related to each other by a sequence of Dehn twists, which affects only the edges crossing the annulus, in the way shown in Figure \ref{fig:annulus-flip}. 
This is the region in which one may simultaneously define FG and FN-type coordinates associated to the non-contractible curve of the ring domain at $\hbar_*$.

\subsubsection{}

Next we consider the limiting behavior of both types of coordinates induced by the Dehn twist dynamics generated by iterating the changes of coordinates associated to the active rays in the vicinity of the accumulation ray.

In particular, we'll be interested in comparing FN-type coordinates $(U,V)$ chosen in a generic, but fixed, angular sector with the limiting values of FG coordinates $(X,Y)$.
As we will see, there is a well-defined limit for this relation, which moreover turns out simpler than the change of variables (\ref{XY-to-UV}) relating FG and FN-type coordinates associated to the same sector.

Assuming that $|U|<1$ we observe that   
\begin{subequations}\label{limit1}
\begin{align}
\tau^n(Y)=\left(\frac{U-U^{-1}}{VU^{-n}+V^{-1}U^{n}}\right)^2\sim U^{2n}(U-U^{-1})^2  V^{-2} ~,
\\
\tau^n(X)=\left(\frac{VU^{1-n}+V^{-1}U^{n-1}}{U-U^{-1}}\right)^2\sim U^{-2n}\frac{(UV)^{2}}{(U-U^{-1})^2} ~.
\end{align}
\end{subequations}
It easily follows that 
\begin{subequations}\label{limit2}
\begin{align}
U^2&=\lim_{n\ra\infty} \tau^n(XY)~, \\
V^2&=(U-U^{-1})^2\lim_{n\ra\infty}U^{2n-1}\tau^n\big(\sqrt{X/Y}\big) ~.
\label{Moeller}\end{align}
\end{subequations}
 Equations \rf{limit1}, \rf{limit2} clarify the relation between 
 the FN-type coordinates $(U,V)$ defined in an angular sector in the $\hbar$-plane chosen in the vicinity of an accumulation ray, and  
 limits \mbox{$n\ra\infty$} of certain functions formed out of the FG coordinates $\tau^n(X)$ and $\tau^n(Y)$.
{A similar limit can be defined assuming $|U|>1$.
 
On the one hand, this relation agrees with an observation of \cite{HN}, stating that FN coordinates arise as \emph{spectral coordinates} (of which $(X,Y)$ are one example) in the limit situation where~$\hbar$ approaches an accumulation ray.
The choice between  $|U|<1$ and $|U|>1$ corresponds to the choice of a resolution in \cite{HN}.}
On the other hand, relation (\ref{Moeller}) also establishes a suggestion from \cite{GMN} on a more rigorous footing.

\subsection{Discussion}

{The reader will have noticed that the FN type coordinates are defined somewhat more indirectly using exact WKB than the FG type coordinates which are literally Borel summations of certain Voros symbols. In the case of the FN type coordinates we did not define the coordinate $\si$ associated to a ring domain directly by Borel summation. Instead we used 
Borel summation to define natural solutions to the differential equation defined by the quantum curve, and defined
$\sigma$ from the holonomy of these solutions around the annulus $A$ defined by the ring domain.

It is of course a  natural question if $\si$ also admits a representation by Borel summation of a Voros symbol.
Interesting results in this direction are provided by the recent work \cite{GGM} devoted to the case $C=C_{0,2}$ 
discussed in this paper.\footnote{The function $a(u,\hbar)$ in \cite{GGM} is almost (up to normalisation) equal to the 
function $\si(U,\la)$ obtained by restricting the coordinate function $\si$ on the space of quantum curves to the 
subspace represented by opers, assuming simple relations between $u$ and $U$, and $\hbar$ and $\hbar$,
respectively.} The extensive numerical studies carried out in \cite{GGM} can be taken as support for 
the conjecture that the Voros symbol associated to $\si$ is Borel summable along the positive real axis,
at least when restricted to the subspace of opers.

The picture changes dramatically when considering a phase of $\la$ away from zero. One finds an
infinite collection of wedges in the $\hbar$-plane in which the Voros symbols associated to $\si$
are again Borel summable. However, according to the results quoted above, one will generically 
find that  the results of the Borel summations are represented by FG type coordinates. 

This may seem counterintuitive. However, one should notice that any 
angular rotation of $\hbar$ away from the positive real axis, as small as it may be, 
will inevitably cross infinitely many active rays. Starting from a wedge around a non-vanishing value 
of $\arg{\hbar}$ one will encounter the infinite sequence of flips discussed in the previous 
subsection when $\arg{\hbar}$ approaches zero. The observation that the FN type coordinates
can be understood as the limits of the sequence of FG type coordinates associated to the 
wedges surrounding the positive real line seems to be perfectly consistent with the 
conjecture motivated by the results of \cite{GGM} that the Borel summation of 
the Voros symbol associated to $\si(U,\la)$ is 
represented by FG coordinates in the wedges that are disjoint from the positive real axis, 
while it reproduces the function $\si(U,\la)$ when the Borel summation is performed along the positive real axis.\footnote{Yet another perspective on this is the observation that both FG and FN type coordinates are special types of `spectral coordinates' \cite{HN}. Spectral coordinates are labelled by homology cycles on $\Sigma$. 
It was argued in \cite{HN} that fixing the cycle to correspond to a purely-electric charge (as is the case for the BPS vectormultiplet), and varying $\hbar$ across a sequence of wedges leading up to the real positive axis, one finds that the corresponding coordinate is FG type away from the real axis, but eventually approaches the (exponentiated) FN coordinate $\sigma$.}
}

\section{Theta series expansions from exact WKB}\label{Rev-C04}
\setcounter{equation}{0}

{Having established how the Exact WKB framework leads to natural choices of Darboux coordinates in different points in the moduli space of quantum curves, we now return to the theta-series expansions induced by these coordinates.
}
On first sight it may seem that the picture we have previously found in the 
case of $C=C_{0,2}$ with two irregular singularities is somewhat different from
the one established for $C=C_{0,4}$ in \cite{CPT}. We will therefore revisit 
the latter case in this section, allowing us to demonstrate that an important basic feature 
is shared by the two cases: In both cases there exists a family of normalised tau-functions
having representations of theta series type in terms of systems of Darboux coordinates
for $\CM_{\rm ch}(C)$ which can be both of Fenchel-Nielsen and of Fock-Goncharov type.

\subsection{Theta series corresponding to FN-type coordinates from exact WKB}

We had seen in Section  \ref{Sec:Theta} that there exists a way to define 
normalised partition functions depending on the choice of a system of 
coordinates for the character 
variety $\CM_{\rm ch}(C)$. It was furthermore shown in Section  \ref{sec:exact-WKB}
that exact WKB allows one to define distinguished sets of coordinates for 
a given region in the space of quantum curves. 
Our next goal is to show that the FN-type coordinates defined from exact WKB are among the 
coordinates $(\si,\eta)$ that appear in the expansions of \emph{theta series type} discussed 
in Section  \ref{Sec:Theta}.

\subsubsection{Case of $C_{0,4}$}

{Let us begin with the case of $C_{0,4}$.} 
Recall that the definition of FN-type coordinates by exact WKB 
depends on the choice of a pair of Stokes graphs on the pairs of pants $C_i$, $i=1,2$, appearing in  
a given pants decomposition of $C_{0,4}$. To discuss a specific example let us consider 
the pants decomposition of $s$-type defined by cutting along a contour separating $0$ and $z$ from 
$1$ and $\infty$.
{Fixing the $s$-channel defines $\sigma$ uniquely, but not $\eta$. In subsection \ref{normTauC04} we discussed several choices of $\eta$ leading to different theta series expansions, and discussed their relation.
We will now show how certain pairs of coordinates $(\sigma,\eta)$ encountered in subsection \ref{normTauC04} are tied to specific types of Stokes graphs by Exact WKB analysis, following observations and definitions of Section \ref{sec:exact-WKB}, and corroborating previous observations of~\cite{CPT}.
}

Let us choose the Stokes graphs to be of type $\CS_2$ (as defined in the discussion around Figure \ref{Stokesgraphs}) on both $C_1$ and $C_2$.
According to the discussion in Section \ref{FN-gen} one may use this set-up to define 
a FN-type coordinate $\eta_{22}$ related to  
$\eta_0$ by a relation of the form  $e^{2\pi\mathrm{i}\,\eta_{22}^{}}\,\mathsf{n}^{(1)}\mathsf{n}^{(2)}=e^{2\pi\mathrm{i}\eta_0^{}}$,
with $\mathsf{n}^{(i)}$ related to the Voros symbol $V^{(p_ib_i)}$ as
$\mathsf{n}^{(i)}=e^{-2V^{(p_ib_i)}}$ for $i=1,2$, respectively. This Voros symbol
has recently been calculated in \cite{ATT1,IKT}.  Theorem 2.2 in \cite{ATT1} immediately implies the
following formula for the corresponding normalisation constants
\begin{align}
&\mathsf{n}^{(2)}=\mathsf{n}_2(\si,\theta_3,\theta_4),\qquad
\mathsf{n}^{(1)}=\mathsf{n}_2(\si,\theta_2,\theta_1),\\
&\mathsf{n}_2(\vartheta_1,\vartheta_2,\vartheta_3)=\frac{2\pi}{\Ga(2\vartheta_1)\Ga(2\vartheta_1-1)}
\frac{\Ga(\vartheta_1+\vartheta_2+\vartheta_3)\Ga(\vartheta_1+\vartheta_2-\vartheta_3)}
       {\Ga(1-\vartheta_1+\vartheta_2+\vartheta_3)\Ga(1-\vartheta_1+\vartheta_2-\vartheta_3)}.
\label{n2-form}\end{align}
{In this way, exact WKB defines a choice of normalisation for the $\eta$-coordinate, and a corresponding expansion for $\CT$. What is not obvious from this definition, is whether or not the expansion of $\CT$ defined by exact WKB takes the form of a generalised theta-series, or not.}

To address this question,
we may recall from Section \ref{normTauC04} that coordinates 
$\eta$ appearing in expansions of generalised theta series type are related to the FN type coordinate 
$\eta_0^{}$ defined in Section \ref{etaC04} by a relation of the form 
\begin{align}\label{eta-CN-remind}
&e^{2\pi\mathrm{i}\,\eta}=e^{2\pi\mathrm{i}\,\eta_0^{}} \,
\frac{\CN^{(\eta)}(\si-1\,;\,\underline{\theta}\,)}{\CN^{(\eta)}(\si\,;\,\underline{\theta}\,)}.
\end{align}
It then suffices to note that the function $N_2$
defined in \rf{Ni-def} 
satisfies
\begin{equation}
\frac{N_{2}(\vartheta_1,\vartheta_2,\vartheta_3)}{N_{2}(\vartheta_1-1,\vartheta_2,\vartheta_3)}=
\mathsf{n}_2(\vartheta_1,\vartheta_2,\vartheta_3)
\end{equation}
in order to see that the coordinate $\eta$ defined by 
setting 
in \rf{eta-CN-remind} 
\begin{equation}
\CN^{(\eta)}\equiv
{N_2(\si,\theta_2,\theta_1)N_2(\si,\theta_3,\theta_4)},
\end{equation} 
is indeed the coordinate $\eta_{22}$ defined from exact WKB above.
The tau function defined by 
this choice of normalisation for $\eta$ 
will be denoted as $\CT^{({22})}$. 
It seems natural to assign 
$\CT^{({22})}$ to the region in the moduli space of quantum curves 
where the pants decomposition of the Stokes graph defined by $(q,\la)$ produces
a pair of Stokes graphs on $C_{0,3}$ of type $(\CS_2,\CS_2)$.

\subsubsection{Case of $C_{0,2}$}

In a similar way one may revisit the case $C=C_{0,2}$, asking
in particular how the choice of normalisation for the FN type coordinates appearing in Section \ref{sec:weak-coupling-expansion} is related to the normalisation defined by Exact WKB. 

To address this, we may start by noting that the weak coupling expansion \rf{GL-exp} can easily be
recast in the form  
\begin{align}\label{thetaser-wc}
&\CT^{(\eta)}(\,\si,\eta\,;\,z\,)=\sum_{n\in\BZ} \,e^{2\pi\mathrm{i}n\,\eta}\,(-1)^n
\CN^{(\eta)}(\si+n)\CF^{}(\,\si+n\,;\,z\,),
\end{align}
with 
\begin{equation}
\CN^{(\eta)}(\si)=\frac{1}{G(1+2\si)G(1-2\si)}.
\end{equation}
The coordinate $\eta$ appearing in \rf{thetaser-wc} has been defined in Section \ref{sec:FN-for-PIII}. In the 
general framework for the definition of FN type coordinates described in Section \ref{FN-gen}
one may characterise the coordinate $\eta$ in terms of the normalisation factors $\mathsf{n}^{(i)}$ 
explicitly given as
{ 
\begin{equation}
\mathsf{n}^{(2)}=\mathsf{n}^{(1)}=\frac{\Ga(1-2\si)}{\Ga(2\si-1)},
\end{equation}  
as follows easily from equation \rf{Phi-Bessel}.}

Turning our attention to the normalisation fixed by Exact WKB,
the relevant Voros symbols can be 
extracted from \cite{IKT}, giving for $\mathrm{Re}(\si)>0$ 
\begin{equation}
\mathsf{n}^{(2)}=
\mathsf{n}^{(1)}=\frac{2\pi}{\Ga(2\si)\Ga(2\si-1)}.
\end{equation}
Let us denote by $\tilde\eta$ the FN type coordinate determined by this choice of normalisations. 
By adapting the discussion in Section \rf{normTauC04} one can easily show that 
$\tilde\eta$ induces a generalised theta series expansion similar to the one in 
\rf{GL-exp}, but with $\mathcal{N}^{(\eta)}(\si)$ replaced by
{\begin{equation}
\mathcal{N}^{(\tilde\eta)}(\si)=
\frac{1}{(G(1+2\si))^2}.
\end{equation}}
It is interesting to note that the partition function $\tilde\CZ(\si,\Lambda)=\CN^{(\tilde\eta)}(\si)\CF(\si,\Lambda)$ 
defined with this choice of 
normalisation coincides with the  topological string partition function computed with the help of 
the topological vertex (see Appendix \ref{TopVertex}). This is not the case if $\CN^{(\tilde\eta)}(\si)$ is replaced by the function 
$\CN^{(\eta)}(\si)$ appearing in \rf{GL-exp}.

\subsection{Changes of FN-type coordinates for fixed pants decomposition}\label{Change-FN}

Our next goal is to discuss how 
the definition for the normalised tau-functions defined by Exact WKB extends
across the loci where the Stokes graphs change topological type.
This will be related to the observations from Section \ref{normTauC04} that there exist several choices of FN-type coordinates inducing generalized theta series expansions associated to a single pants decomposition, as anticipated in \cite{CPT}.

A few basic cases have to be discussed in this regard. 
The first type of changes of Stokes graphs is associated to the appearance of saddle 
trajectories inside one of the pairs of pants defined by the given pants decomposition.  As a first example let
us consider the case where the Stokes graph on $C_2$ changes from type 
$\CS_2$ to $\CS_s$. 

\subsubsection{}

We have given a definition for the case when both Stokes graphs on $C_{i}$, $i=1,2$ 
are of type $\CS_2$ in the previous subsection. In the case where one of the Stokes graphs 
is of type $\CS_s$ one needs to revisit the definition.  Let us assume that the Stokes graph
on $C_2$ has type $\CS_s$. There are now two 
Stokes regions $\CR$, $\CR'$ around the puncture at $x=0$ created by the factorisation of $C_{0,4}$.
One of them is simply the continuation of the Stokes region surrounding $x=0$ through 
the transition from $\CS_2$ to $\CS_s$, the other one is created in this transition, see Figure \ref{fig:transition}.

\begin{figure}[h!]
\begin{center}
\includegraphics[width=0.35\textwidth]{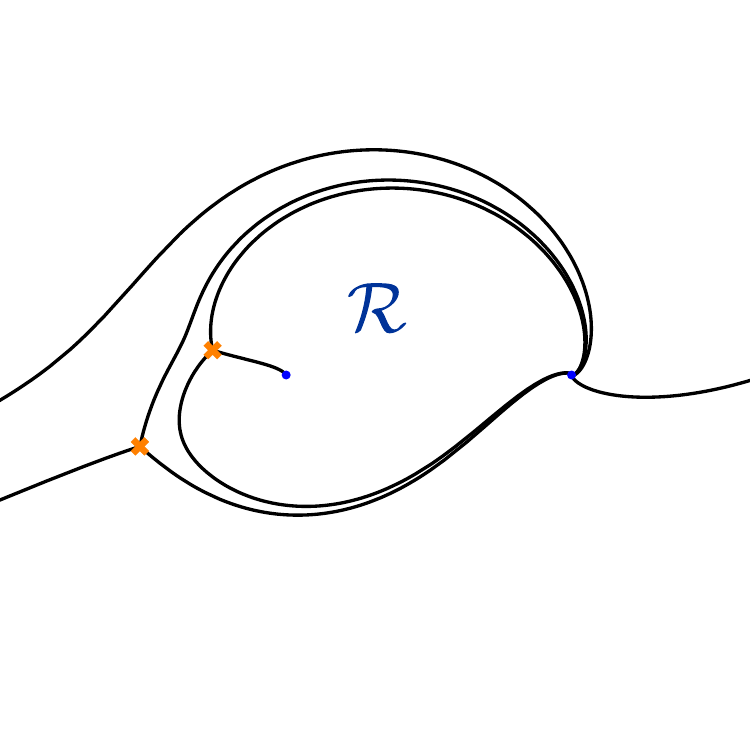}
\hspace*{0.1\textwidth}
\includegraphics[width=0.35\textwidth]{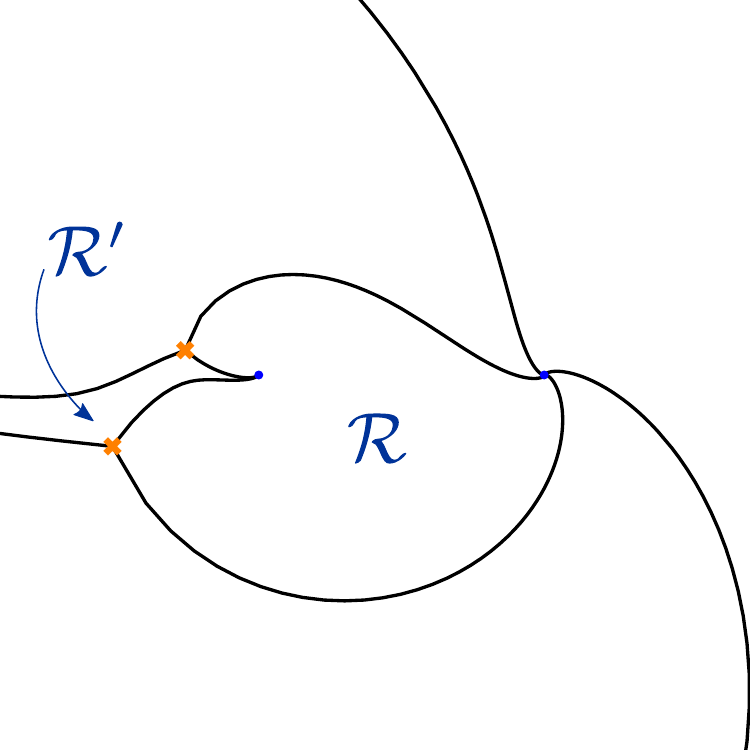}
\caption{Transition from a Stokes graph of type $\CS_2$ (left) to one of type $\CS_s$ (right) on a three-punctured sphere defined by a factorization limit of $C_{0,4}$.}
\label{fig:transition}
\end{center}
\end{figure}

In the definition of  the Voros symbol 
fixing the normalisation, one now encounters an ambiguity in the choice of the branch point 
representing the starting point of the contour of integration. 
We are going to give here a prescription to fix this ambiguity, which is `as canonical as possible'. 
A broader discussion of this point can be found in Appendix \ref{app:different-norm}, where we also present an alternative, less canonical prescription involving more choices.

According to the classification 
of Stokes graphs in \cite{AT} one will find the transition from $\CS_2$ to $\CS_s$ when 
the real part of the parameter $t:=C-B=\theta_3-\theta_4-\si$ becomes negative. 
The Stokes graph 
$\CS_s$ has a pair of Stokes lines connecting the puncture at $x=0$ to two branch points $b_\pm$.
Before the transition from region $\CS_2$ to region $\CS_s$, our general prescription leads us to a choice of normalisation fixed by the Voros symbol $V^{(b_20)}$, defined by an integration contour running from the branch point $b_2$ in $\CS_2$ to the puncture at $0$.
After the transition we face two choices of normalisation, corresponding to Voros symbols $V^{(b_\pm0)}$, whose integration contours run from the branch points $b_\pm$ in $\CS_s$ to the puncture at $0$.

\subsubsection{}

It is important to notice that the transition from $\CS_2$ to $\CS_s$ involves a jump of topology for the Stokes graphs, shown in Figure \ref{fig:transition}, which induces a jump of the Voros symbols.
The Borel summation of the Voros symbol $V^{(b_20)}$ 
has recently been studied in \cite{ATT2}. It was found 
that the Borel summation of the Voros symbol $V^{(b_20)}$ will  jump in the transition from $\CS_2$ to $\CS_s$.
It follows from \cite[Theorem 4.8]{ATT2} that
the analytic continuation of the function $\mathsf{n}^{(2)}_2$ 
defined by the Borel summation of the formal series $e^{-2V^{(b_20)}}$
in the chamber
associated to $\CS_2$ will be related to the functions 
$\mathsf{n}^{(2)}_{s,\pm}$ 
defined by Borel summation of $e^{-2V^{(b_\pm0)}}$ in the chamber
associated to $\CS_s$ by a relation of the form\footnote{Comparing with \cite{ATT2} one should note
that \cite{ATT2} has fixed 
a prescription selecting one of the two possible cases.} 
\begin{equation}\label{Vorosjump}
\frac{\mathsf{n}^{(2)}_{s,\pm}(\si,\theta_3,\theta_4)}{\mathsf{n}^{(2)}_2(\si,\theta_3,\theta_4)}
=\frac{1}{1-e^{2\pi\mathrm{i}\,\ep_{\pm}(\si+\theta_4-\theta_3)}},
\end{equation}
with a sign $\ep_{\pm}\in\{1,-1\}$ depending on
the  path used to 
define the analytic continuation of the Voros symbol $V^{(b_20)}$.
This sign is not arbitrary and can be determined unambiguously by a direct analysis, we will however fix it below taking an alternative route.
The jump associated to the transition from $\CS_2$ to $\CS_s$ at $\mathrm{Re}(t)=0$
is due to an analog of the Stokes phenomenon in the Borel summation of the Voros symbol.\footnote{This is an instance of the Dillinger-Delabaere-Pham formula, see for example \cite{IN} for details.}

\begin{figure}[h!]
\begin{center}
\includegraphics[width=0.7\textwidth]{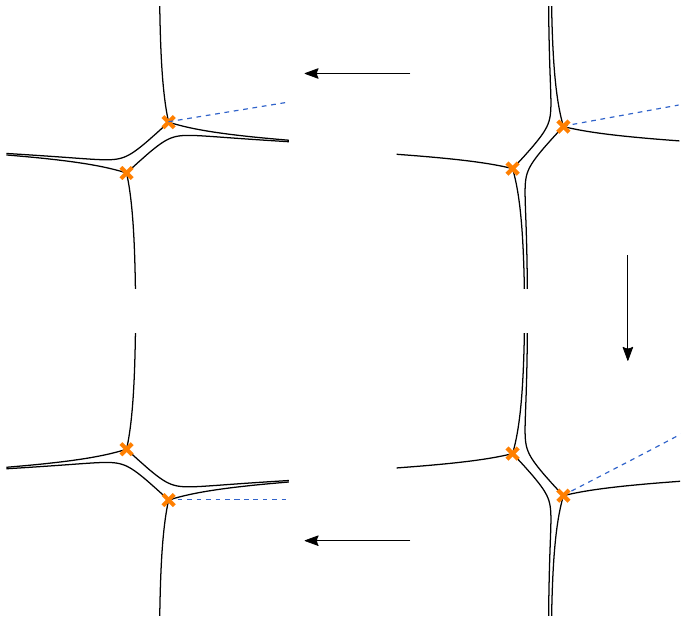}
\caption{Transitions in the type of Stokes graph on $C_{0,3}$, from type $\CS_i$ (top-right) to type $\CS_s$ (top and bottom left). 
The anti-Stokes graph associates contours for the Voros symbol (the dashed blue line) beginning at two different branch points $b_\pm$. For convenience of illustration we depict the transition between regions $\CS_3$ and $\CS_s$ here, with the understanding that the transition between $\CS_2$ and $\CS_s$ is qualitatively identical.}
\label{fig:sym-chambers}
\end{center}
\end{figure}

\subsubsection{}

The difference between the two normalization factors of region $\CS_s$ is due to the change of the choice of 
contour defining the Voros symbols according to our definitions above, see Figure \ref{fig:sym-chambers}. 
In fact, it is evident from (\ref{Vorosjump}) that 
\begin{equation}\label{eq:CS-s-internal-jump}
	\mathsf{n}^{(2)}_{s,+}(\si,\theta_3,\theta_4)
	=
	\mathsf{n}^{(2)}_{s,-}(\si,\theta_3,\theta_4) \cdot (- e^{2\pi\mathrm{i}\,\ep_{-}(\si+\theta_4-\theta_3)})
\end{equation}
Note that the exponent of the factor in brackets is (up to the overall sign) the Voros symbol associated to the shortest closed cycle running between the two branch points. This  should indeed be expected since $V^{(b_-0)}-V^{(b_+0)} = V^{(b_-0)}+ V^{(0b_+)} = V^{(b_- b_+)}$.

The relations between normalisation factors $\mathsf{n}^{(2)}_{2}, \mathsf{n}^{(2)}_{s,+}, \mathsf{n}^{(2)}_{s,-}$ are therefore completely under control, and consistent with each other. We have the two transitions from $\CS_2$ into $\CS_{s,\pm}$ described by (\ref{Vorosjump}) and the transition between $\CS_{s,+}$ and $\CS_{s,-}$ described by (\ref{eq:CS-s-internal-jump}).

With this picture at hand, we are finally led to the following proposal to fix the normalisation of FN-type coordinates in the symmetric chamber $\CS_s$. Instead of choosing between $\mathsf{n}^{(2)}_{s,+}$ and $\mathsf{n}^{(2)}_{s,+}$, we consider their geometric mean. In terms of Voros symbols we define
\begin{equation}\label{eq:Voros-average}
	V^{(s)} : = \frac{1}{2} \left( V^{(0 b_+)} + V^{(0 b_-)} \right)
\end{equation}
and set the corresponding normalisation coefficient to be $\mathsf{n}^{(2)}_s=e^{-2V^{(s)}}$.\footnote{To avoid potential confusion, we stress that this \emph{differs} from $V^{(b_- b_+)} =  V^{(0 b_+)} - V^{( 0b_-)} $.}
This choice of normalisation has the advantage of being analytic throughout $\CS_s$.
It is clearly invariant under (\ref{eq:CS-s-internal-jump}).
With this choice of normalisation we find a unique jump between regions $\CS_2$ and $\CS_s$, replacing (\ref{Vorosjump}) by
\begin{equation}\label{Vorosjump-averaged}
	\frac{\mathsf{n}^{(2)}_{s}(\si,\theta_3,\theta_4)}{\mathsf{n}^{(2)}_2(\si,\theta_3,\theta_4)}
	= - \frac{e^{\pi\mathrm{i}\,(\si+\theta_4-\theta_3)}}{1-e^{2\pi\mathrm{i}\,(\si+\theta_4-\theta_3)}} \,.
\end{equation}
Here we have set $\ep_{\pm}=\pm1$ in (\ref{Vorosjump}), this specialization will be motivated by an analysis of the asymptotic series for the Gamma function, coming up next.
One may observe that (\ref{Vorosjump-averaged}) combines a jump due to the Stokes phenomenon with a jump in the leading term of the Voros symbol.
The former is captured by (\ref{Vorosjump}) and corresponds to the two horizontal arrows in Figure \ref{fig:sym-chambers}, while the latter is captured by (\ref{eq:CS-s-internal-jump}) and arises from the change of basepoint for the integration contour between the left two frames of Figure \ref{fig:sym-chambers}.

\subsubsection{}
Given that formula \rf{n2-form} represents the Borel summation of 
$e^{-2V^{(b_20)}}$ one may alternatively derive \rf{Vorosjump} using the 
Borel summation of 
the asymptotic series of the Gamma functions appearing in formula \rf{n2-form}, as discussed 
in Appendix \ref{BorelGamma}. In this way one finds that the prescription we have used to define the Voros symbols
$V^{(b_\pm0)}$ above is equivalent to defining the leading term $t\log t$ in the 
asymptotic series of $\log\Ga(t)$ to be $t\log(-t)$.
This will lead us to a convenient way to describe our continuation prescription, and to 
extend it to the remaining cases. 
Let us consider the function $\Gamma_B(w)$ defined as follows
\begin{equation}\label{checkGamma}
\Gamma_B(w)=\left\{\begin{aligned}
& \Ga(w)\qquad \qquad \quad \text{for}\;\;\mathrm{Re}(w)>1,\\
&  \frac{2\pi }{\Ga(1-w)} \qquad \quad\text{for}\;\;\mathrm{Re}(w)<1,
\end{aligned}\right.
\end{equation}
The function $\Gamma_B(w)$ is piece-wise analytic on $\BC$, having
jumps only along the imaginary axis. $\Gamma_B(w)$ 
represents the Borel summation
of the asymptotic series for the Gamma-function $\Gamma(w)$ with leading term $w\log w$ defined 
by $w\log(-w)$ (denoted $\check\Gamma_0$ in Appendix \ref{BorelGamma}).
The function $\Gamma_B(w)$ allows us to get the formula for the normalisation factor $\mathsf{n}^{(2)}_s$ 
associated to the quantum curves having Stokes graphs $\CS_s$ on one pair of pants simply by
replacing the function $\Gamma(1-\si-\theta_4+\theta_3)$ in the expression 
for $\mathsf{n}^{(2)}_2$ following from \rf{n2-form} by 
$\Gamma_B(1-\si-\theta_4+\theta_3)$. 


\subsubsection{}

For simplicity in the following discussion we will restrict to cases with 
$\mathrm{Re}(\si)\geq \frac{1}{2}$. 
The conditions for having a Stokes graph of type 
$\CS_2$ on $C_2$ following from \cite{AT} imply that 
$\mathrm{Re}(\theta_3-\theta_4)>0$.  The only  case which is left under these conditions is  associated to the 
Stokes graph $\CS_3$ on $C_2$. Proceeding along similar lines as before one
will find that the normalisation factor $\mathsf{n}^{(2)}_3$ 
associated to the quantum curves having Stokes graphs $\CS_3$ on $C_2$ can be obtained by simply
replacing the function $\Gamma(1-\theta_3-\theta_4+\si)$ in the expression for 
$\mathsf{n}^{(2)}_s$ by 
$\Gamma_B(1-\theta_3-\theta_4+\si)$. The result for all three cases can be represented uniformly as
\begin{equation}
\mathsf{n}^{(2)}(\si,\theta_3,\theta_4)=\frac{2\pi}{\Ga(2\si)\Ga(2\si-1)}
\frac{\Ga(\si+\theta_3+\theta_4)\Ga(\si+\theta_3-\theta_4)}
       {\Ga_B(1-\si+\theta_3+\theta_4)\Ga_B(1-\si+\theta_3-\theta_4)}.
\end{equation}

The cases not covered yet have $\mathrm{Re}(\theta_4-\theta_3)>0$. In this case one may proceed along 
the same lines as before, starting from the Stokes graph $\CS_1$. The result can be represented
in a uniform manner by using the function $\Gamma_B(w)$, as before.

According to the discussion in Section \ref{FN-gen}
there is a corresponding change of coordinates 
\begin{equation}
e^{2\pi\mathrm{i}\eta_{22}^{}}= - e^{2\pi\mathrm{i}\eta_{s2}}
\frac{e^{\pi\mathrm{i}\,(\si+\theta_4-\theta_3)}}{1-e^{2\pi\mathrm{i}\,(\si+\theta_4-\theta_3)}}
\end{equation}
It is not hard to see that the corresponding difference generating function is $\CE_{(2,s)}(\,\si\,,\underline{\theta}\,)$ given as
\begin{equation}
\CE_{(2,s)}(\,\si\,,\underline{\theta}\,)=
\mathcal{E}(\si+\theta_4-\theta_3),
\end{equation}
with $\CE(x)$ being functions constructed
from $G(x)$ as
\begin{equation}
\CE(x)=(-2\pi \mathrm{i})^{-x} \frac{G(1+x)}{G(1-x)}.
\end{equation}
We will interpret  the function $\CE_{(2,s)}$ as a transition function
in a line bundle $\CL_{\Theta}$ on the space $\CZ$ of quantum curves 
associated to thickened neighbourhoods of the loci in $\CZ$
where saddle connections of the type considered above appear. 

With the help of the transition function $\CE_{(2,s)}$
one can define a natural continuation of the tau-function associated 
to the pair of Stokes graphs $(\CS_2,\CS_2)$ on $(C_2,C_1)$
into the region associated to the pair $(\CS_s,\CS_2)$ of Stokes graphs.

Another type of
changes of the Stokes graphs associated
to the appearance of saddle trajectories traversing the annulus
has been discussed in Section \ref{sec:exact-WKB}. In this case 
it is easy to see that the choice of contours defining the solutions $\psi_+^{(i)}$ defining the 
flat sections $Y_i$ can not be continued through the loci where a saddle trajectory appears.
One may note, however, that the contours of integration defining the leading term 
the WKB expansion for the normalisation factor before and after the appearance of 
a saddle trajectory
are related by composition with a cycle 
encircling the annulus, as illustrated in Figure \ref{fig:jugglestart}. This will modify the normalisation factors $\mathsf{n}^{(i)}$ by 
$\mathsf{n}^{(i)}\ra e^{\pm2\pi\mathrm{i}\si}\mathsf{n}^{(i)}$, for $i=1,2$ 
corresponding to a change of coordinates
$(\si',\eta')= (\si,\eta\pm 2\si)$ according to the discussion in Section \ref{Voros-norm}.
The difference generating function $\CE(\si,\underline{\theta})$ associated to this case is
found to be the function
$e^{\mp 2\pi\,\mathrm{i}\,\si^2}$. Such difference generating functions represent transition 
functions of the line bundle $\CL_\Theta$ associated to the loci where 
saddle trajectories traversing the annulus appear.

\begin{figure}[h!]
\begin{center}
\includegraphics[width=0.75\textwidth]{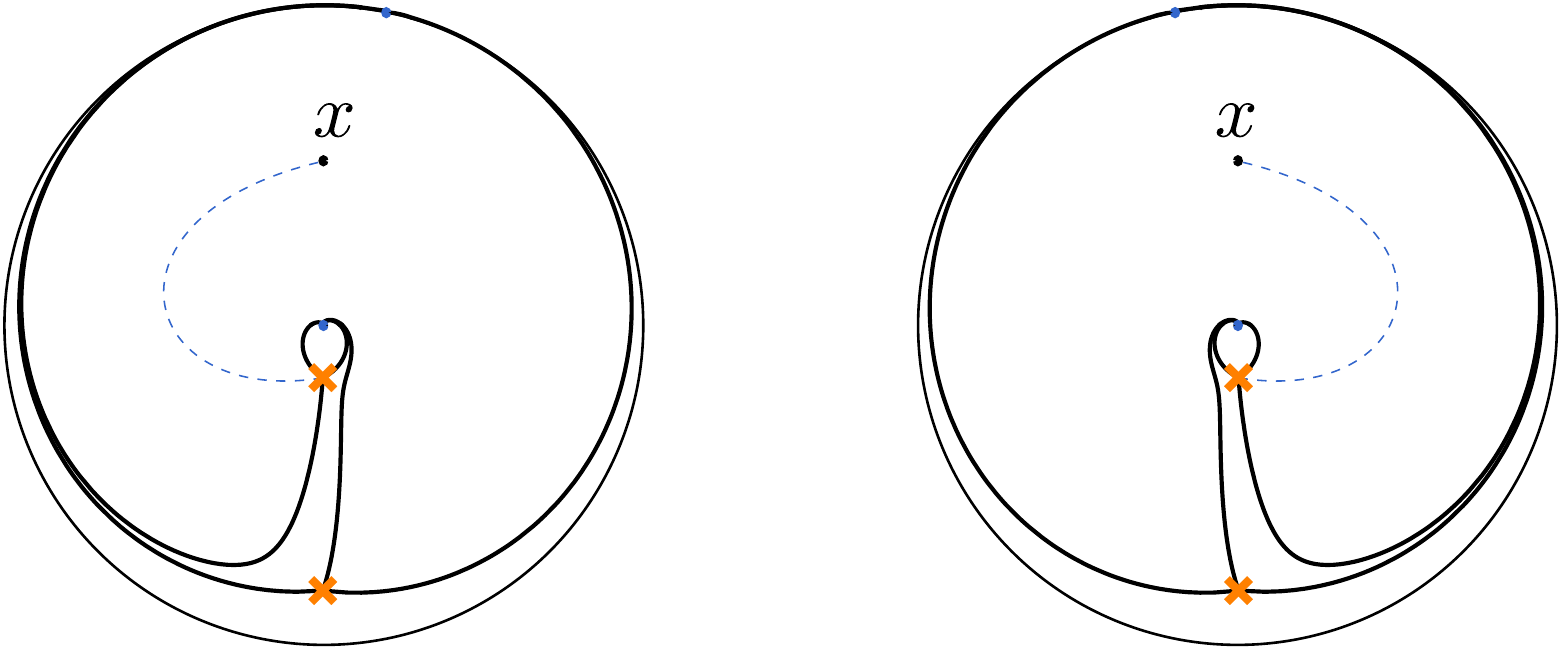}
\caption{Contours of integration before and after the appearance of a saddle trajectory, which are related by composition 
with a cycle encircling the annulus.}
\label{fig:jugglestart}
\end{center}
\end{figure}


\subsection{The real slice}

The space of quantum curves contains an interesting real slice
represented by the so-called Jenkins-Strebel (JS-) differentials, 
quadratic differentials defining Stokes graphs having a maximal number of saddle trajectories.
The Stokes graphs associated to generic JS-differentials decompose the surface $C$ into 
ring domains, consisting either of annuli, or punctured discs around the punctures of $C$. 
A JS differential thereby defines a pants decomposition of $C$. 
Each pair of pants carries a Stokes graph having three edges connecting the two 
branch points. It is therefore necessary that the parameters $a_1$, $a_2$
and $a_3$ associated to any one of the pairs of pants appearing in this pants 
decomposition are purely 
real. 

One should keep in mind that the parameters $\vartheta_i$ 
are related to $a_i$ as $\vartheta_i=\mathrm{i} a_i/\hbar$. 
The differential 
$\la^{-2}q_0(x)$ will then define a JS differential iff $\mathrm{Im}(\la)=0$.
On a four-punctured sphere $C=C_{0,4}$ defined by gluing two such pairs of pants 
one finds that there are accumulation rays at $\mathrm{Im}(\la)=0$.
In order to assign FN type coordinates to JS differentials on $C$ one has to choose
a half-plane from which one approaches the real $\la$-axis.
This corresponds to the choice of a resolution in the abelianisation approach. 
To be specific, we will in the following consider the case where $\mathrm{Im}(\la)=0$
is approached from the upper half-plane. This implies that $\mathrm{Re}(\vartheta_1)>0$.
The other case can be treated in a very similar way.

According to the classification of 
Stokes graphs of \cite{AT} summarised in Section \ref{Stokespants} above one needs to distinguish 
chambers according to the sign of $a_i-a_{i+1}-a_{i+2}$, $i=1,2,3$, defining $a_i\equiv a_{i-3}$ for $i>3$.  
Introducing the notations
 \begin{equation}
 A=\vartheta_1+\vartheta_2-\vartheta_3,\qquad
 B=\vartheta_1+\vartheta_2+\vartheta_3,\qquad
C=2\vartheta_1\,,
\end{equation}
the normalisation factors $\mathsf{n}_i(\vartheta_1,\vartheta_2,\vartheta_3)$ 
associated to the Stokes graphs $\CS_i$ are on the real slice for  $i=1,2,3,s$ respectively given as
\begin{equation}
\begin{aligned}
\mathsf{n}_s &= \frac{1}{\mathrm{i}}
 \frac{\Gamma(A) \Gamma(B) \Gamma (C-A)}{\Gamma(C)\Gamma(C-1)\Gamma(B-C+1)} 
\\
\mathsf{n}_1&= 
\frac{1}{2\pi}  \frac{\Gamma(A) \Gamma(B)\Gamma(C-A) \Gamma (C-B)}{\Gamma(C)\Gamma(C-1)} \\
\mathsf{n}_2&= 
 \frac{2\pi \Gamma(A) \Gamma(B)}{\Gamma(C)\Gamma(C-1)\Gamma(A-C+1) \Gamma (B-C+1)} 
\\
\mathsf{n}_3&= 
 \frac{2\pi \Gamma(C-A) \Gamma(B)}{\Gamma(C)\Gamma(C-1)\Gamma(1-A) \Gamma (B-C+1)}.
\end{aligned}
\end{equation}
This corresponds to assigning the functions $N_i(\vartheta_1,\vartheta_2,\vartheta_3)$ to  
Stokes graphs $\CS_i$, for $i=1,2,3,s$, respectively, with $N_i$ for $i=1,2,3$ defined in \rf{Ni-def}, and 
\begin{align}
N_s(\vartheta_1,\vartheta_2,\vartheta_3) &=G(1+\vartheta_1+\vartheta_2+\vartheta_3)
\frac{\prod_{j=1}^3 G(1+\vartheta_1+\vartheta_2+\vartheta_3 - 2\vartheta_j)}{(2\pi)^{\vartheta_1} \, G(1) \prod_{r=1}^3 G(1+2\vartheta_r)}
\end{align}
Taking the product of the functions $N_i$ assigned to the pairs of 
pants defined by a JS differential in this way yields  normalisation factors defining expansions of 
generalised theta series type for the tau-functions $\CT$.

The tau-functions $\CT$ defined in this way can be compared with 
the topological string partition functions computed in \cite{CPT} with the help of the topological 
vertex.
Picking the resolution specified above,  
{it is easy to see that to each type of toric graph there corresponds a
unique pair of Stokes graphs $(\CS^{(2)},\CS^{(1)})$ and the constraints from 
positivity of the K\"ahler parameters in the toric geometry used for the geometric engineering 
in \cite{CPT} correspond directly to the inequalities \rf{chamberdef} characterising the Stokes graphs on
$C_{0,3}$.}
We find perfect agreement, chamber by chamber.


\section{The space of quantum curves}\label{sp-qcurves}
\setcounter{equation}{0}

Based on the examples studied previously we will in the rest of this paper
formulate a
proposal for the geometric description of the topological string partition functions
in the more general class of cases associated to 
Riemann surfaces $C=C_{g,n}$ of genus $g$ and $n$ punctures.
To simplify the discussion we will only discuss regular singularities at the $n$ punctures.
The first step will be to revisit the notion of the quantum curve, and the representation 
in terms of flat connections in this level of generality.

\subsection{Quantum curves}\label{q-curve}

On a Riemann surface $C_{g,n}$ with genus $g$ and $n$ punctures we will consider 
projective $\la$-connections with $d$ apparent singularities. 
A projective $\la$-connection can 
locally be represented
by differential operators of the form $\la^2\pa_x^2-q_\la(x)$, with 
$q_\la(x)$ transforming under a change of coordinate $\tilde{x}=\tilde{x}(x)$ as 
\begin{equation}
{q}_{\la}({x})=(\tilde{x}'(x))^2\tilde{q}(\tilde{x}(x))-\frac{\la^2}{2}\{\tilde{x},x\},\qquad 
\{f,x\}=\frac{f'''}{f'}-\frac{3}{2}\left(\frac{f''}{f'}\right)^2.
\end{equation} 
We will assume that the functions $q_\la$ have convergent power series expansions
in $\la$ of the form $q_\la(x)=\sum_{k=0}^{\infty}q_k(x)\la^k$. 
If an
apparent singularity is found in a chart with coordinate $x$ at the position $x=u_r$,
it means that $q_\la^{}(x)$ behaves near $x=u_r$ as
\begin{equation}
q_\la^{}(x)=\frac{3\la^2}{4(x-u_r)^2}+\la\frac{v_r}{x-u_r}+{q}_{r}^{}+\CO(x-u_r), 
\end{equation}
with $v_r$ and $q_{r}$ satisfying the system of relations
\begin{equation}\label{appsing}
v_r^2=q_{r}^{}, \qquad r=1,\dots,d,
\end{equation}
ensuring that the monodromy of $\la^2\pa_x^2-q_\la(x)$ is trivial in $\mathrm{PSL}(2,\BC)$.
Such projective $\la$-connections will be called quantum curves.

The difference between two  projective $\la$-connections represented by functions 
$q_\la^{}(x)^{}$ and $q_\la^{\ast}(x)$ defines a quadratic differential. 
A projective $\la$-connection defines  a projective structure on $C$,  a system of local
coordinates on local charts in $C$ with transition functions all represented 
by M\"obius transformations. One simply needs to use the ratio $w(x)=\eta_1(x)/\eta_2(x)$ 
of two linearly independent solutions of $(\la^2\pa_x^2-q_\la^{}(x))\eta(x)=0$ as a new 
local coordinate in a given chart with coordinate $x$. With respect to the coordinate $w$, the 
projective $\la$-connection gets represented by $\la^2\pa_w^2$.
It is often convenient to assume from the outset that the coordinates $x$ on open subsets of $C$
are part of an atlas defining a projective structure.
The functions $q_\la^{}(x)$ appearing in the local representation $\la^2\pa_x^2-q_\la^{}(x)$ of a
generic oper will then represent quadratic differentials $q_\la^{}(x) (dx)^2$. 
One may pick  families of reference projective structures 
varying holomorphically over Teichm\"uller space.
Changing the reference projective connection will modify $q_\la^{}(x)$ by terms of order $\la^2$.

One may note that the quadratic differential $q_\la^{}(x)$ will generically be uniquely determined by 
the data $(\mathbf{u},\mathbf{v})=\{(u_r,v_r);r=1,\dots,d\}$ if $d\geq 3g-3+n$. Since $\mathrm{dim}(H^0(C,K^2))=3g-3+n$,
one then has  sufficiently many 
equations \rf{appsing} in order to fix the freedom to modify $q_\la^{}(x)$ by adding purely 
holomorphic quadratic differentials. We will mostly assume $d=3g-3+n$ in the following.\footnote{This 
corresponds to the minimal choice for the number of apparent singularities that is necessary to cover a patch of $\CM_{{\rm{ch}}}$ around the oper submanifold parametrized by the quantum curve with $\hbar=0$.
}
We will in the following be interested in the space $\CZ$
of quantum curves. We will also be interested in the dependence 
on the choice of complex structure on $C$. Picking local 
coordinates $\mathbf{z}=(z_1,\dots,z_{3g-3+n})$ for the moduli space $\CM(C_{g,n})$ of complex structures on $C_{g,n}$ allows us to
consider the collection $(\mathbf{u},\mathbf{v};\mathbf{z};\la)$ as a system of coordinates for the space $\CZ$
of quantum curves. 

In order to get a concrete parameterisation, 
let us take a  
pair $(\mathbf{u},\mathbf{v})=\{(u_r,v_r);r=1,\dots,d\}$. If $d=3g-3+n$ there generically
exists a unique 
quadratic differential $q_0\equiv Q_0(\mathbf{u},\mathbf{v})$ such that the spectral curve 
$\Sigma=\{(u,v)\in T^\ast C;v^2=Q_0(u)\}$ passes through 
$(u_r,v_r)$ for all $r=1,\dots,d$. One may then define  $Q_{\la;\mathbf{u},\mathbf{v}}$ as
$Q_{\la;\mathbf{u},\mathbf{v}}=Q_0+\la Q_1+\la^2 Q_2$,
where $Q_i\equiv Q_i(\mathbf{u},\mathbf{v})$, $i=1,2$, are the unique quadratic differentials 
having poles of first and second order at $x=u_r$, $r=1,\dots,d$, respectively,  
satisfying\footnote{One should note that the definition of $Q_1$ requires choosing
preferred coordinates around $u_r$, for $r=1,\dots,d$.} 
\begin{equation}\label{Q12-normcond}
\lim_{x\ra u_r}\left(Q_1(x)-\frac{v_r}{x-u_r}\right)=0,
\qquad
\lim_{x\ra u_r}\left(Q_2(x)-\frac{3}{4(x-u_r)}\right)=0,
\end{equation}
for $r=1,\dots,d$. The quadratic differential $Q^{}_{\la;\mathbf{u},\mathbf{v}}$ defined in this way have the
property that the expansion in powers of $\la$ truncates after the second order.
The most general quantum curve
with apparent singularities at $x=u_r$, $r=1,\dots,d$, $d=3g-3+n$, which is 
at most quadratic in $\la$ can be represented in this way. 
We will call this the canonical form 
of the quantum curve.

It may  be useful to consider  $\la$-dependent changes of 
variables of the form $v_r=V_r(\mathbf{u},\mathbf{w};\la)$, $\mathbf{w}=(w_1,\dots,w_d)$.
Functions of the form $V_r= 
w_r+\la \nu_r(\mathbf{u})$ will preserve the feature that the $\la$-expansion 
of $q_\la$ contains only terms up to order $\la^2$.  
There are many different coordinates for spaces of quantum curves, 
in general. Considering only quantum curves which are at most quadratic in $\la$ restricts
possible $\la$-dependent changes of coordinates severely. 


\subsection{Relations with $\la$-connections}\label{q-curves-vs-connections}

A projective $\la$-connection defines an ordinary holomorphic $\la$-connection, locally represented in the
form $\la\pa_x-A(x)$, with $A(x)=\left(\begin{smallmatrix} 0 & q_\la \\ 1 & 0\end{smallmatrix}\right)$. 
$\la$-connections which are gauge equivalent to a connection 
of this form are called opers. It is usually assumed that $q_\la(x)$ is holomorphic in each 
local chart. We are here interested in opers which may have $d=3g-3+n$ apparent singularities, 
referred to as quantum curves.

There is a close relation between quantum curves and
$\la$-connections on $C$, pairs $(\mathcal{E},\nabla_\la)$, where $\CE$ is a holomorphic bundle,
and $\nabla_\la$ is a holomorphic $\la$-connection modulo gauge transformations.
A $\la$-connection can locally be represented as
$\nabla_\la=du(\la\pa_u-A(u))$, transforming under gauge transformations
as $\nabla_\la=h\cdot \tilde{\nabla}_{\la}\cdot h^{-1}$.
Instead of regarding $\la$ as a parameter  we find it useful to consider it as
a coordinate on the moduli space of $\la$-connections, the 
space $\CZ_C$ of gauge equivalence classes $[\la,\mathcal{E},\nabla_\la]$
of triples $(\la,\mathcal{E},\nabla_\la)$ defined by identifying pairs of triples
related by gauge transformations. This space fibers over the moduli space $\CM(C)$
of complex structures on $C$. The space $\CZ$ of our interest will be the total space 
of this fibration.


For a given $\la$-connection one can locally always find gauge transformations
bringing it to oper form.  In order to do this, we may start by 
representing $\CE$ as an extension $0\ra L\ra \CE\ra L^{-1}\Lambda\ra 0$. This 
corresponds to a representation of $\CE$ in terms of transition functions which are
all of the upper triangular form $\big(\begin{smallmatrix} \zeta & \eta \\ 0 & \zeta^{-1}\delta\end{smallmatrix}\big)$,
with $\zeta$ and $\delta$ being transition functions for the line bundles $L$ and $\Lambda$, respectively. 
The local trivializations of such a representation 
allow one to represent  $A\equiv A(x)$ in the form
$A=\left(\begin{smallmatrix} a & b \\ c & -a\end{smallmatrix}\right)$, with
matrix element $c$ being a section of the line bundle $\mathcal{K}=K L^{-2}\Lambda$.
Using such a representation, we can locally find gauge transformations relating 
$\nabla_\la=du(\la\pa_u-A(u))$ to a connection of oper form. 
The gauge transformation to oper form will be singular at the zeros $u_r$ of $c$,
$r=1,\dots,d$, leading to the $d=\mathrm{deg}(\mathcal{K})$ apparent singularities of the resulting oper. It is 
furthermore easy to see that the residue of  an apparent 
singularity at $x=u_r$ is given as $-a(u_r)$, for $r=1,\dots,d$.

This construction associates points $(\mathbf{u},\mathbf{v})$ in the symmetric product $[T^\ast C]^{d}$ to 
$\la$-connections $\nabla_\la$ on extensions $0\ra L\ra \CE\ra L^{-1}\Lambda\ra 0$. 
It can be shown that there exists a local inverse to this construction \cite{DT}, assigning pairs $[\mathcal{E},\nabla_\la]$ to
points in the symmetric product $[T^\ast C]^{d}$. One may use it to define local coordinates 
for the moduli spaces $\CZ_C$.


\subsection{Relations with Higgs pairs}

Higgs pairs are pairs $[\CE,\vf]$, with $\CE$ being a holomorphic $G$-bundle,
and $\vf$ being an element of $H^0(C,\mathrm{End}(\CE)\ot K_C)$.
Let $\CM_{\rm Hit}(C)$ be the space of gauge equivalence classes $[\CE,\vf]$ of
stable Higgs pairs $(\CE,\vf)$. The parameterization of $\la$-connections 
using points $(\mathbf{u},\mathbf{v})\in[T^\ast C]^{d}$
is related to a similar parameterization of Higgs pairs $[\CE,\vf]$, to be outlined next.

To begin, let us note that one may parameterize $\la$-connections  by picking 
families of reference
$\la$-connections $\nabla_\la^\ast$ varying holomorphically with the choice of $\CE$, 
allowing us to define 
$\vf=\nabla_\la^{}-\nabla_\la^\ast$, an element of $H^0(C,\mathrm{End}(\CE)\ot K_C)$. 
The maps $\Pi^{\ast}:(\la,\CE,\nabla_\la)\mapsto (\CE,\vf)$ defined in this way can be 
used\footnote{An important special case  is given by
the conformal limit of the non-abelian Hodge correspondence 
\cite{DFKMMN,CW}. It establishes
bi-holomorphisms between the strata attached to different values of 
$\la$ of the stratification of $\CZ$ obtained 
by gathering the triples having limits $\lim_{\la\ra 0}[\xi\la,\mathcal{E},\xi\nabla_\la]$
in the same connected component \cite{Si}.} 
to define bi-holomorphic maps between open subsets
of  $\CZ_C$ and $\CM_{\rm Hit}(C)\times \BC$, respectively. 

Convenient parameterisations 
for the  classes $[\CE,\vf]$ of Higgs pairs are provided by
the integrable structure of the Hitchin moduli space \cite{Hi}. They can be found
using the correspondence \cite{Hi} between classes $[\CE,\vf]$
of Higgs pairs and pairs $(\Sigma,\CL)$, with $\Sigma$ being the spectral curve defined
by the characteristic equation $\mathrm{det}(v-\vf)=0$, and $\CL$ being 
the bundle of eigenlines of $\vf$ on $\Sigma$. Given 
$(\Sigma,\CL)$ one may reconstruct $[\CE,\vf]$ 
from $(\pi_\ast(\CL),\pi_\ast(y))$, 
with $\pi:\Sigma\ra C$ being the canonical projection,
$\pi_\ast$ being the push-forward, and $y$ being the canonical differential on $\Sigma$.

More explicit parameterizations can be defined as follows.
Given $(\CE,\vf)$ with $\CE$ being an extension $0\ra L\ra\CE\ra L^{-1}\Lambda \ra 0$, 
and $\vf= \left(\begin{smallmatrix} \vf_0 & \vf_+ \\ \vf_- & -\vf_0\end{smallmatrix}\right)$
one may again consider the collection $\{u_r;r=1,\dots,d\}$ 
of zeros of $\vf_-$, and define $v_r=-\vf_0(u_r)$ for $r=1,\dots,d$. 
One thereby assigns  collections of points 
$(\mathbf{u},\mathbf{v})=\{(u_r,v_r);r=1,\dots,d\}$ in $T^\ast C$ to 
Higgs fields $\vf$ on extensions $0\ra L\ra\CE\ra L^{-1}\Lambda \ra 0$. 
Assuming that  $u_r\neq u_s$ for $r\neq s$
one may use $(\mathbf{u},\mathbf{v})$ to define a pair $(\Sigma,\CL)$ by taking $\Sigma$ to be the curve in $T^\ast C$ 
passing through all points $(u_r,v_r)$, $r=1,\dots,d$, and 
$\CL:=\pi^{\ast}(K^{-1}L)\CO_{\Sigma}(D)$, with 
$D$ being the divisor $\sum_{r}\check{u}_r$  on $\Sigma$ satisfying 
$\pi(D)=\sum_{r=1}^du_r$ and $v(\check{u}_r)=v_r$, for $r=1,\dots,d$.
Observing that eigenvectors of $\vf$ can be constructed as 
$\left(\begin{smallmatrix} v+\vf_0 \\ \vf_-\end{smallmatrix}\right)$ one may show
that the pair $(\Sigma,\CL)$
defined in this way agrees with the one defined by the construction above, see \cite{DT} for 
details.  


Comparing with Section \ref{q-curves-vs-connections}
we see that we may parameterise 
families of Higgs pairs $(\CE_{\mathbf{u},\mathbf{v}}^{},\vf_{\mathbf{u},\mathbf{v}}^{})$ 
by collections $(\mathbf{u},\mathbf{v})=\{(u_r,v_r);r=1,\dots,d\}$ of points in $T^\ast C$.
Representing $\CE_{\mathbf{u},\mathbf{v}}^{}$ as extensions, we may introduce a family of reference connections
$\nabla^\ast_{\mathbf{u},\mathbf{v}}=dx(\pa_x-A_\ast(x))$. 
Assuming $A_\ast$ to be of the form
$A_\ast=\left(\begin{smallmatrix} a & b \\ 0 & -a\end{smallmatrix}\right)$ one may then consider the family of 
$\la$-connections $\nabla_{\la;{\mathbf{u},\mathbf{v}}}^{}=
du(\la\pa_u-(\vf_{\mathbf{u},\mathbf{v}}^{}+\la A_\ast))$ associated to these data.
As discussed in Section \ref{q-curves-vs-connections} 
one may gauge transform $\nabla_{\la;{\mathbf{u},\mathbf{v}}}$ to oper form, defining a quantum curve from
$(\mathbf{u},\mathbf{v})$. This quantum curve 
can be represented in the form 
$\la^2\pa^2_x -Q^{}_{\la;\mathbf{u},\mathbf{w}}$ with $\mathbf{w}$ related to $\mathbf{v}$
by a simple change of variables $\mathbf{w}=\mathbf{w}(\mathbf{v})$ which is linear $\la$.

\subsection{Integrable structure}

There is a projection $\Pi_0$ from $\CZ$ to the Hitchin moduli space $\CM_{\rm Hit}(C)$ 
associating to a triple $(\la,\mathcal{E},\nabla_\la)$ the Higgs pair 
$(\mathcal{E},\vf)$ with $\vf=\nabla_0$. 
The moduli space $\CM_{\rm Hit}(C)$ has the structure of a torus fibration defined
by means of the Hitchin map $h:\CM_{\rm Hit}(C)\ra H^0(C,K_C^2)$ sending $[\CE,\vf]$ to 
the quadratic differential $\mathrm{tr}(\vf^2)$.
The equation $\frac{1}{2}\mathrm{tr}(\vf^2)=q$, with $q\in H^0(C,K_C^2)$ being a fixed quadratic differential,
defines submanifolds of $\CM_{\rm Hit}(C)$ which are  abelian 
varieties. The composition $h\circ\Pi_0$ describes $\CZ$ as a fibration over 
$\CB_C\simeq H^0(C,K_C^2)$. We will also consider the  subset $\CB_C'\subset\CB_C$ 
represented by quadratic differentials having simple zeros only, defining the subset 
$\CZ'=(h\circ \Pi_0)^{-1}(\CB_C')\subset\CZ$.

The base
$\CB_C$ carries a geometric structure  called special geometry.
Useful coordinates for the base $\CB_C$ reflecting this structure
 are called homological coordinates,
defined as periods of the 
canonical differential $\sqrt{q}$ on the spectral curve $\Sigma_q$ defined by $q$.
To define a set of homological coordinates one needs to choose
a canonical basis for $
\Gamma=H_1(\Sigma_q^\circ,\BZ)^-$,
where the superscript indicates taking the odd part with respect to the exchange of sheets, 
and $\Sigma_q^\circ\subset \Sigma_q$ is the complement of the inverse image of the poles of $q$.
A canonical basis for $\Ga$ is represented by  a collection of cycles 
$(\al^1,\dots,\al^d;\be_1,\dots,\be_d)$ on $\Sigma_q$ 
having intersection index $\al_r\circ \be_s=\de_{rs}$, 
where $d=\mathrm{dim}(\CB_{C})=3g-3+n$ if $C=C_{g,n}$,
allowing us to define\begin{equation}
a^r=\int_{\al^r}\sqrt{q},\qquad
\ca_r=\int_{\be_r}\sqrt{q}.
\end{equation}
There locally exists a function  $\CF(\sa)$, $\sa=(a^1,\dots,a^d)$,  
such that $\ca_r=\pa_{a^r}\CF(\sa)$. This function is called preportential.
The period matrix $\tau_\sa$ of $\Sigma$ is the matrix
formed out of the
second derivatives $\pa_{a^r}\pa_{a^s}\CF(\sa)$ of $\CF(\sa)$.
The geometric structure called special geometry is encoded in a covering of 
$\CB_C'$ with charts equipped with homological coordinates, and transition functions 
represented by the changes of homological coordinates associated to the elements of
$\mathrm{Sp}(2d,\BZ)$ describing changes of the canonical homology bases.

Canonically associated to the structure of special geometry is a  torus fibration over $\CB_C$ 
\cite{Fr}, with fibre over a 
point in $\CB_C$ having coordinate values $\sa$ being the torus 
$\Theta_\sa=\BC^d/(\BZ^d+\tau_\sa\cdot\BZ^d)$.  The torus fibration defined in this way 
gives a concrete realization of the integrable structure canonically associated to the
special geometry of the base $\CB_C$. The special geometry of the base $\CB_C$
is recovered from the integrable structure by 
representing the fibres over points $b\in\CB_C$  as complex tori of the form
$\Theta_b=\BC^d/(\BZ^d+\tau_b\cdot\BZ^d)$, with $\tau_b$ being a symmetric $d\times d$ matrix. 
It can be shown \cite{Fr} that locally over
$\CB_C$ there exist coordinates $\sa(b)$ and a function $\CF(\sa)$ allowing one to represent the
period matrix $\tau_b$ characterising the tori $\Theta_b$ as the matrix of 
second derivatives $\pa_{a^r}\pa_{a^s}\CF(\sa)|_{\sa=\sa(b)}^{}$ of $\CF(\sa)$. This is how 
the special geometry of the base is recovered 
from the integrable structure of $\CM_{\rm Hit}(C)$.

Recall that a point on $\CB_C$ having coordinates $\sa$ represents
a choice of a spectral curve $\Sigma_\sa$. 
The torus $\Theta_\sa$ represents the Prym 
variety parameterising the choices of line bundles $\CL$ over $\Sigma_\sa$. 
For curves $\Sigma$ which are double coverings of a base curve $C$ as 
considered here, we had seen that there is a natural correspondence
between classes $[\CE,\vf]$ and $d$-tuples $(\mathbf{u},\mathbf{v})$ of 
unordered points in $T^*C$. 
Taken together we
see how the
definition of the quantum curves $\la^2\pa^2_x -Q^{}_{\la;\mathbf{u},\mathbf{v}}$
is related to  the integrable structure of the Hitchin system.

\section{Monodromy of quantum curves}\label{sec:monodromy}
\setcounter{equation}{0}

In the examples studied previously, we had observed the crucial role played by 
distinguished coordinates of FN and FG type for the character varieties representing the  
monodromies of the 
differential operators $\la^2\pa_x^2-q_\la^{}(x)$. As the next step 
in the formulation of our proposal we will now briefly outline how 
such coordinates can be defined in the more general cases 
associated to Riemann surfaces  $C_{g,n}$.

\subsection{Holonomy}

For any $\la\neq 0$ one may consider the holonomy of the $\la$-connection $\nabla_\la$, 
defining a representation $\rho:\pi_1(C)\ra \mathrm{SL}(2,\BC)$. The space of all 
such representations $\rho$ modulo overall conjugation defines the character variety
$\CM_{\rm ch}(C)$. 
In this way we may define a map $\mathsf{Hol}:\CZ^\times\ra \CM_{\rm ch}(C)$,
with $\CZ ^\times$ being the moduli space of triples $[\la,\mathcal{E},\nabla_\la]$, with $\la\in\BC^\times$. 
The character variety has a natural holomorphic symplectic form  $\Omega_{\rm\sst G}$
going back to Goldman.

The space $\CZ$ can be regarded as a fibration over $\BC$ by means of the projection 
$p:\CZ\ra\BC$, $p[\la,\mathcal{E},\nabla_\la]=\la$. The fibers $\CM_{\la}$, $\la\neq 0$, are moduli spaces
of $\la$-connections with fixed parameter $\la$, and $\CM_0\simeq\CM_{\rm Hit}(C)$.
Each fiber $\CM_\la$ has a natural symplectic form $\Omega_\la$. Restricting 
 $\mathsf{Hol}$ to a fibre $\CM_\la$ 
 defines a map $\mathsf{Hol}_\la:\CM_{\la}\ra \CM_{\rm ch}(C)$
 which is locally bi-holomorphic and symplectic \cite{Hi97,AM,KS}.

We will later introduce collections of preferred systems of Darboux coordinates for 
$\CM_{\rm ch}(C)$. Such Darboux coordinates are 
collections of functions $(\bx,\cbx)$, 
$\bx=(x^1,\dots,x^d)$, $\cbx=(\cx^1,\dots,\cx^d)$ from $\CM_{\rm ch}(C)$ to $\BC^\times$
allowing us to represent the symplectic form 
$\Omega_{\rm\sst G}$ as 
\begin{equation}
\Omega_{\rm\sst G}=\sum_{r=1}^d {dx^r}\wedge {d\cx_r}.
\end{equation}
It is often useful to consider the coordinates  $X^r=e^{2\pi\mathrm{i}\,x^r}$, $\check{X}_r=e^{2\pi\mathrm{i}\,\cx^r}$, $r=1,\dots,d$. The coordinates $X^r$ and $\check{X}_r$ can for $r=1,\dots,d$, be represented as  the 
composition of a map $\bX:\CM_{\rm ch}(C)\ra \BT$, with $\BT=(\BC^\times)^{2d}$, 
with the standard coordinate functions on $(\BC^\times)^{2d}$.
The Darboux coordinates considered later will be distinguished from generic sets of Darboux coordinates 
by having important special properties. 

We may finally compose the holonomy map $\mathsf{Hol}$ with the coordinate functions
$\bX$ to get locally defined maps $\CX:\CZ\ra \BT$ which can be regarded as collections
of coordinates $(\CX^r,\check{\CX}_r)$, $r=1,\dots,d$, for $\CZ$. By means of the local isomorphisms
$\CZ^\times\simeq \CM_{\rm Hit}(C)\times\BC^{\times}$ mentioned above one 
gets families of holomorphic functions $(\CX^r(\la),\check{\CX}_r(\la))$, $r=1,\dots,d$, parameterised 
by points in $\CM_{\rm Hit}(C)$. One of the important features of the 
special class of Darboux coordinates to be considered here is the fact that the asymptotic
behavior for $\la\ra 0$ is of the form
\begin{equation}\label{CX-asym}
\log\CX^r\simeq \frac{1}{\la}a^r+\CO(\la^0),\quad
\log\check{\CX}_r\simeq \frac{1}{\la}\ca_r+\CO(\la^0),
\quad r=1,\dots,d,
\end{equation}
with $a^1,\dots,a^d$ and $\ca_1,\dots,\ca_d$ being homological coordinates as introduced 
above. One should note,  in particular, that
equation \rf{CX-asym} establishes a natural correspondence between
the systems $\CX$ of coordinates on $\CZ$ considered later and systems of homological
coordinates for  $\CB_C$.

\subsection{Coordinates associated to spectral networks}

The coordinates $\CX$ on $\CZ$ of our interest are obtained 
from coordinates on $\CM_{\rm ch}(C)$ with the help of the holonomy map $\mathsf{Hol}$. We will
therefore start by discussing very briefly the relevant sets of coordinates on $\CM_{\rm ch}(C)$. 
Two fairly well-known sets of coordinates for $\CM_{\rm ch}(C)$ 
are the Fock-Goncharov (FG) \cite{FG06} and Fenchel-Nielsen (FN) type coordinates. The FG coordinates are associated 
to triangulations of $C$. Coordinates of FN type are relatives of the complexifications of  
the classical Fenchel-Nielsen coordinates on the Teichm\"uller spaces $\CT(C)$,\footnote{With $\CT(C)$  
represented as a connected component of the  character variety 
$\mathrm{Hom}(\pi_1,\mathrm{PSL}(2,\BR))/\mathrm{PSL}(2,\BR)$.} sharing the main feature to 
be associated to pants decompositions of $C$.

While there are obvious differences in the way these types of coordinates 
are defined, they are ultimately more closely related than it may appear. 
An important common feature is the fact that both yield {\it rational} parameterisations of the trace functions,
the generators of the ring of algebraic functions on $\CM_{\rm ch}(C)$. These coordinates therefore
reflect the algebraic structure of $\CM_{\rm ch}(C)$ in a particularly simple way. This feature 
is well-known in the 
case of the FG coordinates, and has been noted in \cite{TV} for the case of the FN-type coordinates.

The relation between FG and FN type coordinates is  closer than it might seem.
It should be possible to show that the FG coordinates can be expressed as {\it rational} functions of 
FN-type coordinates,\footnote{The inverse map will not be rational, in general. This means that
the torus $\BT_{\rm\sst FN}$ containing the image
of the FN-type coordinates is a finite cover of the torus $\BT_{\rm\sst FG}$ to which the 
FG-coordinates map.} generalising the results of Section \ref{Sec:FG-FN}.

A unified framework for the definition of such coordinates is provided by the abelianisation program \cite{HN}.
By introducing a certain graph  $\CS$ on $C$ called spectral network one can set up a one-to-one correspondence between
abelian connections on a cover $\Sigma_\CS$ of $C$, and flat non-abelian connections on $C$. Coordinates for the 
abelian connections on $\Sigma_\CS$ can thereby be used as coordinates for $\CM_{\rm ch}(C)$. 
By specialising the type of network used in this construction one can recover both 
FG- and FN-type coordinates. The coordinates associated to general spectral 
networks by abelianisation are hybrids of FG and FN-type coordinates.

\subsection{Coordinates associated to WKB networks}

As explained above, we are ultimately interested in the coordinates on $\CZ$ obtained 
from coordinates on  $\CM_{\rm ch}(C)$ by composition with the holonomy map
$\mathsf{Hol}$. One of the beautiful features of the coordinates on $\CZ$ obtained
in this way is the existence of natural domains of definition, obtained as follows.
Each class $[\la,\CE,\nabla_\la]$ defines a pair $(q,\la)$ consisting of the quadratic
differential $q=h\circ\Pi_0[\la,\CE,\nabla_\la]$ and $\la$. It is explained in 
\cite{HN} how each pair $(q,\la)$ defines a spectral network called WKB network.
It is then very natural to assign the same type of coordinates 
to all $[\la,\CE,\nabla_\la]$ having a WKB network of the same topological type. 

As discussed in \cite{HN} 
one finds for generic $(q,\la)$ WKB networks defining triangulations of $C$ 
called WKB triangulations in \cite{GMN}. 
Such networks are called Fock-Goncharov (FG) type networks.
At the opposite extreme one finds WKB networks decomposing $C$ into 
a collection of annuli and punctured discs. Such networks are 
called Fenchel-Nielsen (FN) networks, naturally
defining a pants decomposition of $C$.  In between these extremes there
exist several hybrid types of networks having a varying number of ring domains.
The space $\CZ$ can be stratified according to the number of ring domains appearing 
in the Stokes graph.

One may naturally
associate FG coordinates to the subset of $\CZ$ having $(q,\la)$ defining a 
FG network of a fixed topological type. 
It is furthermore natural to associate FN-type coordinates to spectral networks of FN-type.
This program allows one to define coordinates for any saddle-free\footnote{The exists no trajectory connecting branch points. Closed trajectories are allowed.} spectral network $\CW$
defined by a quadratic differential $q$. The resulting coordinates will be of FG-type if 
$\CW$  has no ring domains, and will yield hybrid types of coordinates
combining features of FG- and FN-type when there are ring domains.


The coordinates assigned to a WKB network using abelianisation have 
nice bonus features distinguishing them from generic coordinates for the space $\CZ$.
Most intensively studied is the subspace of $\CZ$ represented by $\la$-opers $\la\pa_u-\big(
\begin{smallmatrix} 0 &q_\la\\ 1 & 0\end{smallmatrix}\big)$
without apparent singularities in the cases where the WKB network defined by $(q,\la)$,  
with $q=h\circ\Pi_0[\la,\CE,\nabla_\la]$, is of FG type. It follows from the results of  \cite{IN,Al19} that
the functions $\CX(\la)$ have asymptotic expansions in powers of $\la$ 
represented by the so-called Voros symbols. These asymptotic expansions
are Borel-summable if the network defining the coordinates $\CX(\la)$ is the FG-type WKB network
defined by $(q,\la)$.

The case of the FN networks has been investigated less, an exception being the discussion 
in \cite[Section 11]{HK}. {However, as discussed in Sections \ref{sec:exact-WKB} 
and \ref{Rev-C04} one may use exact WKB to define canonical bases for the space of 
solutions to $\nabla_{\la}\eta=0$ on each of the three-punctured spheres 
defined by the pants decomposition associated to the given FN network.
Using these canonical bases in the general framework for the definition of FN type coordinates 
described in Section \ref{Sec:Theta} yields a natural way to assign FN type coordinates to 
WKB networks of FN type.}

We conjecture that the coordinates $\CX$ defined in this way all share the important feature
that their leading asymptotics for $\la\ra 0$ is of the form \rf{CX-asym}, establishing a 
correspondence with the set of homological coordinates $(a^r,\ca_r)$, $r=1,\dots,d$, 
appearing on the right side of \rf{CX-asym}.
It seems likely that the systems of coordinates $\CX$ considered here are uniquely 
characterised by the collections of homological coordinates appearing in the 
asymptotics for $\la\ra 0$.

\subsection{Problems of Riemann-Hilbert type for the coordinates $\CX$}

It is often useful to consider the variation of the spectral networks
and the corresponding coordinates with respect to $\la$ for fixed quadratic differential $q$.
This leads to a characterisation of the coordinates of our interest as solutions to 
a  problem of Riemann-Hilbert (RH) type 
related to the ones studied in \cite{GMN} and \cite{Br}.

Considering the locus in $\CZ$ with fixed $q=h\circ\Pi_0[\la,\CE,\nabla_\la]$, one may 
decompose the punctured plane $\BC^\times$ parameterised by the coordinate $\la$
into wedges separated by rays running from the origin to infinity. The wedges are
defined to be loci in $\BC^\times$ where the spectral network has a fixed topological type. 
{Rays separating two such wedges are often called active rays.}{
They are characterised
by the angle they span with $\BR_+^\times\subset\BC^\times$.  The set of angles 
defining the active rays  may have accumulation 
points. $\BC^\times$ thereby gets decomposed as a disjoint union of the collections of wedges, active rays,
and accumulation rays associated to $q\in H^0(C,K_C^2)$.

One may naturally assign coordinates of FG type to the wedges, noting that the corresponding 
spectral networks are of FG type. The coordinates associated to two wedges separated by a ray
will be related by a bi-rational cluster transformation \cite{DDP,IN,Al19}. The collections of these bi-rational transformations
are the key input data  in the RH type problems characterising the FG-type coordinates
according to \cite{GMN,Br}. 

So far we have not assigned coordinates to the
accumulation rays $\BR_+^{\times}\la_{\rm ac}$ yet. However, the 
corresponding spectral networks will have ring domains, defining a decomposition
of $C$ into a collection of annuli and bordered Riemann surfaces of simpler topological type. 
{For $\la$ in a neighbourhood of $\la_{\rm ac}$ one should be able to generalise the 
definition of FN type coordinates from Exact WKB discussed in this paper in order to associate systems of 
hybrid FG-FN-type coordinates to pairs $(q,\la)$ with $\la$ in a neighbourhood of $\la_{\rm ac}$. }

From the point of view of the RH type problems one can view this as
a way to complete the description by assigning coordinates to the accumulation rays, 
{and to some neighbourhood around them.
In order to see why it is natural to assign FN-type or hybrid coordinates to the accumulation rays 
one may recall from Section \ref{sec:juggle} that these 
coordinates can be understood as the {\it limits} of the family of FG coordinates associated to the 
infinite collection of wedges forming the neighbourhood of a given accumulation ray.
It may be useful to 
include the rational transformations from FG to FN-type coordinates into the 
collection of data defining the RH type problems, viewing these transformations
as a renormalised infinite product of the transformations associated to the active rays 
found in the neighbourhood of an accumulation ray.}

\subsection{Choices of polarisation}

As anticipated in the notations $(\CX^r,\check{\CX}_r)$, $r=1,\dots,d$, we are ultimately 
interested in choices of coordinates coming with a choice of {\it polarisation}, expressed by
the splitting into two subsets $\{\CX^r;r=1,\dots,d\}$ and 
$\{\check{\CX}^r;r=1,\dots,d\}$ of equal cardinality. This is  an additional piece of data
not canonically determined by the choice of a set of coordinates itself.
A similar issue occurs in the case of the homological coordinates. 
The choice of a canonical homology basis uniquely determines 
the polarisation of the set of coordinates $(a^r,\ca_r)$, 
$r=1,\dots,d$, indicated by our notations. 

The choice of a polarisation is not canonically determined by the choice of a WKB network
in general. A FG network allows us to define $6g-3+3n$ coordinates $X_e$ associated to the
edges of the corresponding WKB triangulation. There are $n$ monomials in the coordinates
$X_e$ determined by the parameters associated to the punctures. There is no canonical way
to define a system of coordinates with polarisation from the coordinates $X_e$, in general.

The situation is significantly better in the case of WKB networks of FN-type.
The definition of the FN type coordinates given in 
\cite[Section 6.1]{CPT} for the case of $C_{0,4}$ can be generalised straightforwardly 
to general $n$-punctured  Riemann surfaces $C$. It determines unambiguous choices
of FN-type coordinates for certain WKB networks of FN-type. This will be important for the
definition of partition function discussed in the following Section \ref{tau-sec}.

\section{Tau-functions}\label{tau-sec}
\setcounter{equation}{0}

{We finally return to the relation between tau functions and topological string partition functions.
In previous sections we have introduced key ingredients for our proposal, 
the spaces of quantum curves, and certain distinguished systems of coordinates 
on these spaces.  
We will now use these ingredients to propose a definition of fully normalised tau-functions
allowing one to define the topological string partition functions with the help of expansions of the form 
\rf{Ftransf-1}.
}

\subsection{Definition of the tau-functions}

The isomonodromic tau-functions are usually defined as the generating functions 
for the Hamiltonians $H_r$, $r=1,\dots,d$, generating the isomonodromic deformation flows.
We will here propose a generalization of this definition to surfaces $C=C_{g,n}$ of arbitrary genus $g$
which is based on the Riemann-Hilbert correspondence. 

Over open subsets $U$ of the moduli space 
$\CM(C)$ of complex structures on $C$ one may choose
families of reference  projective connections $\pa_x^2-t_U^{}(x)$ with $t_{U}^{}(x)\equiv t_{U}^{}(x;\mathbf{z})$ 
depending holomorphically on the points in $\CM(C)$ represented by coordinates $\mathbf{z}$ on $U$.
To a given projective $\la$-connection $\la^2\pa_x^2-t_\la(x)$ solving the Riemann-Hilbert problem 
one may associate the
quadratic differential $Q_{U,\la}^{}(x)=t_{\la}^{}(x)-\la^2 t_U^{}(x)$. The space of quadratic 
differentials on $C$ is canonically isomorphic to the cotangent 
space $T^*\CM(S)$ of the moduli space $\CM(S)$ of complex structures on a two-dimensional surface $S$. 
A basis $\{\pa_1,\dots,\pa_d\}$ for the tangent space $\CT(C)$ of $\CM(S)$ at $C$  
therefore determines a basis $\{E_1,\dots,E_d\}$ for 
the space of linear functions on $H^0(C,K_C^2)$. 
Assuming that the basis $\{\pa_1,\dots,\pa_d\}$ is defined by a 
choice $\mathsf{z}=(z_1,\dots,z_d)$ of local coordinates on $\CM(S)$ one may define
the tau-function $\CT^{}_U(\mathbf{z})$  as the solution to the system of equations
\begin{equation}\label{SJM-taudef}
\pa_{r}\log\CT^{}_U(\mathsf{z})=E_r(Q_{U,\la}^{}), \qquad r=1,\dots,d.
\end{equation}

The family of reference  projective connections $\pa_x^2-t_U^{}(x)$
can not be extended over all of $\CM(C)$, in general. The reference projective 
connections $\pa_x^2-t_U^{}(x)$ and $\pa_x^2-t_V(x)$ associated to two 
overlapping subsets $U$ and $V$ of  $\CM(C)$ may differ by a quadratic differential $Q_{UV}$
on the overlap. The corresponding tau-functions will therefore be related as
\begin{equation}\label{FSlbundle}
\CT^{}_U(\mathsf{z})=f_{UV}^{}(\mathsf{z})\CT^{}_V(\mathsf{z}), 
\end{equation}
with $f_{UV}^{}(\mathsf{z})$ satisfying $\pa_r\log f_{UV}^{}(\mathsf{z})=\la^2 E_r(Q_{UV}^{})$.

\subsection{Fully normalised tau-functions}

As the monodromy data are, by definition, conserved under the isomonodromic deformation flows
it is natural to consider $\CT$ as a function $\CT(\mu,\mathsf{z})$ 
of (i) the deformation times $\mathsf{z}=(z_1,\dots,z_d)$, and 
(ii) the monodromy data $\mu$ which can be  represented by points 
in $\CM_{\rm ch}(C)$. 
The definition \rf{SJM-taudef} does not fully determine how
$\CT(\mu,\mathsf{z})$ depends on the monodromy data. Multiplying
$\CT(\mu,\mathsf{z})$ with an arbitrary function on $\CM_{\rm ch}(C)$
yields another solution of the defining equations \rf{SJM-taudef}.
An important ingredient of our proposal is a natural way to fix this freedom, 
defining the  fully normalised tau-functions discussed in the following.

Picking Darboux coordinates $x=(\bx,\cbx)$ 
for $\CM_{\rm ch}(C)$ allows one to represent the 
tau-functions 
$\CT(\mu,\mathsf{z})$ in terms of functions 
$\CT(\bx,\cbx;\mathsf{z})\equiv \CT(\mu(\bx,\cbx),\mathsf{z})$ of the Darboux coordinates.
We will propose that to each system of preferred coordinates $x=(\bx,\cbx)$ 
on $\CM_{\rm ch}$ there corresponds 
a   fully normalised tau-function $\CT_x(\bx,\cbx;\mathsf{z})$. Discussing the dependence 
on the monodromy data we will often drop the dependence on $\mathsf{z}$ in the notations,
$\CT_x(\bx,\cbx)\equiv\CT_x(\bx,\cbx;\mathsf{z})$.

The tau-functions ${\CT}_x$ are strongly constrained by the following system of 
difference equations. Note that the definition of the coordinates
$x=(\bx,\cbx)$ involves the choice of a Lagrangian subspace
of $\BT$, parameterised by $x^r$, $r=1,\dots,d$. The difference equations
characterising the tau-functions are then
\begin{subequations}\label{taushift}
\begin{align}
&\CT_x(\bx,\cbx+\de_r)=
\CT_x(\bx,\cbx),\\
&\CT_x(\bx+\de_r,\cbx)=e^{-2\pi\mathrm{i}\,\cx_r}\CT_x(\bx,\cbx).
\end{align}
\end{subequations}
It is clear that multiplication of $\CT_x(\bx,\cbx)$ by a constant will preserve the validity 
of \rf{taushift}. The best we can hope for is therefore to be able to fix the normalisation of the functions
$\CT_x(\bx,\cbx)$ up to a multiplicative constant independent of $\bx$, $\cbx$. 
One may take this freedom into account by considering equivalence
classes $[\CT_x(\bx,\cbx)]$ of functions $\CT_x(\bx,\cbx)$  defined by 
the equivalence relation $\CT_x(\bx,\cbx)\sim \tilde{\CT}_x(\bx,\cbx)$ if 
there exists a constant $\nu_x\in\BC^\times$  such that $\CT_x(\bx,\cbx)=\nu_x \tilde{\CT}_x(\bx,\cbx)$.

The difference equations \rf{taushift}
are equivalent to the fact that the functions $\CT_x$ admit an
expansion in the form of a generalised theta series,
\begin{equation}
\CT_x(\bx,\cbx)=
\sum_{\mathsf{n}\in\BZ^d}e^{2\pi\mathrm{i}(\mathsf{n},\cbx)}
Z_x(\bx+\mathsf{n}).
\end{equation}
We claim that we can associate  fully normalised tau-functions $\CT_x$ to 
any collection of coordinates $x$ associated to spectral networks, be it FG-type, 
FN-type or hybrid type. A very brief outline of our approach can be found in Section \ref{computing} below.

\subsection{Difference generating functions for changes of coordinates}

We will assume having chosen a cover  of
$\CM_{\rm ch}(C)$ with a set of charts $\{\CU_{\imath};\imath\in\CI\}$.
Let  $\CU_\imath$ and $\CU_\jmath$ be overlapping charts  with coordinates
$x_\imath=(\bx_\imath,\cbx^\imath)$ and $x_\jmath=(\bx_\jmath,\cbx^\jmath)$,
respectively.  The coordinates considered in this paper will be such that the
equations $\bx_\imath=\bx_\imath(\bx_\jmath,\cbx^\jmath)$ can be solved for 
$\cbx^\jmath$ in $\CU_\imath\cap\CU_\jmath$,
defining a function $\cbx^\jmath(\bx_\imath,\bx_\jmath)$. 
Having defined tau-functions $\CT_\imath(\bx_\imath,\cbx^\imath)$ and 
$\CT_\jmath(\bx_\jmath,\cbx^\jmath)$ associated to charts $\CU_\imath$ and $\CU_\jmath$, 
respectively, there exist relations of the form
\begin{equation}\label{transition}
\CT_\imath(\bx_\imath,\cbx^\imath)=F_{\imath\jmath}(\bx_\imath,\bx_\jmath)
\CT_\jmath(\bx_\jmath,\cbx^\jmath),
\end{equation}
on the overlaps $\CU_{\imath\jmath}=\CU_\imath\cap\CU_\jmath$.
In order to ensure that 
both  $\CT_\imath$ and $\CT_\jmath$ satisfy relations of the form \rf{taushift} 
the function $F_{\imath\jmath}(\bx_\jmath,\bx^\jmath)$ must satisfy the two relations
\begin{subequations}\label{dGenFct}
\begin{align}
&F_{\imath\jmath}(\bx_\imath+\de_r,\bx_\jmath)= e^{-2\pi\mathrm{i}\,{\cx}^\imath_r}
F_{\imath\jmath}(\bx_\imath,\bx_\jmath),\\
&F_{\imath\jmath}(\bx_\imath,\bx_\jmath+\de_r)=
e^{+2\pi\mathrm{i}\,{\cx}^\jmath_r}
F_{\imath\jmath}(\bx_\imath,\bx_\jmath).
\end{align}
\end{subequations} 
We will call functions $F_{\imath\jmath}(\bx_\jmath,\bx^\jmath)$
associated to 
a change of coordinates $x_{\imath}=x_{\imath}(x_{\jmath})$
satisfying the relations \rf{dGenFct}  
{\it difference generating functions}.

The relations between the function $\CT_{\imath}$ and $\CT_{\jmath}$ associated to 
different charts   $\CU_\imath$ and $\CU_\jmath$ takes a slightly different form 
in the cases where $\bx_\imath=\bx_\jmath$, $\cbx_\imath=\cbx_\jmath+f_{\imath\jmath}(\bx_{\imath})$.
In this case one needs to modify \rf{transition} to 
\begin{equation}\label{transition2}
\CT_\imath(\bx_\imath,\cbx^\imath)=F_{\imath\jmath}(\bx_\imath)
\CT_\jmath(\bx_\jmath,\cbx^\jmath),
\end{equation}
with a function $F_{\imath\jmath}(\bx_\imath)$ satisfying
\begin{equation}
F_{\imath\jmath}(\bx_\imath+\de_r)=
e^{-2\pi\mathrm{i}\,(\cbx_\imath-\cbx_\jmath)}
F_{\imath\jmath}(\bx_\imath)=
e^{-2\pi\mathrm{i}\,f_{\imath\jmath}(\bx_{\imath})}
F_{\imath\jmath}(\bx_\imath).
\end{equation}
Examples for relations of the form \rf{transition2} have been presented in \cite{CPT}.

\subsection{Computing the difference generating functions}\label{computing}

We propose that there exists a cover $\{\CU_{\imath};\imath\in\CI\}$ of
$\CM_{\rm ch}(C)$ having coordinates 
$x_\imath=(\bx_\imath,\cbx^\imath)$ associated to the subsets $\CU_{\imath}$ for 
$\imath\in\CI$, together with a set of difference generating 
functions $F_{\imath\jmath}$  defined on the intersections $\CU_{\imath\jmath}=\CU_\imath\cap\CU_\jmath$ 
as introduced  above. 
In Section \ref{sec:relations-theta-series} we discussed several examples of this structure, and computed the corresponding difference generating functions explicitly.
We will now outline an approach to compute these in general.

We plan to give a detailed proof  elsewhere. For the moment let us note 
that a related 
problem has been encountered in the context of quantum Teichm\"uller 
theory. This theory was defined using coordinates of FG type in \cite{CF}.
In order to give an unambiguous definition of the unitary operators representing the 
changes of coordinates associated to different triangulations one needs to choose polarisations.
An elegant formalism for doing this was the basis of Kashaev's approach to the 
quantisation of the Teichm\"uller spaces \cite{Ka}. In order to establish the relation to
conformal field theory it was necessary to construct quantised analogs of the 
changes of coordinates between coordinates of FG and FN type \cite{T07}.
The  operators representing the changes between different systems of FN coordinates 
defined thereby are the characteristic data of 
the generalised modular functor associated to the quantum Teichm\"uller theory \cite{T07}.
The computation of 
explicit integral representations for these operators was completed in \cite{TV}.

One may in our case proceed similarly. 
{Instead of infinite-dimensional Hilbert spaces associated to 
systems of FN-coordinates one will here find one-dimensional spaces, with tau-functions 
representing local bases. The 
Moore-Seiberg groupoid characterising the modular functor has vertices associated to pants decompositions,
and edges corresponding to changes of pants decompositions. It will be represented by the 
difference generating functions associated to the changes of FN-coordinates.} 
The use of a formalism like the one used in \cite{T07,TV}
reduces the computation of the difference generating functions associated to the changes of 
coordinates to a small number of basic cases. For the passage from 
FG to FN type coordinates one may find\footnote{In many cases there is a triangulation of $C$
admitting a restriction to a ring domain isotopic to the triangulation of the twice punctured sphere
defining the FG type coordinates relevant for the Painlev\'e III case.
The remaining cases can be covered with the trick used in \cite{NT}.} 
 the result with the help
of  \cite[Proposition 4.3]{ILTy2}. The difference generating functions
for changes between two systems of FN type coordinates can be 
reduced to the cases where $C=C_{0,4}$ and $C=C_{1,1}$, respectively.
It has been found for $C=C_{0,4}$ in \cite{ILTy,ILP}.
The case $C=C_{1,1}$ can  be reduced to $C=C_{0,4}$ by using the 
main idea from \cite{HJS}.

\subsection{Partition functions associated to solutions of the Riemann-Hilbert problem}

Let $\CV_w\subset \CZ'$ be the set of all points in $\CZ'$ such that the 
WKB network $\CW_{q,\la}$ has topological type $w$. 
Let $x_w=(\bx_w,\cbx_w)$ be coordinates on $\CM_{\rm ch}(C)$ 
associated to $\CW_{q,\la}$ by 
abelianisation. One should keep in mind  that the definition of $x_w$ involves an additional choice
not canonically determined by the topological type $w$, the choice of a polarisation
represented by the splitting as $x_w=(\bx_w,\cbx_w)$.
The corresponding coordinates on $\CV_w\subset\CZ'$ are denoted by $\CX_w$.

Whenever the WKB network has FN type it defines both a polarisation and a pants 
decomposition. We can then use the gluing construction from CFT, reviewed and adapted to 
the cases of our interest in Appendix \ref{FF-CFT}, in order to define 
fully  normalised tau-function $\CT_{x_w}(\mu)$ for this case. 
In many cases there should also exist natural choices of polarisation 
for the FG coordinates associated to  WKB triangulations, allowing us
to define  fully  normalised tau-function $\CT_{x_w}(\mu)$ for such cases as well.
The Painlev\'e III case discussed in this paper is such a case. 

Having associated a  fully  normalised tau-function $\CT_{x_w}(\mu)$ 
to the coordinate system $x_w$ on 
$\CM_{\rm ch}(C)$ one may  define a function $\Theta_w:\CV_w\ra\BC$ such that
\begin{equation}
\Theta_w(\la,\CE,\nabla_\la)=\CT_{x_w}\big(\mathsf{Hol}[\la,\CE,\nabla_\la];\mathbf{z}\big).
\end{equation}
We conjecture that
the coordinates $\CX_w$ and the corresponding functions $\Theta_w$
can be  analytically continued to coordinates 
defined in larger regions $\hat{\CV}_w$ 
such that the collection of  $\hat{\CV}_w$ associated to all possible topological 
types $w$ of spectral networks provides a cover of  $\CZ$.

One may then define 
the topological string partition function 
$Z_w^{\rm top}$ on $\CB_C\times\BC^\times$ 
from  the coefficients $Z_w(\bx)$ appearing in the theta-series expansions 
\begin{equation}\label{theta-series++}
\CT_{x_w}(\bx,\cbx)=
\sum_{\mathsf{n}\in\BZ^d}e^{2\pi\mathrm{i}(\mathsf{n},\cbx)}
Z_{w}(\bx+\mathsf{n}).
\end{equation}
To this aim let us recall that the coordinates $\CX$ are related to 
a system of homological coordinates $(\mathsf{a},\check{\mathsf{a}})$ 
via \rf{CX-asym}. We may then define
\begin{equation}
Z_w^{\rm top}(\sa,\la)=Z_w^{}(\fr{1}{\la}\sa_w).
\end{equation}
We are here taking advantage of the fact that there is a natural 
one-to-one correspondence between the homological coordinates $\sa$
and the coordinates $\sx$ parameterising Lagrangian subspaces of $\CZ$ defined by
fixing $\check{\sx}$.

\subsection{A holomorphic line bundle over $\CZ'$}

Let $\{\hat{\CV}_{\imath};\imath\in\CI\}$ be a cover $\CZ'$ having charts $\hat{\CV}_\imath\equiv\hat\CV_{w_\imath}$ 
equipped with Darboux coordinates   $\CX_\imath=\CX_{w_\imath}$, projecting to subsets $U_\imath\subset \CM(C)$
equipped with 
reference projective connections $\pa_x^2-t_\imath(x)$, and let the 
corresponding  partition functions be $\Theta_{\imath}\equiv\Theta_{w_\imath}$ for $\imath\in\CI$.
Whenever there are charts $\hat{\CV}_\imath$, $\hat{\CV}_\jmath$ such that 
$\hat{\CV}_{\imath\jmath}={\hat{\CV}}_{\imath}\cap {\hat{\CV}}_{\jmath}\neq \emptyset$ one may define on
$\hat{\CV}_{\imath\jmath}$ the  function
$\Phi_{\imath\jmath}=\Theta_{\imath}/\Theta_{\jmath}$.
The function $\Phi_{\imath\jmath}$ factorises as
\begin{equation}\label{FS-factor}
\Phi_{\imath\jmath}=f_{\imath\jmath}^{}(\mathbf z)F_{\imath\jmath}(\mathsf{Hol}[\la,\CE,\nabla_{\la}]),
\end{equation}
where  $f_{\imath\jmath}(\mathbf{z})\equiv f_{U_\imath U_\jmath}^{}(\mathbf{z})$,
with $f_{UV}^{}(\mathbf{z})$
 being defined in \rf{FSlbundle}, and 
$F_{\imath\jmath}(\mathsf{Hol}[\la,\CE,\nabla_{\la}])$  being the difference generating function 
associated to the change of coordinates
between the spectral coordinates associated to $\hat{\CV}_\imath$ and $\hat{\CV}_\jmath$, respectively.

We claim that the collection of all such functions $\Phi_{\imath\jmath}$ defines  
a holomorphic  line bundle $\CL_\Theta^{}$ on $\CZ'$. 
This line bundle comes with a collection of preferred
holomorphic sections $\{\Theta_{\imath},\imath\in\CI\}$, defining a connection $\nabla_\Theta$
on $\CL_\Theta^{}$. 

{One should note that this claim is realized in a somewhat subtle way. 
The relation between tau-functions and free fermion conformal
blocks outlined in Appendix \ref{FF-CFT} allows us to take advantage of some insights and results from 
conformal field theory (CFT).
Recall that the changes of pants decompositions defining FN-type coordinates represent
the Moore-Seiberg groupoid. The representations of this groupoid coming from 
CFT have changes of pants decompositions realized by the products of  two  operations. 
One of them is independent of the complex structure, in our case represented by the
functions $F_{\imath\jmath}$ in \rf{FS-factor}.
The definition of conformal blocks, on the other hand, 
involves choices of projective structures on Riemann surfaces, leading to 
complex-structure dependent factors 
\cite{FS,TV,T17},
 in \rf{FS-factor} represented by the functions $f_{\imath\jmath}^{}$.
A discussion quite similar to the one in \cite{FS,TV} shows that 
neither $f_{\imath\jmath}^{}$ nor $F_{\imath\jmath}^{}$ would 
define line bundles by themselves, as the respective cocycle conditions will be violated on Riemann surfaces 
of higher genus.
The violation of the cocycle conditions by the complex structure independent 
factors $F_{\imath\jmath}^{}$ is universal and completely determined by the central charge \cite{FS}. 
It must therefore coincide with the one calculated  in \cite{TV}, see equations
(6.34e) and (8.22).  However, the existence
of local sections defined by the gluing construction of conformal blocks implies that 
the violations of the cocycle conditions for $f_{\imath\jmath}^{}$ and $F_{\imath\jmath}^{}$ 
must exactly cancel each other.} 

We  now see why it  was important to complete the solution to the 
Riemann-Hilbert problem defined by the FG-coordinates by 
including FN- and hybrid type coordinates.  Without this completion there would be 
no natural way to assign topological string partition functions to the neighbourhoods in 
$\CZ$ associated to spectral networks of FN-type. In the application to 
topological string theory these regions appear to be of particular interest, corresponding to 
regions admitting weak-coupling expansions. Let us furthermore recall that networks of FN type 
determine a canonical polarisation for the coordinates associated to them. This is important for 
having an  unambiguous definition of the corresponding partition functions $\Theta_w$.
For FG type networks one does not have a canonical polarisation, forcing us to make additional 
choices.

\section{Summary and relations to other lines of research}\label{Sec:summary}
\setcounter{equation}{0}

We shall here offer a brief summary, and a discussion of relations
to other lines of research.

\subsection{Summary}

We observed a direct correspondence between choices of coordinates and possible
definitions for fully normalised free fermion partition functions. When applied to the FN type 
coordinates associated to weak coupling regions by exact WKB, we recovered the partition 
functions computed by the topological vertex. Weak coupling regions in the space of quadratic 
differentials are characterised 
by the appearance of ring domains for certain critical phases $\theta_\ast$ 
of the parameter $\hbar$. 
Given a quadratic differential $q$ 
in the weak coupling region there exists a finite neighbourhood of the phase $\theta_\ast$
within which we can use the ring domains defined by $(q,\rho e^{i\theta_\ast})$, $\rho\in\BR_+$, 
to define coordinates of FN type with the help of  the decomposition 
of the surface $C$ defined by the ring domains appearing at $\mathrm{arg}(\la)=\theta_\ast$.
It may then seem natural to start by defining the partition functions in a given
weak coupling region, and extend the definition to a larger domain using the changes
of normalisation induced by  the coordinate transformations associated to changes of 
the Stokes graph. 

One may note, however, that there are regions in the space of quadratic differentials 
where no ring domain occurs. Such regions are called strong coupling regions. The
results of \cite{ILTy2} give a candidate for the definition of the partition functions in this
region. We had seen that there exist changes of Fock-Goncharov coordinates
preserving the feature that there exist expansions of generalised theta-series type.
In this way we can generate other candidates for partition functions in the strong 
coupling region. A general issue to be addressed in any attempt to assign partition 
functions to coordinates of Fock-Goncharov type is the need to choose a polarisation.

The changes
of normalisation induced by  the changes of coordinates
have been used to define the 
line bundle $\CL_{\Theta}$. 
Having defined  a partition function in a weak coupling region gives a
distinguished holomorphic section $\Theta$ of the line bundle $\CL_{\Theta}$. It does not
matter which weak coupling region is used 
since the relations between the partition functions assigned to different weak 
coupling regions are defining the corresponding 
transition functions of $\CL_{\Theta}$. The functions $\Theta_\imath$ representing the 
section $\Theta$ in the weak coupling region with label $\imath$ are 
identified with the partition functions associated to the same region.

In order to define actual partition {\it functions} in strong coupling regions one would 
need to make additional choices. For each topological type of Stokes graph
one would need to choose a polarisation for the Fock-Goncharov coordinates 
associated to it.
It is not clear to us at the moment if there is a
natural way to do this. 

{ There are, in the end of the day, always certain ambiguities in the definition of partition functions in
quantum field theory and (topological) string theory. 
These ambiguities include, on the one hand, choices
entering the
definition of the classical part represented by the prepotential, in our case  
concretely represented by the 
necessity to choose a canonical basis for the homology of the spectral curve.
Other choices can be related to the choice of a UV duality frame, here represented by the 
choice of a pants decomposition. 
Physically relevant information appears to be encoded in the 
transition functions between different regions in the moduli space. These transition functions
in particular encode the spectrum of BPS states, as will be discussed more in the following.
From this point of view it seems tempting to view the partition functions associated to FN and FG type coordinates as 
different ways to package the same physical information.} 

{In view of the intended application to topological string theory
one should keep in mind that the A-model topological string (Gromov-Witten (GW)) theory
comes with preferred reality conditions, given by the reality of the
K\"ahler parameters.  The topological
string partition functions should certainly be defined on this real slice, possibly allowing 
analytic continuations to the complexified K\"ahler cones. 
This real slice gets represented by the JS-differentials in the
cases studied in this paper. All our discussions indicate that the FN-type coordinates are relevant in this case.  
It is furthermore important in this context that there is an analytic continuation away from 
the real slice represented by JS differentials allowing us to extend the 
definitions of the partition functions to the complexified K\"ahler cones. 

The role of the FG coordinates needs to be
better understood from this point of view. This type of coordinates might play a role 
in other representations of topological string theory related to the 
GW theory by string dualities. Dual representations of the topological string 
relate it to the counting of BPS degeneracies. 
Some of these connections will be discussed
next. }

\subsection{Relations to the spectrum of BPS states} 

A key role in  our paper is played by the changes of coordinates associated 
to non-generic topological types of the Stokes graphs.  The relevant non-generic types
of Stokes graphs include the appearance of saddle trajectories, and the appearance of closed
trajectories. Work initiated by \cite{GMN} has revealed a profound relation between 
Stokes graphs of these types and BPS states in the class $\CS$-theories associated to
the Riemann surface $C$ and the Lie algebra $A_1$.

A general framework for the description of the spectrum of BPS states in $d=4$, $\mathcal{N}=2$ 
supersymmetric field theories has previously been proposed in \cite{GMN08}.
A key role in the approach 
of \cite{GMN08} is played by certain Darboux coordinates for the moduli space of vacua 
of the three-dimensional field theories obtained by circle compactification of the
$d=4$, $\mathcal{N}=2$ supersymmetric field theory of interest. These Darboux coordinates can be
characterised as solutions of a problem of Riemann-Hilbert (RH) type, the key data in the 
formulation of the relevant RH type problem being the changes of coordinates generated by
changes of the phase of the hyperk\"ahler parameter $\zeta$.

The work \cite{GMN} offers a large class of interesting examples for the description of the 
spectrum of BPS states previously proposed in \cite{GMN08}. 
The Darboux coordinates considered in \cite{GMN08} have been identified with Fock-Goncharov
coordinates for the case of class $\CS$-theories considered in \cite{GMN}. These  coordinates are related to
the coordinates considered in our 
paper in the conformal limit \cite{Ga14,DFKMMN,CW}. This induces a direct relation between
the data defining the RH type problems arising in the study of BPS spectra with
the data defining the line bundle $\CL_{\Theta}$ introduced in our paper.

The relation between the spectrum of BPS states and Darboux coordinates can be made 
particularly transparent using 
the nonlinear integral equations of TBA type satisfied by the Darboux coordinates  
\cite{GMN08}. The conformal limit of these
integral equations first discussed in \cite{Ga14} should be relevant for the cases studied 
in this paper. For the case $C=C_{0,2}$ considered above such integral equations have been 
studied in particular in \cite{Ga14,HNb,GGM}. Both \cite{HNb} and \cite{GGM} propose nonlinear 
integral equations for FN type coordinates.\footnote{This is not explicitly formulated in this way in \cite{GGM}, 
but this follows from the identification of the quantum periods $\Pi^{\rm ex}_{A,B}$ with FN type 
coordinates mentioned in Section \ref{Rel:specdet} together with the results of \cite[Section 4.2]{GGM}.}


One may expect that the generalisations of Donaldson-Thomas (DT) theory 
developed by Kontsevich, Soibelman, Joyce and others
give a widely applicable  framework for the mathematical description of 
BPS states in string theory which should be  related to the one in SUSY field 
theory by geometric engineering \cite{CDMMS}. 
RH type problems solved by Darboux coordinates on the
space of stability conditions have been formulated  in this context  in \cite{Br}.
Projective connections on Riemann surfaces provide large families of examples for the solutions
to these RH type problems, as conjectured in \cite{Br}, and more 
recently proven in \cite{Al19b}. Exact WKB relates the transitions between
coordinate systems representing solutions to these RH type problems to the DT invariants
associated to spaces of quadratic differentials on Riemann surfaces $C$ \cite{BrS}.
The quantum curves studied in this paper generalise the projective connections 
considered in \cite{Br,Al19b} by allowing apparent singularities. We conjecture that 
the results of \cite{Br,Al19b} can be generalised to this case. This
would represent important groundwork for realising some of  the proposals made in this paper
in the case of general Riemann surfaces $C$. 

{In most of the above-mentioned studies of the spectrum of BPS states the Fock-Goncharov coordinates
have played a key role. Such coordinates are assigned to certain wedges in the $\hbar$-plane, 
in the solution 
provided by projective connections represented by the Borel summations of Voros symbols. 
A variant of the Stokes phenomenon makes the Borel summations jump across the rays in the $\hbar$-plane
separating the wedges \cite{DDP}. The coordinate transformations representing 
these jumps represent key data entering the formulation of the RH-type problems
introduced in \cite{GMN08,Br}.
Our results suggest that there exist natural completions of the
solutions to the RH type problems proposed in \cite{GMN,Br} assigning FN-type coordinates to the accumulation 
rays. The transformations between coordinates of FN and FG type can be thought of as infinite products 
of the coordinate transformations associated to the infinite sequence
of rays surrounding an accumulation ray.}

{
Bridgeland has recently proposed geometric structures encoding DT invariants 
called Joyce structures \cite{Br19}.  Joyce structures can be expressed in terms of complex hyperk\"ahler 
structures on the total spaces of the tangent bundles on the spaces of stability conditions \cite{BrSt}. 
There exist natural generating functions for these geometric structures called tau-functions in \cite{Br23}.}

{
The stability conditions on spaces of quadratic differentials defined in \cite{BrS} have been shown
in \cite{Br22} to define Joyce structurse closely related to isomonodromic deformations of bundles with 
connections. The tau-functions associated to these Joyce structures coincide with the isomonodromic 
tau-functions \cite{Br23}. These results exhibit the the theory of Joyce structures as a natural
geometric framework for the relations between spectra of BPS states and the 
objects discussed in this paper. The relation between Joyce structures 
and the generalized theta series expansions underlying the approach discussed in this paper
should be clarified.}

{String theory suggests that the topological string
partition functions studied in this paper are related to 
 generating functions for the degeneracies of bound states of D0, D2 and D6-branes
 \cite{MNOP,DVV}. It seems natural to expect that such degeneracies should 
display wall-crossing phenomena somewhat analogous to the 
phenomena studied in the case of toric CY in \cite{JM}, and to 
the wall-crossing of 
 framed BPS states studied in \cite{GMN10,CDMMS}.
 From this point of view one might suspect that the relations predicted  
 in \cite{DHSV} only hold in weak coupling chambers of the space of stability data, 
 being replaced by more subtle relations in other chambers. However, in view of
 the results presented in this paper  
 we feel tempted 
 to speculate that the jumping of the free fermion partition functions $Z_{\rm ff}$ discussed
 here represents the framed wall-crossing behaviour for the cases at hand
 in all chambers.}

{Of particular interest in this regard seem to be the transitions between partition functions 
associated to FG and FN type coordinates, respectively. While it is not clear if the 
coefficients of FG-type expansions 
will correspond to topological string partition functions in the sense of GW theory, 
it still seems reasonable to suspect that the coefficients of the 
strong coupling expansions might have an interpretation as generating functions for 
degeneracies of BPS states.

Among the ingredients in our proposal it seems to us that the characterisation of the partition 
functions with the help of RH type problems solved by distinguished coordinates on the spaces of stability 
data proposed above
has a particularly high potential for generalisations beyond the family $Y_\Sigma$ of local CY, 
and possibly even beyond the case of local CY.}

\subsection{Relation to spectral determinants} \label{Rel:specdet}

Another approach to the definition of a non-perturbative completion of the topological string 
partition function is based on the spectral determinants of the quantum curve \cite{GHM}. 
This approach has been successfully 
applied to various toric CY.  The geometric engineering of SUSY field theories in
string theory suggests that the resulting partition functions might be related to the 
partition functions studied in this paper.  {In \cite{BGT} it has indeed been shown 
that the limit used in the geometric engineering of pure SU(2) SYM relates 
the spectral determinant associated to the relevant toric CY to the 
tau-function of Painleve III at $\eta=0$. 

The spectral determinant associated to the quantum curve relevant for the case $C=C_{0,2}$
has recently been studied in \cite{GGM}. 
The result is given by an explicit expression (formula (5.6) in \cite{GGM}) in terms of  quantum periods 
denoted $\Pi^{\rm ex}_{A,B}$ in \cite{GGM}. It is possible to show that these quantum 
periods are closely related to the FN type coordinates introduced for the case $C=C_{0,2}$
in our paper. It seems intriguing to observe that the formula for the 
spectral determinant proposed in \cite{GGM} resembles strongly
our formula \rf{eq:Y-FG} expressing the Fock-Goncharov coordinates 
in terms of FN type  coordinates.} 

In this regard it seems interesting to observe that the function 
$D(U,\hbar)=e^{\pi\nu(U,\hbar)}$ defined by evaluating
the FG coordinate ${Y}^{-1}$ introduced in Section \ref{sec:FG} 
on the restriction of the monodromy map to the submanifold represented by opers
coincides with the spectral determinant of the differential operator 
$\CD_{\rm\sst M}=\la^2\pa_w^2-2\cosh(w)$. This differential operator is
related to the quantum curve considered above for the 
case $C=C_{0,2}$ by dropping the apparent singularities, setting $U=0$, and performing
the change of coordinate $x=e^w$.
To establish the relation between $D(U,\hbar)$ and a spectral 
determinant one can proceed along the lines of \cite{BLZ},
the key observation being that zeros of the function $D(U,\hbar)$ 
are in one-to-one correspondence with eigenfunctions of the operator
$\CD_{\rm\sst M}$ which decay for $w\ra \pm\infty$, as follows easily from
\rf{eq:conn-mat} using equation \rf{eq:Y-FG} which relates the diagonal elements of the matrix $E$
in \rf{eq:conn-mat} to  $D(U,\hbar)$.

One should note, however, that the spectral determinant studied in \cite{GGM} is related to 
 the partition functions studied in our paper in a somewhat intricate way.
 It should  be interesting to clarify the relation between
 spectral determinants and topological string partition functions for this class of 
 cases.

\subsection{Relation to topological recursion}

The topological recursion \cite{EO1} offers a rather flexible scheme for defining formal series expansions
of various objects arising in mathematical physics  including  the large $N$ expansion of matrix models, 
and the topological string partition functions. It is known that the topological 
recursion can in particular be used to describe the genus expansion of the topological 
string partition functions associated to  toric Calabi-Yau manifolds \cite{BKMP,EO2}.
One may therefore expect that the expansion of the partition functions studied in this paper in powers of $\hbar$ 
can be described with the help of topological recursion. 

An  encouraging step in this direction has recently been made in \cite{Iw}. This paper studies the 
tau-function of the Painlev\'e I equation, and shows how to describe the expansion in a parameter $\hbar$ 
which is a direct analog of our parameter $\hbar$  with the 
help of the topological recursion. Given that all Painlev\'e equations are obtained from 
Painlev\'e VI by certain collision limits, one expects to find a similar picture for 
all Painlev\'e equations. The results from \cite{Iw} suggest that the objects most straightforwardly 
defined by topological recursion are the partition functions playing the role of 
coefficient functions in the expansions of generalised theta series type considered. 
In \cite{Iw} it is shown for the case of Painlev\'e I 
that the object {\it defined} by these expansions satisfies the defining equation 
for the corresponding tau-function. It is furthermore shown in \cite{Iw} how the Stokes data 
of Painlev\'e I are related to the variables appearing in the generalised theta series.
It is clear that the generalisation to the cases studied in our paper would be very interesting.

A particularly interesting question is to what extent the non-perturbatively defined partition 
functions defined here 
are determined by the formal series expansions provided by the topological recursion.
We believe that our work offers a few encouraging hints in this direction. We have observed a
direct correspondence between certain distinguished sets of coordinates for the space of 
monodromy data and non-perturbatively defined partition functions. We have furthermore seen
that the Borel summation of exact WKB allows us to define distinguished coordinates for the space
of quantum curves in many cases. These coordinates can be seen as a generalisation of the 
objects frequently called quantum periods away from the locus of opers. This means that we 
have given a prescription how to assign non-perturbatively defined partition functions to 
(generalised) quantum periods.
We have furthermore confirmed by explicit
computations that the partition functions associated to the quantum periods 
give us exactly the topological string partition functions computed with the help of the topological 
vertex. We take this as a piece of evidence for the conjecture 
that the partition functions of our interest are determined by 
exact WKB and/or topological recursion in a way which is "as canonical as possible".

A general framework relating topological recursion to structures from conformal field theory was recently 
proposed in \cite{BE} building upon previous work \cite{BER}. Although there are some suggestive 
similarities to the formalism used here, it is currently not clear to us how these approaches are 
related in detail.

\subsection{Relations to hypermultiplet moduli spaces}

The geometry of hypermultiplet moduli spaces in string compactifications has
been studied intensively, see \cite{AMPP} for a review. The main challenges are 
to understand the corrections to the geometry associated to D-instantons 
and NS5-branes.  There is a systematic procedure to compute
D-instanton corrections from the spectrum 
of stable D-branes in the given string compactification \cite{APP2}. 
{ Computing NS5-brane corrections is less well-understood, but it has been 
suggested in \cite{APP1}  that the twistor space partition function  
associated to a single NS5-brane 
admits a representation in the form of a generalised theta series
with expansion coefficients given by the topological string partition functions (see equation (5.11)
in \cite{AMPP}). The similarity between the generalised theta series appearing in \cite{APP1} 
and the ones studied in our paper is probably not an accident. It has been proposed 
in \cite{AMPP} that the above-mentioned twistor space partition function 
may represent a section of a pre-quantum line bundle over a cluster variety. This suggestion 
has been developed further in \cite{AP}.
The transition functions defining this line bundle are 
representable in terms of the generating functions for the cluster mutations
describing transitions between different systems of Darboux coordinates
which can be represented explicitly with the help of the 
Rogers dilogarithm function. A closely related line bundle has been studied in \cite{Nei}.}

We  find it intriguing to observe that the line bundle $\CL_{\Theta}$ defined in this paper 
from the difference generating functions describing changes of Darboux-coordinates of FG-FG,
FG-FN, FN-FG, and FN-FN type also admits a description using transition functions 
constructed from the Rogers dilogarithm. To make this concrete let us recall that the 
basic building block for all of these changes of variables is the change of coordinates 
associated to the flip discussed in Section \ref{FG-to-FG}. In Appendix \ref{app:Rogers} we show that a small 
change of normalisations yields an alternative representation for the transition function
associated to the flip which { can be recognised as an example of the transition functions considered in \cite{AP}.
This suggests that the line bundles  $\CL_{\Theta}$ introduced in our paper are in fact related
to the line bundles denoted  $\CL_{\Theta}$ in \cite{AP}.

We  believe that a detailed investigation of the relations between these two 
lines of research should be worthwhile. This might in particular suggest how to characterise the 
topological string partition functions for more general Calabi-Yau manifolds within a similar framework.}

\subsection{Relation to a variant of quantum Teichm\"uller theory}

The difference generating functions associated to changes of coordinates are  more than just analogs of the operators representing the changes of coordinates
in quantum Teichm\"uller theory. The observations made in \cite[Section 7.2]{ILT} indicate that the
theta series expansions  \rf{theta-series++} 
relate a representation of the quantised algebra $\CA$ of functions on 
$\CM_{\rm ch}(C)$ in terms of finite difference operators acting on 
wave functions $Z_{w}(\bx)$ 
to an equivalent representation on sections $\CT_{x_w}(\bx,\cbx)$ 
of the line bundle on $\BT$ defined by the relations \rf{taushift} in which all elements of
$\CA$ are represented as multiplication operators.
The 
difference generating functions represent the
operators associated to changes of coordinates in a 
variant
of the quantum Teichm\"uller theory. 

The relevant version of  quantum Teichm\"uller theory is probably related 
to the theory developed in unpublished work of 
Fock and Goncharov \cite{FG}, see \cite[Section 5.3]{AMPP} for a brief discussion. 
Key ingredients of the relevant variant of quantum Teichm\"uller theory have been proposed in 
\cite{AP}. 
The relation to the usual 
quantum Teichm\"uller theory is subtle. Quantum Teichm\"uller theory 
has a parameter $b$. The case discussed here would formally 
correspond to the value $b=\mathrm{i}$ for which the usual
quantum Teichm\"uller theory is no longer defined.

{ We'd finally like to express our expectation that our results can be recognised as
a concrete realisation of fairly old ideas going back to \cite{Wi93,ADKMV}
relating topological string partition functions to the quantisation of the moduli spaces 
of complex structures  in string theory. One may hope that generalisations of the emerging picture
may lead to a unified understanding of various aspects of topological string theory.}

\bigskip

{\bf Acknowledgements:} 
While carrying out this research, the work of IC is supported by the 
ERC starting grant H2020 ERC StG No.640159, while 
the work of JT was supported 
by the Deutsche Forschungsgemeinschaft under Germany's Excellence Strategy 
EXC 2121 ``Quantum Universe'' 390833306. 
The work of PL is supported by a grant from the Swiss National Science foundation. 
He also acknowledges the support of the NCCR SwissMAP that is also funded by the 
Swiss National Science foundation. The work of PL was also supported by the National 
Science Foundation under Grant No. DMS- 1440140 while the author was in residence 
at the Mathematical Sciences Research Institute in Berkeley, California, during the Fall 2019 semester.


\appendix

\section{The quantum curve for the case $C_{0,4}$}\label{appendix:quantum-curves}
\setcounter{equation}{0}

We focus here on quantum curves which correspond to curves $\Sigma$ 
of the type  
\begin{equation} 
\Sigma=\{(x,y)\subset T^*C_{0,4} \, | \, y^2 + q(x) = 0 \}~ 
\end{equation}
defined as double covers of $C_{0,4}=\mathbb{P}^1\backslash \{0,z,1,\infty \}$, 
with $q(x)$ being of the form
\begin{equation} \label{standardformC04qd}
q(x) 
 = \frac{d_1}{x^2} + \frac{d_2}{(x-z)^2} + \frac{d_3}{(x-1)^2} + \frac{d_1+d_2+d_3-d_4}{x(1-x)} + \frac{z(z-1)}{x(x-1)}\frac{H}{(x-z)} ~,
\end{equation}
using the notations $d_i=-a_i^2$, $i=1,2,3,4$.

The quantisation of the curve $\Sigma$ produces differential operators of the form 
$\la^2\pa_x^2-q_\la(x)$, with 
\begin{align} \label{eq:quantumcurve0ccccc}
q_\hbar (x) &= 
 \frac{\delta_1}{x^2}+\frac{\delta_2}{(x-z)^2} + \frac{\delta_3}{(x-1)^2}+\frac{\de_1+\de_2+\de_3-\de_4}{x(1-x)}+
 \frac{z(z-1)K}{x(x-1)(x-z)} 
 \nonumber \\
& - \hbar\left( \frac{u(u-1) }{x(x-1)(x-u)} + \frac{2u-1}{x(x-1)}\frac{u-z}{x-z} \right)v + \frac{3}{4} \frac{\hbar^2}{(x-u)^2} 
 ~.
\end{align}
We assume that the parameters in \rf{eq:quantumcurve0ccccc} coincide with the corresponding 
parameters in \rf{standardformC04qd} up to terms which vanish when $\la$ tends to zero.
The condition that the singularity at $x=u$ is an apparent singularity determines $K$ uniquely,
\begin{equation} 
v^2= 
\frac{\delta_1}{u^2}+\frac{\delta_2}{(u-z)^2} + \frac{\delta_3}{(u-1)^2}+\frac{\de_1+\de_2+\de_3-\de_4}{u(1-u)}+
\frac{z(z-1)K}{u(u-1)(u-z)} 
~. 
\end{equation}
$q_\la(x)$ therefore depends on the six parameters $(u,v;\de_1,\dots,\de_4)$.

We now demonstrate that it is indeed possible to obtain any quantum curve \eqref{eq:quantumcurve0ccccc} from
a holomorphic $\hbar$-connection $\nabla_\la$  
\begin{equation} 
\nabla_\la=dx(\hbar\partial_x +A(x)), \qquad
A (x) = \left( \begin{matrix} A_0 & A_+ \\ A_- & - A_0 \end{matrix} \right) \in sl_2 (\mathbb{C}) ~,
\end{equation}
by a gauge transformation. 

To begin with, we shall introduce a parameterisation of the holomorphic $\hbar$-connections $\nabla_\la$
on  $C=C_{0,4}$.
The matrix $A(x)$ representing $\nabla_\la$ can be written in the form
\begin{equation} \label{eqgen:connectionmatrix}
A (x) = \sum_{k=1}^4 \frac{A_k}{x-z_k} ~,  \qquad \mathrm{with} \quad 
A_k= \left( \begin{matrix} r_k p_k-l_k & -r_k^2 p_k+2 r_k l_k \\ p_k & -r_k p_k + l_k \end{matrix} \right) 
\in sl_2 (\mathbb{C})~,
\end{equation}
Regularity of $A(x)$ at infinity imposes the following constraints 
\begin{align} \label{eq:physConstraints}
\sum_{k=1}^4 p_k=0~, \qquad \sum_{k=1}^4 r_k p_k-l_k =0 ~, \qquad \sum_{k=1}^4 r_k^2 p_k - 2 r_k l_k =0 ~.
\end{align} 
These equations determine three out of the four parameters $p_k$, $k=1,2,3,4$.
It is furthermore
natural to identify tuples of parameters $(r_1,r_2,r_3,r_4)$ related by 
M\"obius transformations. This allows one to fix $r_1=0,r_3=1$ and send 
$r_4\to\infty$. Equations \eqref{eq:physConstraints} then determine $p_k$ to be 
\begin{align}
p_1 = -\kappa+(x-1)p ~, \qquad p_2 = p ~, \qquad p_3 = \kappa -xp ~, \qquad p_4 =0 ~,
\end{align}
where $r\equiv r_2$ and $\kappa=l_1+l_2+l_3-l_4$. 
Choosing the locations of the punctures to be at 
$(z_1,z_2,z_3,z_4)=(0,z,1,\infty)$ then completely determines the matrix $A (x)$.  
We thereby find that the matrices $A(x)$ are determined by the six parameters $(p,r;l_1,l_2,l_3,l_4)$.




As recalled in Section \ref{Sec:QCurves}, there exists a gauge transformation $h$ such that
\begin{equation} \label{eqdef:singgaugetransformation}
h^{-1} \cdot \nabla_\la \cdot h 
= dx\left(\hbar \partial_x + \left( \begin{matrix} 0 & Q_\hbar (x) \\ 1 & 0 \end{matrix} \right)\right) ~.
\end{equation}
The function $Q_\la(x)$ defined in this way can be represented as 
\begin{equation} \label{eq:quantumcurve00}
Q_\hbar (x) = A_0^2 + A_+ A_- + \hbar\left(A_0' - \frac{A_0 A_-'}{A_-}\right)+
\hbar^2\bigg(\frac{3}{4}\left(\frac{A_-'}{A_-}\right)^2- \frac{A_-''}{2A_-}\bigg) ~.
\end{equation}
By means of a straightforward computation one may then show that $Q_\la(x)=q_\la(x)$ 
iff the parameters $(p,r;l_1,l_2,l_3,l_4)$ and $(u,v;\de_1,\dots,\de_4)$ are related as
\begin{subequations}\label{par-rels}
\begin{align} \label{esdef:u}
&u=z\frac{\kappa + p (1-r)}{ \kappa + p (z-r)},\\
&v =  
- \frac{k_1}{u}-\frac{k_2-p r}{u-z} - \frac{k_3+p r-\kappa}{u-1},\qquad k_i=l_i-\frac{\la}{2},\quad i=1,2,3,\\
&\delta_i=l_i^2- \frac{\hbar^2}{4} ~, \qquad i=1,2,3 ~; \qquad 
\delta_4= l_4^2-\la l_4.
\end{align}
\end{subequations}
It is furthermore straightforward  to show that for given $(u,v;\de_1,\dots,\de_4)$ there always 
exist parameters $(p,r;l_1,l_2,l_3,l_4)$ such that the relations \rf{par-rels} are satisfied. 
This proves that all quantum curves of the form \rf{eq:quantumcurve0ccccc} can be obtained
from holomorphic $\la$-connections $\nabla_\la$.

\section{Relation with conventions used in the literature}\label{appendix:consistency-check}
\setcounter{equation}{0}

Here we demonstrate that the theta series expansion \eqref{thetaser-gen}
presented in Section \ref{Sec:Theta} 
is equivalent to
 the result of \cite{ILT}, which was using the function 
\begin{align} \label{normalisationfactorILTeV2B}
& N_\text{ILTe}(\vartheta_3,\vartheta_2,\vartheta_1) = \\
& \quad 
=\frac{G(1+ \vartheta_3+\vartheta_1+\vartheta_2)G(1+ \vartheta_3-\vartheta_1-\vartheta_2)G(1- \vartheta_3-\vartheta_1+
\vartheta_2)G(1- \vartheta_3+\vartheta_1-\vartheta_2)}{G(1)G(1+2\vartheta_3)G(1-2\vartheta_1)G(1-2\vartheta_2)} ~
\nonumber \end{align}
instead of $N(\vartheta_3,\vartheta_2,\vartheta_1)$,  
and coordinates $(\sigma,\eta_\text{ILTe})$ defined by the following parameterisation of the trace functions
 \begin{align}
L_s = 2\cos (2\pi \sigma)~, \qquad  L_u = C_+ e^{ i\eta_\text{ILTe}} + N_0 + C_- e^{- i \eta_\text{ILTe}} ~,
\end{align}
with
\begin{align} \label{factorstraceparametrtisationILTe}
& C_+ = - \frac{4 \sin \, \pi (\sigma +\theta_1-\theta_2) \, \sin \, \pi (\sigma -\theta_1+\theta_2) \, 
\sin \, \pi (\sigma +\theta_3-\theta_4) 
\, \sin \, \pi (\sigma -\theta_3+\theta_4) \, }{(\sin (2\pi\sigma))^2} \nonumber\\
& C_- = - \frac{4 \sin \, \pi (\sigma +\theta_1+\theta_2) \, \sin \, \pi (\sigma -\theta_1-\theta_2) \, 
\sin \, \pi (\sigma +\theta_3+\theta_4) 
\, \sin \, \pi (\sigma -\theta_3-\theta_4) \, }{(\sin (2\pi\sigma))^2} \nonumber\\
 & (\sin (2\pi\sigma))^2 N_0 = -2\left[\cos 2\pi\theta_1 \cos 2\pi\theta_4 + \cos 2\pi\theta_2 \cos 2\pi\theta_3 \right] 
 \nonumber\\
& \qquad\qquad\qquad\qquad\qquad 
+ 2\cos 2\pi\sigma \left[ \cos 2\pi\theta_1 \cos 2\pi\theta_3 + \cos 2\pi\theta_2 \cos 2\pi\theta_4 \right] ~.
\end{align}
For convenience let us recall that the   normalisation factor appearing in \eqref{thetaser-gen}
\begin{align} \label{normalisationfactorCompare1}
& N(\sigma,\theta_2,\theta_1) = \\
& \qquad =
\frac{G(1+\sigma+\theta_1+\theta_2)G(1+\sigma-\theta_1+\theta_2)G(1+\sigma+\theta_1-\theta_2)
G(1+\sigma-\theta_1-\theta_2)}{(2\pi)^\sigma \, G(1)G(1+2\sigma)G(1+2\theta_1)G(1+2\theta_2)} ~
\nonumber \end{align}
corresponds to coordinates $(\sigma,\eta)$ parameterising the trace functions as follows,
\begin{equation} \label{eq:TraceM23-A}
L_s = 2\cos (2\pi \sigma)~, \qquad L_u = e^{2\pi i \eta} + N_0 + e^{-2\pi i \eta} N ~, 
\end{equation}
with $N$ in equation \rf{eq:TraceM23-A} related to $C_\pm$ in 
\rf{factorstraceparametrtisationILTe} as $N=C_+C_-$. 

The ratio of the two normalisation factors \eqref{normalisationfactorILTeV2B} and \eqref{normalisationfactorCompare1} 
is given as
\begin{align*}
\frac{N (\sigma,\theta_2,\theta_1)}{N_\text{ILTe} (\sigma,\theta_2,\theta_1)  } &=  
\frac{G(1+(\sigma+\theta_1-\theta_2))}{G(1-(\sigma+\theta_1-\theta_2))}
\frac{G(1+(\sigma-\theta_1+\theta_2))}{G(1-(\sigma-\theta_1+\theta_2))} 
\frac{G(1-2\theta_1)}{G(1+2\theta_1)}\frac{G(1-2\theta_2)}{G(1+2\theta_2)}(2\pi)^{-\sigma} 
\end{align*}
and thus it follows from the identity 
\begin{equation}
S(x\pm 1)=\mp(2\sin\pi x)^{\mp 1}S(x), \qquad S(x):=\frac{G(1+x)}{G(1-x)},
\end{equation}
that  
\begin{align*} 
\frac{N(\sigma +n,\theta_2,\theta_1)}{N(\sigma,\theta_2,\theta_1)} = 
\frac{N_\text{ILTe} (\sigma +n,\theta_2,\theta_1)}{N_\text{ILTe}  (\sigma,\theta_2,\theta_1)} 
\left( - \frac{\pi}{2 \, \sin \pi (\sigma+\theta_1-\theta_2)  \sin \pi (\sigma-\theta_1+\theta_2  )} 
\right)^n ~.
\end{align*}
We similarly find for the normalisation factors with arguments $\theta_3,\theta_4$ that  
\begin{align*}  
\frac{N(\sigma +n,\theta_3,\theta_4)}{N(\sigma,\theta_3,\theta_4)} = 
\frac{N_\text{ILTe} (\theta_4,\theta_3,\sigma +n)}{N_\text{ILTe} (\theta_4,\theta_3,\sigma)} 
\left(\frac{(\sin (2\pi\sigma))^2 }{2\pi \sin \pi (\sigma+\theta_3-\theta_4)  \sin \pi (\sigma-\theta_3+\theta_4  )} 
\right)^n ~.
\end{align*}
It follows that the expansion found in \cite{ILT} is equivalent to the one
presented in Section \ref{Sec:Theta} 
if $\eta_\text{ILTe}$ is related to $\eta$ such that
\begin{align*}
&e^{2\pi i \eta} = \\
&\;\;- \, e^{2\pi i \eta_\text{ILTe}} \, \frac{ 4 \,  \sin \pi  (\sigma+\theta_1-\theta_2) \sin \pi  (\sigma-\theta_1+\theta_2)
\sin \pi  (\sigma+\theta_3-\theta_4) \sin \pi  (\sigma-\theta_3+\theta_4)}{(\sin (2\pi\sigma))^2} ~.
\end{align*}
Comparing the parameterisations of the trace functions through
 $\eta$ and $\eta_\text{ILTe}$, we indeed find the same relation between these two variables.

\section{Traffic rules and Fock-Goncharov coordinates}\label{app:FG}
\setcounter{equation}{0}

In \cite{FG}, Fock and Goncharov constructed coordinate systems on $\mathcal{M}_{flat}^{PSL_N(\mathbb{C})}(C_{g,n})$, 
assuming that the surface is hyperbolic, $\chi(C_{g,n})=2-2g+n$, and with at least one puncture $n\geq 1$. The \emph{traffic 
rules} allowing to determine the parallel transport matrices associated to paths on $C_{g,n}$ are particularly simple for the 
case where $N=2$. The starting point is an ideal triangulation of the Riemann surface, which has the vertices of the 
triangles located at the punctures. These furthermore have a point marked onto each edge, the collection of which forms 
a lattice and to which coordinates are associated. To each such triangulation it is possible to assign a dual graph, like in 
Figure \ref{FG1}, to which corresponds a fat graph. This is composed of three types of segments: \emph{i)} those intersecting 
an edge of the triangulation, \emph{ii)} those intersecting the dual graph and \emph{iii)} those contained inside a face of the 
triangulation but not intersecting any other edges. The prescription described in \cite{FG} associates in the case $N=2$  
the matrices 
\begin{equation}
\mathsf{s}=  \left( \begin{matrix} 0 & 1 \\ -1 & 0\end{matrix} \right)~,  \quad 
\mathsf{v}=   \left( \begin{matrix} 1 & 0 \\ 1 & 1\end{matrix} \right)~, \quad 
\mathsf{e}(x)= \left( \begin{matrix} 1/\sqrt{x} & 0 \\ 0 & \sqrt{x} \end{matrix} \right)~,
\end{equation}
to the three types of segments of a fat graph like in Figure \ref{FG1}, where $x$ is the FG-coordinate associated to the 
edge of the triangulation 
intersected by the segment $\mathsf{e}$. This recipe allows to calculate monodromy matrices on $C_{g,n}$ by picking 
a base point for the corresponding paths on the Riemann surface, projecting a path onto the fat graph, and computing 
the overall monodromy from the building blocks $\mathsf{s},\mathsf{v}$ and $\mathsf{e}$ corresponding to the segments 
of the fat graph onto which the path is projected.

\begin{figure}[h]
\centering
\includegraphics[width=0.60\textwidth]{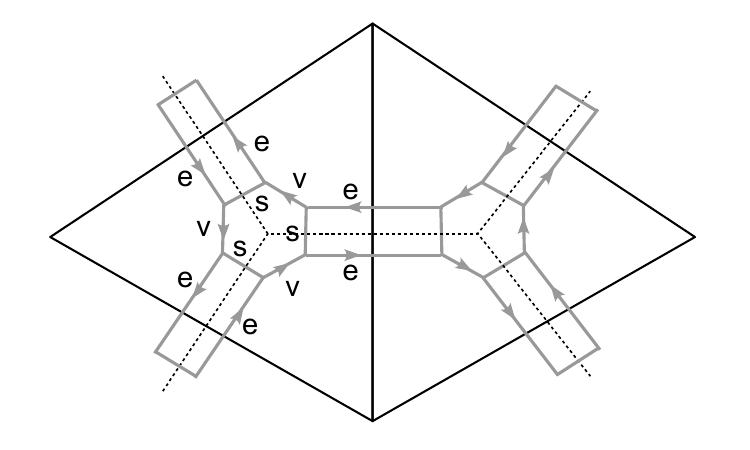}
\caption{\it Two triangles that are part of an ideal triangulation, with edges connecting the punctures. The dual graph is 
drawn as a dashed line, with the associated fat graph, depicted in grey. The outer edges of the fat graph are assigned 
a clockwise orientation around the nearest puncture. There are three types of segments forming a fat graph: \emph{i)} 
intersecting an edge of the triangulation, \emph{ii)} intersecting the dual graph and \emph{iii)} contained inside a face 
of the triangulation.}
\label{FG1}
\end{figure}
%


\section{Borel summation of the asymptotic series for the Gamma function}\label{BorelGamma}
\setcounter{equation}{0}

The Gamma function has the asymptotic expansion
\begin{equation}\label{Gammaexp-gen}
[\log\Gamma(w+s)]_{\rm f}^{}=
\bigg(w+s-\frac{1}{2}\bigg)\log(w)-w+\frac{1}{2}\log(2\pi)+\sum_{n=2}\frac{(-)^{n} B_{n}(s)}{n(n-1)}w^{1-n},
\end{equation}
for $|w|\ra\infty$, $|\mathrm{arg}(w)|<\pi$,
where $B_n(s)$ are the Bernoulli polynomials defined by
\begin{equation} 
\frac{te^{ts}}{e^t-1}=\sum_{n=0}^{\infty}B_n(s)\frac{t^n}{n!}.
\end{equation}
In the case $s=0$ one can use that $B_n\equiv B_n(0)$ is non-vanishing only for even $n$ to 
get a slightly simpler formula,
\begin{equation}\label{Gammaexp}
[\log\Gamma(w)]^{}_{\rm f}=\bigg(w-\frac{1}{2}\bigg)\log(w)-w+\frac{1}{2}\log(2\pi)+\sum_{g=1}\frac{B_{2g}}{2g(2g-1)}w^{1-2g}.
\end{equation}
This expansion is not convergent, but Borel summable. Binet's first formula for the Gamma function can be written as
\begin{equation}\label{Binet}
\log\Ga(w)=\bigg(w-\frac{1}{2}\bigg)\log(w)-w+\frac{1}{2}\log(2\pi)+
\int_0^{\infty}\frac{dt}{t^2}\,e^{-tw}\left(\frac{t}{e^{t}-1}-1+\frac{t}{2}\right),
\end{equation}
for $\mathrm{Re}(w)>0$. It is
easy to see that 
the right hand side of \rf{Binet} 
is the Borel sum of the divergent series \rf{Gammaexp}.

Our main goal in this appendix is to discuss the Borel summation of the formal series \rf{Gammaexp}
in the 
case  $\mathrm{Re}(w)<0$. The first thing that needs to be defined is clearly the factor $\log(w)$ in the first 
term on the right hand side of \rf{Gammaexp}. Three options appear to be fairly natural: 
One could define $\log(w)$ for $\mathrm{Re}(w)<0$ to be equal to one of the three functions
\begin{equation}
\log_\ep(w)= \ep \,\pi\mathrm{i}+\log(-w), \qquad \ep=\pm 1, 0.
\end{equation}
The functions $\log_\ep(w)$, $\ep=\pm 1$, clearly correspond to defining $\log(w)$ for $\mathrm{Re}(w)<0$ by
analytic continuation through the upper or lower complex half-planes, respectively. Choosing 
$\log_0(w)$ appears to be most natural when there is motivation to preserve the reality properties of the 
function $\log(w)$ or if one is mainly interested in real values of $w$.

Having specified a definition of the factor $\log(w)$ in \rf{Gammaexp} for $\mathrm{Re}(w)<0$
one may ask for a Borel summation 
of the resulting formal series. In order to find it, we may conveniently use \rf{Gammaexp-gen}, which 
can be specialised to 
\begin{equation*}
[\log\Gamma(1-w)]^{}_f=
\bigg(\frac{1}{2}-w\bigg)\log(-w)+w+\frac{1}{2}\log(2\pi)+
\sum_{n=2}\frac{(-)^{n} B_{n}(1)}{n(n-1)}(-w)^{1-n},
\end{equation*}
valid for $|w|\ra\infty$, $|\mathrm{arg}(-w)|<\pi$.
Using $B_{n}(1-s)=(-1)^nB_n(s)$ and $B_n(0)=0$ for $n$ odd we find that 
$- \log\Gamma(1-w)$ has a formal series which is almost identical to the 
one for $\log\Gamma(w)$,
which is very similar to the one for 
\begin{equation*}
-[\log\Gamma(1-w)]^{}_f=
\bigg(w-\frac{1}{2}\bigg)\log(-w)-w-\frac{1}{2}\log(2\pi)+
\sum_{g=1}\frac{B_{2g}}{2g(2g-1)}w^{1-2g}.
\end{equation*}
It immediately follows from this observation  that the functions 
$\check{\Ga}_\ep(w)$ defined as 
\begin{equation}
\check{\Ga}_\ep(w)=2\pi\frac{e^{\ep\pi\mathrm{i}(w-\frac{1}{2})}}{\Ga(1-w)},
\end{equation}
have the asymptotic expansions
\begin{equation}\label{Gammaeps-exp}
[\log\check\Gamma_\ep(w)]^{}_{\rm f}=\bigg(w-\frac{1}{2}\bigg)\log_\ep(w)-w+\frac{1}{2}\log(2\pi)+\sum_{g=1}\frac{B_{2g}}{2g(2g-1)}w^{1-2g}.
\end{equation}
for $|w|\ra \infty$, $\mathrm{Re}(w)< 0$.
We may note that
\begin{equation}
\frac{\check{\Ga}_\ep(w)}{\Ga(w)}=
\frac{2\pi\,e^{\ep \pi\mathrm{i}(w-\frac{1}{2})}}{\Ga(1-w)\Ga(w)}= 2
\sin(\pi w)e^{\ep \pi\mathrm{i}(w-\frac{1}{2})}=\left\{
\begin{aligned} 1-e^{\pm 2\pi \mathrm{i} w} \;\;\text{for $\ep=\pm 1$},\\
2\sin(\pi w)\;\;\text{for $\ep=0$}.
\end{aligned}
\right.
\end{equation}
We  conclude the following: The asymptotic series for the Gamma function $\Ga(w)$ is Borel summable 
for $\pm \mathrm{Re}(w)>0$. For $\mathrm{Re}(w)<0$ there is an ambiguity in the definition 
of the leading term of the asymptotic series, in general. The three natural ways to fix this
ambiguity by continuation in the upper or lower half planes, or along the real axis define
three functions $\check{\Ga}_\ep(w)$ representing reasonable definitions for the Borel summation 
of the asymptotic series for the Gamma function in the domain $\mathrm{Re}(w)<0$.
Among the three cases one may regard the choice $\ep=0$ as the most canonical. It allows one 
to define the Borel summation for the whole half-plane $\mathrm{Re}(w)<0$.

\section{Pure SU(2) SYM and the topological vertex}\label{TopVertex}
\setcounter{equation}{0}
The topological strings partition function for $\cN=1$ pure SU(2) SYM in 5d on a circle $S^1_R$ of radius $R$ can 
be computed using the topological vertex \cite{AKMV}. It is thus a function of the K\"ahler parameters 
$\{Q_B,\ Q_F\}$ of the toric CY3 manifold mirror dual to the local CY3 which enters the geometric engineering realisation. 
It is also a function of the string coupling $\lambda= - \mathrm{i} \hbar$ through the 
parameter $q=e^{-R\lambda}$. The procedure 
to compute topological string partition functions from the toric diagram of the CY3, like that depicted in Figure \ref{fig:branes} 
for the pure SU(2) theory, follows a set of rules. The precise 
conventions that we use 
have been summarised in 
\cite{CPT}. Associating \emph{edge} and \emph{vertex} functions of the K\"ahler parameters $Q_e$ and Young diagrams 
$\mu_e$ to the edges and vertices of the toric diagram, the topological string partition function is the product of these functions 
summed over all possible Young diagrams associated to the internal edges $e$
\begin{equation}
\mathcal{Z}^{\text{top}}=\sum_{\{ \mu_e\} } \  
\prod_{\text{edges}}\text{edge factor}\times \prod_{\text{vertices}}\text{vertex factor} .
\end{equation}

\begin{figure}[h!]
   \centering
\includegraphics[scale=0.6]{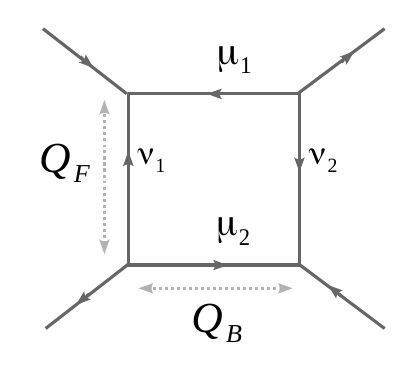}
\caption{ \it The toric diagram associated to the $\cN=1$, pure $SU(2)$ SYM in 5d on $S^1$.}
\label{fig:branes}
\end{figure} 

For pure SU(2) SYM, the topological vertex yields the topological string partition function
\begin{equation} 
\label{eq:boxpartitionfunction3ppp}
\mathcal{Z}_{\text{box}}^{\text{top}} = \mathcal{M}(Q_F)^2 \sum_{\boldsymbol{\mu}} \prod_{i=1}^2  
(2^{-4} Q_B/Q_F)^{|\mu_i|} \frac{1}{N^R_{\mu_i\mu_i}(1)N^R_{\mu_1\mu_2}(Q_F)N^R_{\mu_2\mu_1}(Q_F^{-1})} ~,
\end{equation}
where the functions $N^R_{\mu_i\mu_j}(e^{-Ra}) $ are defined by 
$N^R_{\mu\nu}(e^{-Ra}) = N^R_{\mu\nu}(e^{-Ra};\epsilon, -\epsilon)$ with 
\begin{align}\label{eqdef:Nek2}
N^R_{\mu\nu}(a;\epsilon_1,\epsilon_2)
&=\prod_{(i,j)\in \mu}2\sinh\frac{R}{2}\left[a+\epsilon_1(\mu_i-j+1)+\epsilon_2(i-\nu^t_j)\right]\nonumber\\
&\times \prod_{(i,j)\in \nu}2\sinh\frac{R}{2}\left[a+\epsilon_1(j-\nu_i)+\epsilon_2(\mu^t_j-i+1)\right]~
\end{align}
and 
$Q_F=q^{2\sigma}$. The function $\mathcal{M}(q^u)$ is related to the q-Barnes function $G_q(u)$ defined as \cite{UN}
\begin{equation}
G_q(u+1)=\frac{  (q;q)_\infty^{u} }{ (q;q,q)_{\infty}   \mathcal{M}(q^u) }  (1-q)^{-\frac{1}{2}u(u-1)}~,
\end{equation}
where the shifted factorial is
\begin{equation}
\label{eq:shiftedfactorials}
(x;q_1,\dots,q_r)_{\infty} = \prod_{n_1,\dots,n_r=0}^{\infty} 
\left( 1-x q_1^{n_1}\cdots q_r^{n_r} \right) \, , \quad \forall  \quad   |q_i|<1 ~.
\end{equation}

\subsection*{The 4d limit}

The 4d limit $R\to 0$ where $S^1_R$ vanishes is under control for $N^R_{\mu\nu}(e^{-Ru})$ \eqref{eqdef:Nek2} \cite{PM}
\begin{equation}
\lim_{R\to 0} N^R_{\mu\nu} (e^{-Ru}) = R^{|\mu|+|\nu|} N_{\mu\nu} (u) ~,
\end{equation}
where 
\begin{equation}\label{eqdef:4dNek}
N_{\mu\nu} (u;\epsilon_1,\epsilon_2) = \prod_{(i,j)\in\mu} (u +\epsilon_1(\mu_i-j+1) + \epsilon_2 (i-\nu_j^t) ) 
\prod_{(i,j)\in\nu} (u+ \epsilon_1 (j-\nu_i) + \epsilon_2 (\mu_j^t-i+1) ) ~.
\end{equation}
So the sum over 
$\{\boldsymbol{\mu}\}$ in equation \eqref{eq:boxpartitionfunction3ppp}  matches the Nekrasov instanton sum in 4d, 
with the factor $(2^{-4} Q_B/Q_F)$ replaced by $\Lambda^4$ in the limit $R\to 0$. 
Looking at the factor $\mathcal{M}(Q_F)^2$, through the relation of $\mathcal{M}(q^u)$ to the q-Barnes function 
$G_q(u+1)$, it is clear that the former diverges in the limit $q\to 1$. The latter however has a well-defined limit 
$\lim_{q\to 1}G_q(u)=G(u)$ giving the Barnes function $G(u)$ \cite{UN}. Therefore in the 4d limit, the regular part of 
$\mathcal{M}(Q_F)^2$ in equation \eqref{eq:boxpartitionfunction3ppp} is  
\begin{equation}
\mathcal{M}(Q_F)^2 ~~ \stackrel{\longrightarrow}{_{R\to 0}} ~~ 
\frac{1}{(G(1+2\sigma))^2} ~.
\end{equation}
Plugging 
$\mathcal{N}^{(\eta')}(\sigma)\equiv (G(1+2\sigma))^{-2}$ into equation \eqref{eta-CN-remind} 
thus gives the relative factor 
\begin{equation}
\frac{\mathcal{N}^{(\eta')}(\sigma-1)}{\mathcal{N}^{(\eta')}(\sigma)} = \left( \Gamma(2\sigma) \Gamma(2\sigma-1) \right)^2
\end{equation}
between the parameters $\eta'$ and $\eta_0$ in the series expansions of the Painlev\'e III tau-function. Let us finally note 
that the $R\to 0$ limit of the topological strings partition function $\lim_{R\to 0} \mathcal{Z}_{\text{box}}^{\text{top}}$ is related 
to the factor appearing in the series expansions of the Painlev\'e III tau-function by a factor $\Lambda^{4\si^2}$.

\section{A different choice of normalisation in symmetric chambers}\label{app:different-norm}
\setcounter{equation}{0}

In section \ref{Change-FN} we discussed how to extend the  given 
definition for the normalised tau-functions across the loci where the Stokes graphs change topological type, focussing on the case in which the Stokes graph on $C_2$ changes from type $\CS_2$ to $\CS_s$. 
In this appendix we explore a different choice of normalisation, and comment on its relation to the choice adopted in the main body of the paper.

\subsection{} 

Let us assume that the Stokes graph
on $C_2$ undergoes a transition from type $\CS_2$ to type $\CS_s$, as in Figure \ref{fig:transition}. There are now two 
Stokes regions $\CR$, $\CR'$ around the puncture at $x=0$ created by the factorisation of $C_{0,4}$.
One of them is simply the continuation of the Stokes region surrounding $x=0$ through 
the transition from $\CS_2$ to $\CS_s$, the other one is created in this transition, see Figure~\ref{fig:transition}.


The issue that we wish to address is the fact that one now encounters an ambiguity in the choice of the branch point 
representing the starting point of the contour of integration that defines the Voros symbol fixing the normalisation of the tau function. 
To fix this ambiguity we may consider the 
following rule: let us consider the two 
edges of the {\it Anti-}Stokes graph emanating from the branch points of the Stokes region $\CR'$
created in the  transition from $\CS_2$ to $\CS_s$. Truncating these edges when they leave the 
Stokes region of interest for the first time will generically define a decomposition of the Stokes region $\CR'$
into three connected components, see Figure {\it Anti-Stokes}.
For each puncture $p$ in the boundary of $\CR'$ there is a unique 
branch point $b(p)$ such that $p$ and $b(p)$ are contained in  the
boundary of the same connected component defined by the Anti-Stokes graph. 
We will use the branch point $b(p)$ associated to the 
puncture $p$ at $x=0$ in this way as the initial point of the contour defining the
Voros symbol. 

\begin{figure}[h!]
\begin{center}
\includegraphics[width=0.4\textwidth]{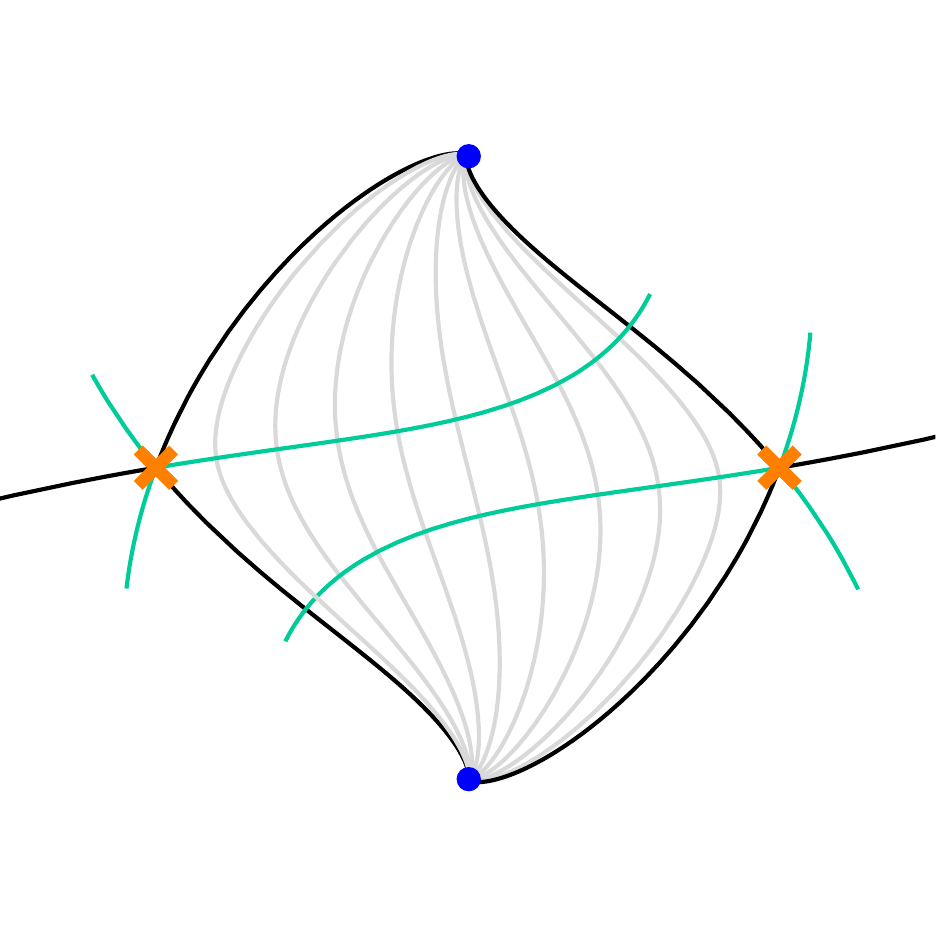}
\caption{Decomposition of a Stokes region induced by the anti-Stokes graph, shown in green.}
\label{fig:anti-stokes}
\end{center}
\end{figure}

\subsection{}

We are next going to note that the coordinates defined by making these choices are related in the 
transition from $\CS_2$ to $\CS_s$
in a particularly natural way. To this aim let us begin by observing that the location of  the 
branch point representing the initial point of the contour defining the
Voros symbol in the case of the Stokes graph $\CS_2$ can be analytically 
continued through the transition from $\CS_2$ to $\CS_s$ as long as one avoids the 
singularity where the two branch points located on $C_2$ collide. According to the classification 
of Stokes graphs in \cite{AT} one will find the transition from $\CS_2$ to $\CS_s$ when 
the real part of the parameter $t:=C-B=\theta_3-\theta_4-\si$ becomes negative. 
The Stokes graph 
$\CS_s$ has a pair of Stokes lines connecting the puncture at $x=0$ to two branch points $b_\pm$.
Depending on the sign of $\mathrm{Im}(t)$ one will find that either $b_+$ or $b_-$ represent
the analytic continuation of the branch point connected to $x=0$ by a Stokes line on $\CS_2$.
It is possible to show that the branch point defined by the analytic continuation 
will coincide with the branch point 
associated to $x=0$ with the help of the Anti-Stokes graph by the prescription
above, as also illustrated by Figure \ref{fig:transition}.

In a similar way one may 
analytically continue the Voros symbol $V^{(b_20)}$ 
determining the formal WKB expansion $\mathsf{n}^{(2)}\simeq e^{-2V^{(b_20)}}$ of the normalisation factors 
$\mathsf{n}^{(2)}$
through the transition 
from $\CS_2$ to $\CS_s$. 
The 
result  will depend on the path  of  analytic continuation. 
It may be represented  as one of the Voros symbols 
$V^{(b_+0)}$ or $V^{(b_-0)}$
defined by integrals over contours starting either at $b_+$ or
at $b_-$, depending on the choice of path defining the analytic continuation, respectively. 
It is possible to continue the Voros symbol $V^{(b_20)}$ regarded as a function of $t$ along a 
path going around $t=0$ in the upper half plane and returning to the region $\mathrm{Re}(t)>0$ through the lower 
half plane. The result of this analytic continuation  
will differ from the Voros symbol $V^{(b_20)}$ by the Voros symbol
associated to the contour connecting the two branch points on $C_2$. It can no longer be
represented as a Voros symbol associated to a path located in a single Stokes region. 

\subsection{}

In order to define an unambiguous assignment of coordinates to the regions in the space of
quantum curves characterised by the type of Stokes graphs considered here it then seems 
natural to decompose the region in which the Stokes graph on $C_2$ is of type $\CS_s$ into 
two subregions distinguished by the branch point associated to $x=0$ with the help of the rule
defined using the Anti-Stokes graph above, see Figure \ref{fig:sym-chambers}. By applying the classification of 
\cite{AT} to Anti-Stokes graphs instead of Stokes graphs, which is equivalent to exchanging 
$\mathrm{Re}(t)$ with $\mathrm{Im}(t)$, one may easily show that the resulting 
decomposition of the considered region is equivalent to the decomposition defined by
the sign of $t$.
The Voros symbol defined in this way has a jump when there
is a change of topological type of the Anti-Stokes graph. This happens when 
$\mathrm{Im}(t)=0$, in which case the Anti-Stokes graph has a saddle trajectory
connecting the two branch points $b_+$ and $b_-$ considered above. 

The formula for this jump was given in (\ref{Vorosjump}).
The region characterised by $\mathrm{Re}(t)<0$ 
where the Stokes graph has type $\CS_s$
decomposes in two subregions distinguished by the topological type of 
Anti-Stokes graph, depending on the sign of $\mathrm{Im}(t)$. 
The two choices of sign in \rf{Vorosjump} are attached to these two subregions, respectively. 
As noted in (\ref{eq:CS-s-internal-jump}),
the relative change in the Voros symbol between the two sub-regions of $\CS_s$, is due to the change of the choice of 
contour defining the Voros symbols, and corresponds to multiplication by the Voros symbol associated to the shortest closed cycle running between the two branch points. 
From the viewpoint of the present choice of normalisation, this change is induced by a change in topology of the anti-Stokes graph.

In comparing to the prescription of Section \ref{Change-FN}, the present one based on anti-Stokes graphs induces a more refined chamber structure, dividing $\CS_s$ into two sub-regions with different normalizations. In \ref{Change-FN} we opted instead for taking the geometric average of these two, extending the definition to the whole region $\CS_s$, this is consistent because such a choice is invariant under the change  (\ref{eq:CS-s-internal-jump}) induced by the jump in the anti-Stokes graph.



\section{Free fermions twisted by flat bundles}\label{FF-CFT}
\setcounter{equation}{0}

{This appendix outlines how CFT techniques allow one to represent the tau-functions as 
free fermion conformal blocks twisted by flat bundles.} 

\subsection{Flat bundles}\label{flatbd}

We are going to define conformal blocks for a system of free fermions on a Riemann surface $C$ 
twisted by a flat bundle. To begin with, we shall consider a Riemann surface $C=C_{g,n}$ with 
genus $g$ and $n$ punctures. 
Recall that the structure of a flat bundle can be 
described by a sufficiently fine cover $\{\CU_\al;\al\in \CA\}$ of $C$ and a vector bundle $E$ with 
a preferred system of trivialisations such that the transition functions $g_{\al\be}:\CU_\al\cap\CU_{\be}\ra 
\mathrm{GL}(N,\BC)$ are constant. 
Flat bundles differing by changes of local 
trivialisations represented by constant functions 
$h_\al:\CU_{\al}\ra\mathrm{GL}_n(\BC)$ are 
identified with each other.
Flat bundles are in a canonical one-to-one correspondence with representations
$\rho:\pi_1(C)\ra \mathrm{GL}(N,\BC)$. We will use this correspondence to label flat bundles 
by the corresponding representations $\rho$ of $\pi_1(C)$. 

These data define a 
Riemann-Hilbert problem. 
The solution to the Riemann-Hilbert problem defined by a flat bundle on $C$
is a pair $(\CE,\nabla)$ of a holomorphic bundle $\CE$ and a holomorphic connection $\nabla$. 
One may equivalently consider the row vector $\Psi=(\psi_1,\dots,\psi_N)$ formed out of $N$
flat sections of $\nabla$ forming a frame. In a local trivialization of $\CE$ one may represent $\Psi$ by a matrix
$\Psi(x)$ of holomorphic functions such that the analytic continuation $\Psi(\ga.x)$ of $\Psi(x)$
along a closed curve $\ga$ starting and ending at $x$ satisfies $\Psi(\ga.x)=\Psi(x)\rho(\ga)$.
The solution to this Riemann-Hilbert problem becomes unique after imposing additional 
conditions on the behaviour of $\Psi(x)$ at the $n$ punctures.

We will mostly consider the case $N=2$ in this paper. In this case one can always 
solve the Riemann-Hilbert problem by meromorphic opers, pairs $(\CE_{\rm op},\nabla_{\rm op})$,
where $\CE_{\rm op}$ is the unique (up to isomorphism) extension
\begin{equation}\label{opbundle}
0 \ra K^{\frac{1}{2}}\ra \CE_{\rm op}\ra K^{-\frac{1}{2}}\ra 0,
\end{equation}
and $\nabla_{\rm op}$ can be represented with respect to the frame associated to the representation as
\begin{equation}
\nabla_{\rm op}=\pa+\left(\begin{matrix} 0 & -t \\ 1 & 0\end{matrix}\right)dx.
\end{equation}
The matrix element $t$ is allowed to have apparent singularities. Generically one needs a minimum of 
$3g-3+n$ apparent singularities in order to solve the Riemann-Hilbert problem defined on 
a Riemann surface of genus $g$ and $n$ punctures at which $t$ has poles of second order. 
The matrix $\Psi$ formed out of the flat sections of $\nabla_{\rm op}$ takes the form 
$\Psi=\left(\begin{smallmatrix} \eta_1' & \eta_2' \\ \eta_1 & \eta_2\end{smallmatrix}\right)$,
where $\eta_1$, $\eta_2$ are two linearly independent solutions of $(\pa_x^2+t(x))\eta=0$.

\subsection{Free fermions}

The free fermion super VOA is generated by fields $\psi_s(z)$, $\bar{\psi}_s(z)$, $s=1,\dots,N$,
The fields $\psi_s(z)$ will be arranged into a row vector $\psi(z)=(\psi_1(z),\dots,\psi_N(z))$, 
while $\bar\psi(z)$ will be our notation for the column vector with components $\bar{\psi}_s(z)$. 
The modes of $\psi(z)$ and $\bar\psi(z)$, introduced as
\begin{equation}
\psi(z)=\sum_{n\in\BZ}\psi_{n} z^{-n-1}\,,\quad
\bar\psi(z)=\sum_{n\in\BZ} \bar\psi_{n} z^{-n}\,,
\end{equation}
are row and column vectors 
with components $\psi_{s,n}$ and $\bar{\psi}_{s,n}$, respectively,  satisfying 
\begin{equation}
\{\,\psi_{s,n}\,,\,\bar\psi_{t,m}\,\}=\de_{s,t}\de_{n,-m}\,,\qquad
\{\,\psi_{s,n}\,,\,\psi_{t,m}\,\}=0\,,\qquad\{\,\bar\psi_{s,n}\,,\,\bar\psi_{t,m}\,\}=0\,.
\end{equation}
The Fock space $\CF$ is a representation generated from a highest weight vector $\mathfrak{f}_{0}^{}$ 
satisfying  
\begin{equation}
\psi_{s,n}\cdot\mathfrak{f}_{0}^{}=0\,,\quad n\geq 0\,,\qquad
\bar\psi_{s,n}\cdot\mathfrak{f}_{0}^{}=0\,,\quad n>0\,.
\end{equation}
$\CF$ is generated from $\mathfrak{f}_{0}^{}$ by the action of the modes $\psi_{s,n}$, $n<0$, 
and $\bar\psi_{s,m}$, $m\leq 0$. 

We will also consider the conjugate representation $\CF^*$, a {\it right} module generated 
from a highest weight vector $\mathfrak{f}_{0}^{*}$ 
satisfying  
\begin{equation}
\mathfrak{f}_{0}^{*}\cdot\psi_{s,n}=0\,,\quad n< 0\,,\qquad
\mathfrak{f}_{0}^{*}\cdot\bar\psi_{s,n}\,=0\,,\quad n\leq 0\,.
\end{equation}
The Fock space $\CF^*$ is generated from $\mathfrak{f}_{0}^{*}$ by the right action of the modes $\psi_{s,n}$, $n\geq 0$, 
and $\bar\psi_{s,m}$, $m> 0$.  A natural bilinear form $\CF^*\otimes\CF\ra \BC$ is defined
by the expectation value, 
\begin{equation}
\langle \,\mathfrak{f}_0^*\cdot\mathsf{O}_{\mathfrak{f}^*}\,,\,\mathsf{O}_{\mathfrak{f}}\cdot\mathfrak{f}_0\,\rangle= \Omega
(\mathsf{O}_{\mathfrak{f}^*}\mathsf{O}_{\mathfrak{f}}\cdot \mathfrak{f}_0),
\end{equation}
where $\Omega(\mathfrak{f})=c$ if $\mathfrak{f}= c\,\mathfrak{f}_{0}^{}+\sum_{s=1}^N(\sum_{n<0} \psi_{s,n}\mathfrak{f}_{s,n} +
\sum_{m\leq 0}\bar{\psi}_{s,m}\mathfrak{f}_{s,m})$.

\subsection{Free fermion states from the Riemann-Hilbert correspondence}\label{FFfromD}

We are now going to outline how to modify the formalism used in \cite{CPT} to the higher genus 
situation, generalizing the approach of \cite{AGMV} appropriate for the case  $N=1$.

A simple and natural way to characterise a state $\mathfrak{f}\equiv \mathfrak{f}_G\in\CF$ is through the 
matrix $G(x,y)\equiv G_\mathfrak{f}(x,y)$ of two-point  functions  having matrix elements
\begin{equation}\label{FFtwopt}
G_\mathfrak{f}(x,y)_{st}=\big\langle \,\bar{\psi}_s(x)\psi_t(y) \,\big\rangle_{\mathfrak{f}} \equiv
\frac{\langle \,\mathfrak{f}_{0}^{\ast}\,,\,\bar{\psi}_s(x)\psi_t(y)\,\mathfrak{f}\,\rangle}{\langle \,
\mathfrak{f}_{0}^{\ast}\,,\,\mathfrak{f}\,\rangle}.
\end{equation}
Indeed, given a function $G(x,y)$ such that 
\begin{equation}\label{A-exp}
G(x,y)=\frac{{1}}{x-y}+A(x,y)\,,
\qquad
A(x,y)=\sum_{l\geq 0} y^{-l-1}\sum_{k>0}x^{-k}A_{kl},
\end{equation}
there exists a state $\mathfrak{f}_G$ such that
its two-point function $G_{\mathfrak{f}_G}$ is given by $G(x,y)$. 
States $\mathfrak{f}_G^{}$ having this property can be constructed as
\begin{equation}\label{CBfromR}
\mathfrak{f}_G^{} 
=N_G^{}\exp\bigg(\!-\sum_{k>0}\sum_{l\geq 0} \psi_{-k}\cdot A_{kl}\cdot \bar{\psi}_{-l}\bigg)\, \mathfrak{f}_{0}^{}\,,
\end{equation}
with  $N_G\in\BC$ being a normalisation constant.

We are interested in two-point functions $G(x,y)$ having a multi-valued analytic continuation 
to the Riemann surfaces $C$
with given monodromies with respect to both $x$ and $y$. The monodromies describing the analytic
continuation in $x$ are required to act on $G(x,y)$ from the left, while the analytic continuation in $y$ generates
monodromies acting from the right,
\begin{equation}
G(x,\ga_r.y)=G(x,y)\cdot M_r,\qquad G(\ga_r.x,y)=M_r^{-1}\cdot G(x,y). 
\end{equation}
This characterizes $G(x,y)$ as a solution to 
a variant of the Riemann-Hilbert problem formulated above where one allows a first order pole at 
$y=x$. 

In order to construct solutions to this problem, let 
us choose a reference connection $\nabla_0$ on the 
holomorphic bundle $\CE$ furnished by the solution to the Riemann-Hilbert problem
discussed in Section \ref{flatbd}.
Choosing flat sections $\Psi$ and $\Psi_0$ of $\nabla$ and $\nabla_0$, respectively, 
one can form the combination
$\hat{\Psi}(x)=(\Psi_0(x))^{-1}\Psi(x)$.
$G(x,y)$ can then be constructed in the following 
form
\begin{equation}\label{GfromPsi}
G_{\Psi}(x,y)=(\hat{\Psi}(x))^{-1}S(x,y)\hat{\Psi}(y),
\end{equation}
with $S(x,y)$ being the Szeg\"o-kernel \cite{Fay}\footnote{The precise definition depends 
on the choice of the spin structure needed for defining the fermions on $C$.}. One may, in particular, choose 
$\CE=\CE_{\rm op}$ as above, relating the choice of $\nabla_0$ to the choice of a reference oper
defining a projective structure on $C$.

The
construction of the fermionic states $\mathfrak{f}_{G}$ described above therefore 
gives us a natural way to assign  fermionic states $\mathfrak{f}_\Psi^{}\equiv \mathfrak{f}_{G_\Psi}^{}$ to 
solutions $\Psi$ of the Riemann-Hilbert problem.

\subsection{Conformal blocks}

We may associate a free fermion conformal block to a Riemann surface $C$ with a flat bundle as follows.
Let $P$ be a point on $C$, and $\CU_P$ be an open neighbourhood of $P$ which is part of the cover 
defining the given flat bundle. Let $x$ be a local coordinate in $\CU_P$ vanishing at $P$, and 
let $\Psi_P(x)$ be the function of a complex variable $x$ 
representing the solution to the Riemann-Hilbert problem on $\CU_P$.
From these data one may then define states $\mathfrak{f}_\Psi^{}\in\CF$ and 
$\mathfrak{f}_\Psi^\ast\in\CF^\ast$ characterised by infinite sets of linear equations of 
the form 
\begin{equation}\label{FFWI-0}
\psi[g]\cdot\mathfrak{f}_\Psi^{}=0,\qquad
\bar{\psi}[\bar{f}]\cdot\mathfrak{f}_\Psi^{}=0,\qquad
\mathfrak{f}_\Psi^\ast\cdot\psi[g]=0,\qquad
\mathfrak{f}_\Psi^\ast\cdot\bar{\psi}[\bar{f}]=0,
\end{equation}
where 
\begin{equation}
\psi[{g}]\,=\,\frac{1}{2\pi\mathrm{i}}\int_{\CC} dx\; \psi(x)\cdot {g}(x)\,,\qquad
\bar{\psi}[\bar{f}\,]\,=\,\frac{1}{2\pi\mathrm{i}}\int_{\CC} dx\; \bar{f}(x)\cdot \bar{\psi}(x)\,,
\end{equation}
with $\CC\subset \CU_P$ being a sufficiently small circle around $P$, and $g\in{{W}}_{P}(\Psi)$, 
$\bar{f}\in\bar{W}_{P}(\Psi)$, with 
\begin{equation}
\begin{aligned}
&g\in {{W}}_{P}(\Psi)=\Big\{\,\hat{\Psi}^{-1}(x)\cdot v(x);\;\,{v}(x)(dx)^{\frac{1}{2}}\in \BC^N\otimes 
H^0\big(C\!\setminus\!\{P\},K^{\frac{1}{2}}\big)
\,\Big\}\,,\\
&\bar{f}\in\bar{W}_{P}(\Psi)=\Big\{\,\bar{v}(x)^t\cdot\hat{\Psi}(x);\;\,\bar{v}(x)(dx)^{\frac{1}{2}}
\in  \BC^N\otimes 
H^0\big(C\!\setminus\!\{P\},K^{\frac{1}{2}}\big)\,\Big\}\,.
\end{aligned}
\end{equation}
The linear equations \rf{FFWI-0}, in the following referred to as free fermion Ward identitites, determine 
$\mathfrak{f}_\Psi^{}$ and $\mathfrak{f}_\Psi^\ast$ up to multiplicative constants. The normalisation 
is determined by fixing the corresponding partition functions
$Z_C(\Psi)=\langle \mathfrak{f}_0^\ast,\mathfrak{f}_\Psi^{}\rangle$ and 
$Z_C^\ast(\Psi)=\langle \mathfrak{f}_\Psi^\ast,\mathfrak{f}_0^{}\rangle$. Normalising 
 $\mathfrak{f}_\Psi^{}$ and $\mathfrak{f}_\Psi^\ast$ such that $Z_C^{}(\Psi)=Z_C^{\ast}(\Psi)$,
 one sees that $\mathfrak{f}_\Psi^{}$ and $\mathfrak{f}_\Psi^\ast$ are two different ways to package
 the same information.

There is a generalisation of this construction assigning vectors $\mathfrak{f}_{\Psi,n}\in\CF^{\ot n}$ to 
$n$-tupels of points $P_1,\dots,P_n$. However, in the same way as in the case $n=1$ considered before 
one finds that the vectors $\mathfrak{f}_{\Psi,n}$ are uniquely characterised by the multi-point analogs of 
\rf{FFWI-0} together with the corresponding 
partition function $Z_C(\Psi)=\langle \mathfrak{f}_0^{\otimes n},\mathfrak{f}_{\Psi,n}^{}\rangle$, a function 
which can be chosen to be $n$-independent. 
This means that the information defining a conformal block, in our case represented
by $\Psi$ and $Z_C(\Psi)$, can be ``localised'' at arbitrary $n$-tuples of points. This is a special case of a phenomenon
often referred to as ``propagation of vacua'' in the CFT literature.

Infinitesimal reparameterisations of the coordinate $x$  are generated by the 
free fermion energy-momentum tensor $T(x)$ constructed in the standard way as a bilinear
expression in the fermions.
Conformal blocks of the free fermion algebra are automatically conformal blocks for the 
Virasoro algebra. This means that the identities
\begin{equation}\label{Virblock}
T[v]\,\mathfrak{f}_{\Psi}=0,\qquad T[v]=\int_{\CC} dx \,v(x)T(x),
\end{equation}
hold for all vector fields $v=v(x)\pa_x$ on $\CU_P\!\setminus\!\{P\}$ that extend holomorphically to  the rest of $C$,
and vanish at the punctures.

There is a natural connection on the line bundle of free fermion conformal blocks over the moduli 
space $\CP_{g,n}$ of triples $(C,P,x)$. 
In order to define it, let us note that the tangent space $T\CP_{g,n}$ can be represented  
as the space of equivalence classes $[v]$ of 
vector fields $v$ on the annulus modulo vector fields that extend holomorphically
to the rest of $C$. 
The connection on the line-bundle of conformal blocks over $\CP_{g,n}$ is then defined by
having flat sections $\mathfrak{f}_\Psi$ satisfying
\begin{equation}
\de_{[v]}^{}\mathfrak{f}_{\Psi}=T[v]\,\mathfrak{f}_{\Psi},
\end{equation}
where $v$ is a vector field on the annulus representing an element of $\CP_{g,n}$. The partition functions $Z_C(\Psi)=\langle\mathfrak{f}_0^\ast, \mathfrak{f}_{\Psi}^{}\rangle$ 
associated to $\mathfrak{f}_{\Psi}$
will then
be a section
of the corresponding line bundle over $\CM_{g,n}$ which is flat with respect to the connection
defined by 
\begin{equation}\label{FSconn}
\de_{[\![v]\!]}^{}Z_{C}(\Psi)=t[v]\,Z_C(\Psi), \qquad
t[v]:=\frac{\langle\mathfrak{f}_0^\ast, T[v]\mathfrak{f}_{\Psi}^{}\rangle}
{\langle\mathfrak{f}_0^\ast, \mathfrak{f}_{\Psi}^{}\rangle},
\end{equation}
where $[\![v]\!]$ is an equivalence class in the double quotient defined by 
considering vector fields $v$ on the annulus modulo vector fields that extend holomorphically
to the rest of $C$ on the one hand, and modulo vector fields that extend holomorphically to 
the disc $D_P$, on the other hand. This double quotient is according to the Virasoro uniformisation 
theorem isomorphic
to the Teichm\"uller space $T\CM_{g,n}$, see \cite[Section 17]{FB} or \cite[Section 2.4]{Du} for 
useful reviews.

An observation going back to \cite{Mo} identifies $t[v]$ as the Hamiltonian generating the 
isomonodromic deformation of the connection $\nabla=\pa+A(x)dx$ induced by a
variation of the complex structure of $C$ in the direction $[\![v]\!]$. One thereby identifies
the flat section $Z_C(\Psi)$ of the connection defined by \rf{FSconn} with the isomonodromic
tau-function. 

The connection defined in \rf{FSconn} is not flat, but only projectively flat. 
A concise discussion of this issue can be found in \cite[Section 1.8]{T17}. The curvature can 
be trivialised locally. The data one needs to pick in order to achieve this are equivalent to 
a family of reference opers varying holomorphically over a given subset of 
the moduli space $\CM_{g,n}$ of complex structures of $C_{g,n}$.

\subsection{Flat bundles with regular singularities}

We will say that
the solution $\Psi$ of the Riemann-Hilbert problem has regular singular behaviour at  $P\in C$
if the preferred system of local trivialisations can be chosen such that the corresponding 
solution $\Psi$ of the Riemann-Hilbert problem is on $\CA_P\equiv \CU_P\setminus\{P\}$ represented
by a function $\Psi_P(x)$ having the form
$\Psi(x)=\Phi(x)x^{\sd}$, with $\Phi(x)$ regular at $x=0$, 
and $\sd=\mathrm{diag}(\si_1,\dots,\si_N)$. As explained in \cite{CPT} one may represent the conformal block
in terms of a vector 
$\mathfrak{f}_{\Psi,\si}^\ast$ in a twisted representation $\CF^\ast_\si$ of the free fermion algebra 
in which the fermions $\psi(x)$, $\bar{\psi}(x)$ have diagonal monodromy around $x=0$.
The vector
$\mathfrak{f}_{\Psi,\si}^\ast$
is characterised by a system of  equations very similar to \rf{FFWI-0}.
The case where $\Psi$ is regular at $P$ is  the special case $\sd=0$.

When $\Psi$ has two regular singular points $P_i$, $i=1,2$,
corresponding coordinates $x_i$, and choices of frames on $D_i$ such that 
$\Psi(x_i)=\Phi(x_i)x^{\sd_i}$, 
one can also define an 
operator $Y_{\Psi}:\CF_{\si_1}\ra\CF_{\si_{2}}$ satisfying
\begin{equation}\label{FFWI}
\psi_2[g]\, Y_{\Psi}=Y_{\Psi} \, \psi_1[g], \qquad
\bar{\psi}_2[\bar{f}]\,Y_{\Psi}=Y_{\Psi} \, \bar{\psi}_1[\bar{f}].
\end{equation}
where 
\begin{equation}
\psi_i[{g}]\,=\,\frac{1}{2\pi\mathrm{i}}\int_{\CC_i} dx\; \psi(x)\cdot {g}(x)\,,\qquad
\bar{\psi}_i[\bar{f}\,]\,=\,\frac{1}{2\pi\mathrm{i}}\int_{\CC_i} dx\; \bar{f}(x)\cdot \bar{\psi}(x)\,,
\end{equation}
for $i=1,2$, with $\CC_i$ being a small circles around $P_i$, for $i=1,2$, respectively, and
\begin{equation}
\begin{aligned}
&g\in {{W}}_{P_2P_1}(\Psi)=\Big\{\,\hat{\Psi}^{-1}(x)\cdot v(x);\;\,{v}(x)(dx)^{\frac{1}{2}}\in \BC^N\otimes 
H^0\big(C\!\setminus\!\{P_2,P_1\},K^{\frac{1}{2}}\big)
\,\Big\}\,,\\
&\bar{f}\in\bar{W}_{P_2P_1}(\Psi)=\Big\{\,\bar{v}(x)^t\cdot\hat{\Psi}(x);\;\,\bar{v}(x)(dx)^{\frac{1}{2}}
\in  \BC^N\otimes 
H^0\big(C\!\setminus\!\{P_2,P_1\},K^{\frac{1}{2}}\big)\,\Big\}\,.
\end{aligned}
\end{equation}
Acting with the operator $Y_{\Psi}$ on $\mathfrak{f}_{\si_1}$ defines a vector 
$\mathfrak{f}_{\si_2}\in\CF_{\si_2}^{}$. This vector  satisfies the same system 
of linear equations as $\mathfrak{f}_{\Psi,\si_2}^{}$. As the solution to this system of equations
is unique up to normalisation, we can normalise $Y_{\Psi}$ such that 
we have $Y_{\Psi}^{}\mathfrak{f}_{\si_1}^{}=\mathfrak{f}_{\Psi,\si_2}^{}$.
The operator $Y_{\Psi}$ may in this sense be regarded as a repackaging
of the information contained in $\mathfrak{f}_{\Psi,\si_2}^{}$. For $\si_1=\si_2=0$
one recovers $\mathfrak{f}_{\Psi}^{}$ and gets another manifestation
of the propagation of vacua phenomenon.

The natural generalisation of \rf{FSconn} represents
variations of  $(C,P_2,x_2,P_1,x_1)$  in the form 
\begin{equation}\label{FSconn-2}
\de_{[v]}Y_{\Psi}=T_2[v_2]\,Y_{\Psi}-Y_{\Psi}\,T_1[v_1],
\end{equation}
with $(v_2,v_1)$ being a pair of vector fields on $\CU_{P_i}\!\setminus \!\{P_i\}$
representing the variation $[v]$ of $(C,P_2,x_2,P_1,x_1)$ by a  generalisation of 
the Virasoro uniformisation theorem.

One may note that the fermionic Ward identities defining the states $\mathfrak{f}_{\Psi}^{}$, 
$\mathfrak{f}_{\Psi}^{\ast}$ do not depend on the type of singular behavior imposed at the 
$n$ punctures of $C_{g,n}$. Given that irregular singularities can be defined by
suitable collision limits from collections of regular singularities \cite{GT} one may therefore 
use the same formalism in order to define conformal blocks for
free fermion systems twisted by flat bundles having irregular singularities 
at some points. 

\subsection{Gluing}

There are well-known operations on the conformal blocks associated to the geometric operations 
of cutting a Riemann surface $C$ along a simple closed curve $\ga$, and gluing of bordered Riemann surfaces 
by identifying annuli around boundary components. 

\subsubsection{Gluing of Riemann surfaces}

Let $C_i$, $i=1,2$, be two Riemann surfaces with punctures $P_i$ having neighbourhoods with 
coordinates $x_i$ vanishing at $P_i$, for $i=1,2$, respectively.
Out of these data one may define families of Riemann surface $C$ by considering the 
annuli $A_i=\{Q_i\in D_i;r_i<|x_i(Q_i)|<R_i\}$, and identifying pairs of points 
$(Q_1,Q_2)$ satisfying $x_1(Q_1)x_2(Q_2)=q$ to get an annulus $A\subset C$.

One may note that a change of the parameter
$q$ is equivalent to a rescaling of either $x_1$ or $x_2$. It is a matter of convenience if 
one wants to use the parameter $q$, or prefers to set $q=1$ using one of the parameters
$r_i$ or $R_i$, for $i=1,2$ instead. This is  a simple example for a more general 
issue. Gluing two surfaces $C_i$ with choices of local coordinates around $P_i$ yields a surface 
$C$ having a complex structure which depends only very little on the choices we had made. Most of the possible
changes of the local coordinates 
around the points $P_i$ will not change the complex structure of the glued surface $C$.
Some of them will, as could be parameterised either by the parameter $q$, or left implicit 
in the choices of the radii $r_i$, $R_i$, for $i=1,2$.

This construction is easily extended to surfaces with flat bundles. If $\rho_i:\pi_1(C_i)\ra\mathrm{GL}(N,\BC)$
are the representations of $\pi_1(C_i)$ characterising flat bundles on 
$C_i$ for $i=1,2$, respectively, we may construct a representation $\rho:\pi_1(C)\ra\mathrm{GL}(N,\BC)$
as follows. Let us 
pick a base point in the annulus 
$A$, and pick systems of generators for $\pi_1(C_i)$, $i=1,2$. The union of the sets of generators
for  $\pi_1(C_i)$ forms a set of generators for $\pi_1(C)$. Without essential loss of generality 
we may assume that the sets of generators for  $\pi_1(C_i)$ contain curves $\ga_i$ contained in 
$A_i$ which get identified to the non-contractible curve $\ga$ around $A$ by gluing. 
The given representations $\rho_i$ will allow us to define a representation 
$\rho:\pi_1(C)\ra\mathrm{GL}(N,\BC)$ whenever the elements $\rho_1(\ga_1)$ and $\rho_2(\ga_2)$ coincide,
defining $\rho(\ga)=\rho_1(\ga_1)=\rho_2(\ga_2)$.

The resulting representation will, of course, not only depend on the 
conjugacy class of representations $\rho_i$, but on the choice of the representations $\rho_i$ itself. 
Restricting attention to representations $\rho_i$ for which $\rho(\ga)\equiv\rho_i(\ga_i)$,  is
diagonal for $i=1,2$, we get a residual freedom in the choice of  $\rho_i$ represented by 
conjugating $\rho_i(\de)$ for all $\de\in\pi_1(C_i)$ with the same diagonal matrix. The resulting ambiguity
can be described equivalently in terms of the solutions $\Psi_i(x)$, $x\in A$, of the 
Riemann-Hilbert problems defined by the flat bundles 
on $C_i$, $i=1,2$, by multiplying $\Psi_i(x)$ from the right by diagonal matrices. This amounts to
a change of normalisation of the elements $\psi_{ik}$, $k=1,\dots,N$, of a basis for the space of  solutions to 
$(\pa_x+A_i(x))\psi_i(x)=0$ associated to $\Psi_i(x)=(\psi_{i1}(x),\dots,\psi_{iN}(x))$, for $i=1,2$, respectively. 
One may notice that the issue of picking preferred trivialisations of the flat bundle
around the points $P_i$ is somewhat analogous to the issue of choosing local coordinates around these
points. 

\subsubsection{Gluing of conformal blocks}

The gluing operation has a simple 
counterpart on the level of the conformal blocks. Instead of describing it 
in full generality let us focus on the case of conformal blocks associated to two surfaces 
$C_1$ and $C_2$ each twisted by flat bundles $\rho_1$ and $\rho_2$. 
As explained above, we may describe conformal blocks on $C$
in terms the one-point localisation picture in terms of vectors 
$\mathfrak{f}_\Psi\in\CF$ attached to a point $P_{\infty}\in C$ and a 
local coordinate $x_{\infty}$ on the disc $D_{\infty}$ around $P_\infty$. Without essential loss of generality 
let us  consider the case where $P_{\infty}\in C_2$. As explained above, we may
represent the conformal block attached to $(C_2,\rho_2)$ in a two-point localisation 
in terms of an operator
$Y_{\Psi_2}:\CF_\si\ra \CF$ satisfying  the free fermion Ward identities
\rf{FFWI}. The conformal block attached to $(C_1,\rho_1)$ can be represented by a 
vector $\mathfrak{f}_1\in\CF_{\si}$. To the surface $C$ defined in the gluing 
construction we may then associate the states
$
Y_{\Psi_2}\,\mathfrak{f}_1\in\CF
$.
It is not hard to show that $Y_{\Psi_2}\,\mathfrak{f}_1$ satisfies the free fermion Ward identities defining 
conformal blocks on $C$ with flat bundle constructed by the gluing
construction described above. 

The behaviour under variations of 
$(C,P_\infty,x_\infty)$ can be found as follows.
Let $v$ be a vector field on $D_{\infty}$ representing a variation of $(C,P_\infty,x_\infty)$.
Our goal is to compute $\de_{[v]}Y_{\Psi_2}\mathfrak{f}_1^{}$. Having constructed $C$ by the gluing
construction allows us to represent any variation of $(C,P_\infty,x_\infty)$ as  the sum
of a variation $\de_{[v],2}$ of $(C_2,P_{\infty},x_{\infty},P_2,x_2)$ and
a variation $\de_{[v],1}$ of $(C_1,P_1,x_1)$, leading to 
$\de_{[v]}Y_{\Psi_2}\mathfrak{f}_1^{}=
(\de_{[v],2}Y_{\Psi_2})\mathfrak{f}_1^{}+Y_{\Psi_2}(\de_{[v],1}\mathfrak{f}_1^{})$.
There is a pair of vector fields $(v_\infty,v_2)$ which represents the deformation 
$\de_{[v],2}$ of $(C_2,P_{\infty},x_{\infty},P_2,x_2)$, and there is a vector field 
$v_1$ representing the deformation $\de_{[v],1}$ of 
$(C_1,P_1,x_1)$. We may assume that $v_2=v_1\equiv v_A$. Indeed, 
it follows from the Virasoro uniformisation that the analytic continuation 
of the vector field $v$, restricted to $A$ can be used to define both  $(v_\infty,v_2)$ and $v_1$.
This allows us to compute 
\begin{align}
\de_{[v]}Y_{\Psi_2}\mathfrak{f}_1^{}&=
(\de_{[v],2}Y_{\Psi_2})\mathfrak{f}_1^{}+Y_{\Psi_2}(\de_{[v],1}\mathfrak{f}_1^{}) \notag\\
&=\big(T_\infty[v]Y_{\Psi_2}-Y_{\Psi_2}T_A[v_A]\big)\mathfrak{f}_1+Y_{\Psi_2} 
\big(T_A[v_A]\mathfrak{f}_1\big)
=T_\infty[v]Y_{\Psi_2}\,\mathfrak{f}_1.
\end{align}
This shows that the vector $Y_{\Psi_2}\mathfrak{f}_1^{}$ is flat with respect to the canonical connection
on the line bundle of conformal blocks associated to $(C,P_\infty,x_\infty)$.

\subsection{Cutting}

For an oriented  simple closed curve $\ga$ on a Riemann surface $C$ let $A\equiv A_\ga\subset C$ be an annulus 
in $C$ containing $\ga$ in its interior bounded by curves $\ga_1$, $\ga_2$ isotopic to $\ga$. The orientation of $\ga$ allows us to distinguish an ``incoming'' boundary, w.l.o.g. represented by $\ga_1$, from the ``outgoing'' boundary  $\ga_2$.
We may define $C_1'$ to be the connected component of the 
surface obtained by cutting along $\ga_2$ which contains $A$. $C_2'$ will be defined in the same way 
with $\ga_1$ taking the role of $\ga_2$. A suitable choice of local coordinate $x$ in $A$ allows us to represent 
$A$ w.l.o.g. as $A\simeq \{x\in\BC; \rho_2>|x|>\rho_1\}$.
From $C_i'$ one may define punctured Riemann surfaces $C_i$, $i=1,2$,
by gluing punctured discs 
to the boundary components that have been created by the cutting operation. 

It is easy to extend the cutting operation to surfaces $C$ with flat bundles. 
To this aim one may define flat bundles on $C_i$, for $i=1,2$, from the representations of $\pi_1(C_i)$
defined from the given representation $\rho:\pi_1(C)\ra \mathrm{GL}(N,\BC)$
by choosing a base-point in $A$, and restricting $\rho$ to generators of $\pi_1(C)$ 
represented by curves  in 
$C_i'$, for $i=1,2$, respectively. We will restrict attention to flat bundles having diagonalisable 
monodromy along the non-contractible cycle in $A$, with eigenvalues being 
$e^{2\pi \mathrm{i}\si_k}$, $k=1,\dots,N$.  The resulting flat bundles on $C_i$, $i=1,2$, will then admit solutions
of the Riemann-Hilbert problem having a
regular singularity at the punctures $P_i$ contained in $D_i$, characterised 
by the diagonal matrix $\mathsf{d}=\mathrm{diag}(\si_1,\dots,\si_N)$, with $0\leq\mathrm{Re}(\si_k)<1$ for $k=1,\dots,N$.

It is clear that cutting represents an inverse to the gluing operation. 
We may thereby represent the vector $\mathfrak{f}_\Psi$ as 
$\mathfrak{f}_\Psi=Y_{\Psi_2}\mathfrak{f}_{\Psi_1}$, leading 
to the following representation of the partition function
\begin{equation}\label{partfactor}
Z_C(\Psi)\,=\,\langle\,\mathfrak{f}_{\Psi_2}^{\ast}\,,\,\mathfrak{f}_{\Psi_1}^{}\rangle\,.
\end{equation}
As explained in \cite{CPT}, one may represent the right hand side of \rf{partfactor}
as a Fredholm determinant of the following form
\begin{equation}\label{Fredholm-appendix}
Z_C(\Psi)\,=\,\langle\,\mathfrak{f}_{\Psi_2}^{\ast}\,,\,\mathfrak{f}_{\Psi_1}^{}\rangle
=\mathrm{det}_{\CH_+}^{}(1+\mathsf{B}_{\Psi_2}\mathsf{A}_{\Psi_1}),
\end{equation}
where $\mathsf{A}_{\Psi_1}$ and $\mathsf{B}_{\Psi_2}$ are integral 
operators constructed from solutions $\Psi_1$ and $\Psi_2$ to the Riemann Hilbert problems
defined by flat bundles on the surfaces $C_1$ and $C_2$ used in the gluing construction, 
respectively. It is in this way becoming clear that the compositions 
appearing in the gluing construction of free fermion conformal blocks are analytically 
well-defined. Whenever there is an explicit solution to the 
Riemann Hilbert problems
on $C_1$ and $C_2$, one may use \rf{Fredholm-appendix} to get combinatorial expansions 
of the isomonodromic tau-function including the expansions
computed explicitly in \cite{GL,GL17}.

Recursive use of this construction yields Fredholm determinant representations for 
the  free fermion conformal blocks associated to any
pants decomposition. Conformal blocks associated to different pants decompositions are related by
analytic functions of the complex structure moduli of $C$, and of the flat bundle $\rho$ defining 
the conformal block.

\section{Relation to the Rogers dilogarithm}\label{app:Rogers}
 \setcounter{equation}{0}
 
As remarked above, the FG-FG transition has a kind of universal status, underlying FG-FN transitions as well as FN-FN transitions.
This makes it especially suited to discuss generalizations of the paradigm relating theta series expansions to specific choices of coordinates.
Here we consider a generalization induced by a change of normalisation for functions $\CT(x, y)$ defined by $\mathfrak{T}(x, y)=e^{\pi\mathrm{i}x y}\CT(x, y)$.
This choice will turn out to give hints in the direction of possible relations to the 
description of NS5-brane corrections to the geometry of 
 hypermultiplet moduli spaces of type II string theory reviewed in \cite{AMPP}.

The functions $\mathfrak{T}(x, y)$ have the following quasi-periodicity properties 
\begin{equation}
\mathfrak{T}(x+m, y+ n)=e^{\pi\mathrm{i}(n x - m y + mn)}\,\mathfrak{T}(x, y).
\end{equation}
The transition functions induced by changes of coordinates are then 
\begin{equation}
\mathfrak{T}(x, y)=\Phi(x,x')\,\mathfrak{T}'(x', y'),
\end{equation}
where 
\begin{equation}
\Phi(x,x')=e^{\pi\mathrm{i}(x y-x' y')}F(x,x').
\end{equation}
The resulting expression for $\Phi(x,x')$ can be simplified considerably by using \cite{Ad}
\begin{equation}
\log\frac{G(1+z)}{G(1-z)}=-\frac{1}{2\pi}\mathrm{Cl}_2(2\pi z)-z\log\frac{\sin(\pi z)}{\pi},
\end{equation}
together with Kummer's relation 
\begin{equation}
\mathrm{Cl}_2(2\pi z)=\frac{1}{\mathrm{i}}\big(\mathrm{Li}_2(e^{2\pi\mathrm{i}\,z})-\zeta(2)+\pi^2 z(1-z)\big),
\end{equation}
and finally the relation\footnote{It is important to note that this equation holds for $(x,y)$ and $(x',y')$ related by the change of variables in Section \ref{FG-to-FG}, in equations \eqref{Dehn} and \eqref{Logs}.
}
\begin{equation}
e^{\pi\mathrm{i}(x y-x' y')}=\exp(-2\pi\mathrm{i}xx'+x'\log(1-e^{2\pi \mathrm{i}x'})),
\end{equation}
following from \rf{Dehn}, \rf{Logs}. Taken together we arrive at 
\begin{equation}
\log\Phi(x,x')=-\frac{1}{\pi\mathrm{i}}L(e^{2\pi\mathrm{i}\,x'})-\frac{\pi\mathrm{i}}{6},
\end{equation}
where $L(z)$ is the Rogers dilogarithm
\begin{equation}
L(z)=\mathrm{Li}_2(z)+\frac{1}{2}\log(z)\log(1-z).
\end{equation}
Transition functions of this type have appeared in \cite{AP}. We take this observation as a hint that 
our line bundle $\CL_\Theta$ might be related to the one discussed in \cite[Section 5.3]{AMPP}.



\end{document}